\documentclass[aps,pre,reprint,superscriptaddress,showpacs,amsmath,floatfix]{revtex4-1} 
\usepackage{mathtools}
\usepackage{amssymb}
\usepackage{mathrsfs}
\usepackage{bm}
\usepackage{bbm}
\usepackage{graphicx}
\usepackage{multirow}
\usepackage{color}	
\usepackage{booktabs}
\usepackage{enumitem}
\usepackage{nameref}
\usepackage[colorlinks=true,linkcolor=blue,urlcolor=blue,citecolor=blue,anchorcolor=blue]{hyperref}
\usepackage{algorithm}
\makeatletter
\renewcommand{\ALG@name}{Procedure}
\makeatother
\usepackage{algpseudocode}

\newcommand{\gv}[1]{\ensuremath{\mbox{\boldmath$ #1 $}}} 

\makeatletter
\g@addto@macro\normalsize{%
  \setlength\abovedisplayskip{5pt}
  \setlength\belowdisplayskip{5pt}
  \setlength\abovedisplayshortskip{5pt}
  \setlength\belowdisplayshortskip{5pt}
}
\makeatother

\setlist[itemize]{noitemsep, topsep=0pt}
\setlist[enumerate]{noitemsep, topsep=0pt}

\makeatletter
\newcommand{\labitem}[2]{%
\def\@itemlabel{\textcolor{black}{#1}}
\item
\def\@currentlabel{#1}\label{#2}}
\makeatother

\begin{document}

\title{\texorpdfstring{Threefold way to the dimension reduction of dynamics on networks:\\ an application to synchronization}{Lg}}

\author{Vincent Thibeault}\email[]{vincent.thibeault.1@ulaval.ca}
\affiliation{D\'epartement de physique, de g\'enie physique et d'optique,
Universit\'e Laval, Qu\'ebec (Qu\'ebec), Canada G1V 0A6}
\affiliation{Centre interdisciplinaire en mod\'elisation math\'ematique, Universit\'e Laval, Qu\'ebec (Qu\'ebec), Canada G1V 0A6}
\author{Guillaume St-Onge}
\affiliation{D\'epartement de physique, de g\'enie physique et d'optique,
Universit\'e Laval, Qu\'ebec (Qu\'ebec), Canada G1V 0A6}
\affiliation{Centre interdisciplinaire en mod\'elisation math\'ematique, Universit\'e Laval, Qu\'ebec (Qu\'ebec), Canada G1V 0A6}
\author{Louis J. Dub\'{e}}
\affiliation{D\'epartement de physique, de g\'enie physique et d'optique,
Universit\'e Laval, Qu\'ebec (Qu\'ebec), Canada G1V 0A6}
\affiliation{Centre interdisciplinaire en mod\'elisation math\'ematique, Universit\'e Laval, Qu\'ebec (Qu\'ebec), Canada G1V 0A6}
\author{Patrick Desrosiers}\email[]{patrick.desrosiers@phy.ulaval.ca}
\affiliation{D\'epartement de physique, de g\'enie physique et d'optique,
Universit\'e Laval, Qu\'ebec (Qu\'ebec), Canada G1V 0A6}
\affiliation{Centre interdisciplinaire en mod\'elisation math\'ematique, Universit\'e Laval, Qu\'ebec (Qu\'ebec), Canada G1V 0A6}
\affiliation{Centre de recherche CERVO, Qu\'ebec (Qu\'ebec), Canada G1J 2G3}


\begin{abstract}

Several complex systems can be modeled as large networks in which the state of the nodes continuously evolves through interactions among neighboring nodes, forming a high-dimensional nonlinear dynamical system. 
One of the main challenges of Network Science consists in predicting the impact of network topology and dynamics on the evolution of the states and, especially, on the emergence of collective phenomena, such as synchronization.  
We address this problem by proposing a Dynamics Approximate Reduction Technique (DART) that maps high-dimensional (complete) dynamics unto low-dimensional (reduced) dynamics while preserving the most salient features, both topological and dynamical, of the original system. 
DART generalizes recent approaches for dimension reduction by allowing the treatment of complex-valued dynamical variables, heterogeneities in the intrinsic properties of the nodes as well as modular networks with strongly interacting communities. 
Most importantly, we identify three major reduction procedures whose relative accuracy depends on whether the evolution of the states is mainly determined by the intrinsic dynamics, the degree sequence, or the adjacency matrix.
We use phase synchronization of oscillator networks as a benchmark for our threefold method. We successfully predict the synchronization curves for three phase dynamics (Winfree, Kuramoto, theta) on the stochastic block model. 
Moreover, we obtain the bifurcations of the Kuramoto-Sakaguchi model on the mean stochastic block model with asymmetric blocks and we show numerically the existence of periphery chimera state on the two-star graph. 
This allows us to highlight the critical role played by the asymmetry of community sizes on the existence of chimera states. Finally, we systematically recover well-known analytical results on explosive synchronization by using DART for the Kuramoto-Sakaguchi model on the star graph.
Our work provides a unifying framework for studying a vast class of dynamical systems on networks.

\end{abstract}


\maketitle




\section{Introduction}

Complex systems are characterized by the emergence of macroscopic phenomena that cannot be explained by the properties of its constituents taken independently \cite{Mitchell2009, Charbonneau2017}.
Synchronization is an archetypal example where collective movements emerge from the interactions between the oscillators of a system \cite{Boccaletti2018, Pikovsky2003}.
The relationship between the interactions in a complex system and its capacity to synchronize was found to be rich and subtle in many fields of applications, including physics \cite{strogatz2003, Turtle2017}, neurosciences \cite{Izhikevich2007, laurent2002, Fell2011, DiSanto2018, Lynn2019}, and ecology \cite{potts1984, Vasseur2009, Vicsek2012, Deacy2017, Couzin2018}. Phase dynamics on networks give insights on this complex relationship: they model the oscillations in a simple way and networks encode the underlying structure of the systems \cite{Newman2018}.

However, the large number of dimensions $N \gg 1$ of the system, the non-linearity of phase dynamics, and the lack of symmetries in complex networks often prevent any thorough mathematical exploration of the coupled differential equations.
Moreover, it is often more desirable to describe the emergent behavior of the system with a certain observable rather than the microscopic states.
For these reasons, it is often more practical and intuitive to find a dynamical system with a reduced number of dimensions $n \ll N$ that describes the temporal evolution of the synchronization observable of the system. By doing so, the mathematical analysis, the computational cost, and the interpretation of the behavior of the complex system are simplified.

Multiple attempts have been made to obtain such lower-dimensional descriptions. In 1994, Watanabe and Strogatz successfully transformed a $N$-dimensional Kuramoto dynamics with identical oscillators and all-to-all coupling into an exact 3-dimensional reduced dynamics with $N-3$ constant of motions \cite{Watanabe1994}. 
In 2008, Ott and Antonsen introduced an Ansatz which allowed them to write a reduced dynamics in the limit $N \to \infty$  for the original Kuramoto model and some of its extensions (e.g., with external driving, communities of oscillators, and time-delayed coupling) \cite{Ott2008}. 
It was later shown that the Ott-Antonsen Ansatz can be applied to other phase dynamics, such as the theta model \cite{Luke2013,Bick2020}, and extended, with a circular cumulant method, to noisy phase dynamics \cite{Tyulkina2018}. 

Despite these advances, reducing the number of dimensions of a dynamical system while preserving its network's structural properties remains a very challenging task. Real complex networks are finite and it is known that finite-size effects shape synchronization transitions \cite{Rodrigues2016}. Moreover, their properties, such as the degree of their nodes, are often heterogeneously distributed. Therefore, all-to-all couplings are not suited to describe the rich behavior of synchronization dynamics on complex networks.

Attempts have been made to reduce the dimensions of dynamical systems by considering only structural features such as degrees \cite{Gao2016, Jiang2018}, symmetries \cite{Cho2017}, and spectrum \cite{Jiang2018, Laurence2019}. However, they generally do not consider variations in the intrinsic dynamics of the nodes or they are only suited for dynamics with specific coupling functions. There are also multiple approaches to coarse-grain synchronization dynamics involving the Laplacian matrix \cite{Moon2006, *Rajendran2011, Gfeller2007, Gfeller2008}. However, the methods (equation-free) are not mathematically systematic in general, i.e., they use computer algorithms to get reduced graphs instead of getting a reduced dynamical system. Besides, the whole field of model order reduction or reduced order models is very mature. Yet, this field mainly focus on engineering and linear control problems, the dynamics generally do not involve networks, and the reduction methods often require information on the time series of the complete dynamical system (e.g., with proper orthogonal decomposition) \cite{Antoulas2005, Kramer2019}.

The purpose of this paper is to develop a Dynamics Approximate Reduction Technique (DART) and to use it for predicting phase synchronization regimes. We first present the structural and dynamical setup for the paper (Sec.~\ref{sec:struct-dyn}). We then describe DART along with target procedures and we apply it to phase dynamics on networks to clarify the effects of reducing the number of dimensions (Sec.~\ref{sec:dart}). Finally, we use the reduced dynamics to analyze the effects of the graph structure on the synchronization regimes of the Kuramoto-Sakaguchi model, such as chimera states and explosive synchronization (Sec.~\ref{sec:chimeras}). We gather our conclusions in Sec.~\ref{sec:conclusion} and add a number of appendices to provide details on the analysis, the numerical implementations, and the application of DART to the case of real dynamics on networks with
more than one equation per node.

\section{Structural and dynamical setup}
\label{sec:struct-dyn}

In this section, we introduce preliminary definitions and notations. In particular, we define the modular networks and the phase dynamics that will be used throughout the paper.  Note that from now on,  the words ``network'' and ``graph'' will be considered as \mbox{exact~synonyms}.  

\subsection{Modular graphs}
\label{subsec:modular_graph}
Let us consider the graph $\mathcal{G} = (\mathcal{V},\mathcal{E})$, where \mbox{$\mathcal{V}=\{1,2,\ldots N\}$} is the set of vertices (nodes) and $\mathcal{E}$ is the set of edges (links or connections).
The corresponding adjacency matrix, \mbox{$A= (A_{jk})_{j,k=1}^N$}, is such that the element $A_{jk}$ equals one if node $k \text{ connects to node }j$ and zero otherwise. All graphs considered in the paper are undirected, so all adjacency matrices are symmetric.  

In this paper, a graph is said to be modular when its vertices are partitioned into disjoint blocks (groups, communities, or modules). Each block is a set that contains vertices with similar properties and each vertex belongs to one and only one block. Let $q$ be the number of blocks.  Moreover, for $\mu\in\{1,\ldots, q\}$, let $B_\mu$ denote the $\mu$-th block. Then, the ordered set  $\mathscr{B} = \big(B_1,...,B_q\big)$ provides a partition of $\mathcal{V}$ that unequivocally describes the modular structure of the graph. Moreover, $\mathscr{B}$ induces the surjection \mbox{$s: \mathcal{V} \to \{1,..., q\}$} that assigns each vertex to its corresponding block. Note that if $N_{\mu}$ is equal to the size of the $\mu$-th block, then $N = \sum_{\mu=1}^{q} N_{\mu}$. 

\begin{figure}[t] 
 \centering    
   \includegraphics[width=1\linewidth, clip=true]{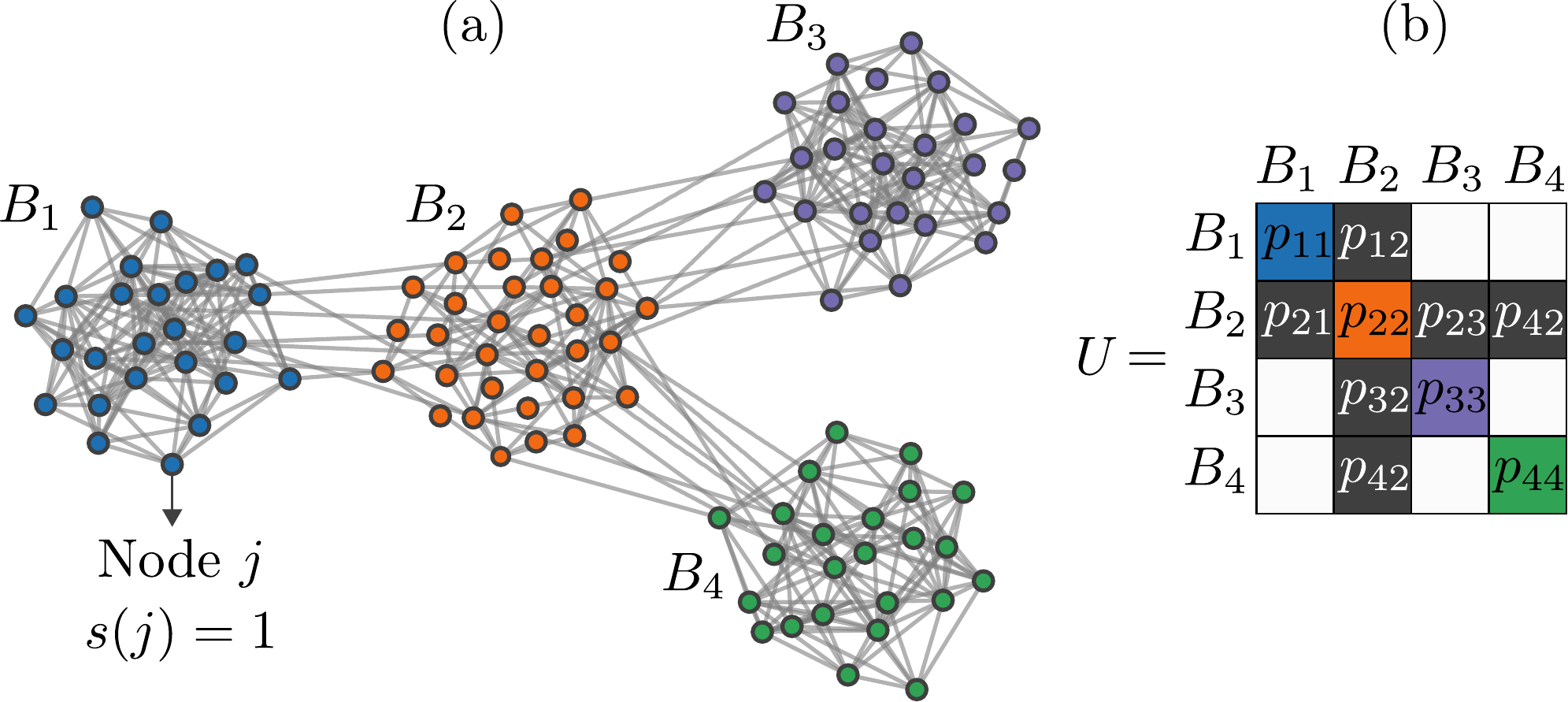}
\caption{(Color online) (a) A modular graph with $q = 4$ blocks. (b) The matrix of probabilities $U=(p_{\mu\nu})_{\mu,\nu=1}^q$ for the modular graph in (a). The white squares indicate matrix elements with value 0.}
   \label{fig:modulargraph}
 \end{figure}

An example of a modular graph with $q=4$ blocks is displayed in Fig.~\ref{fig:modulargraph}~(a). The vertices inside each block are densely connected, while the pairs of nodes belonging to different blocks are sparsely connected.  The opposite situation also occurs, as in star and two-star graphs.  Indeed, in a star graph with $N$ nodes, one node (the core) is connected to all the other $N-1$ nodes (the periphery) while the latter do not connect among themselves.  Yet, this corresponds to a modular structure with $q=2$ blocks: the core $B_{1} =\{1\}$ and the periphery $B_{2} =\{2,\ldots, N\}$.  Now, let $\mathcal{G}_1$ and $\mathcal{G}_2$ be two separate star graphs with  $N_{p_1}+1$ and $N_{p_2}+1$ nodes, respectively. Connecting the cores of $\mathcal{G}_1$ and $\mathcal{G}_2$ produces a two-star graph that is partitionable into $q=4$ blocks, with two blocks for cores, $B_{1} =\{1\}$ and $B_{3} = \{N_{p_1}+2\}$,  and two blocks for peripheries,  $B_{2} =\{2, ..., N_{p_1}+1\}$ and $B_{4} = \{N_{p_1}+3, ..., N\}$.  

A random modular graph is a set of modular graphs equipped with a probability distribution. A \textit{stochastic block model} (SBM) is a type of random modular graph whose probability distribution depends on a set of probabilities related to the block structure and that guarantees the independence of all possible edges. To be more specific, let $\mathscr{B}$ be as above and let  \[\mathscr{P} = \Big(p_{\mu\nu}\in [0,1]\:\Big|\:1 \leq \mu \leq \nu \leq q\Big),\] which is an ordered set of probabilities.  Then, SBM$(\mathscr{B}, \mathscr{P})$ is such that the probability of drawing a graph with adjacency matrix $A$ is
\begin{equation}
P(A)=\prod_{1\leq \mu\leq\nu\leq q}\,\,\prod_{\substack{i\in B_\mu, \, j\in B_\nu\\i<j}}p_{\mu\nu}^{A_{ij}}(1-p_{\mu\nu})^{1-A_{ij}}.
\label{eq:probA}\end{equation}
The parameter $p_{\mu\nu}$ can thus be interpreted as the probability for a node in $B_{\mu}$ to  connect with a node in $B_{\nu}$. Note that it is often more suggestive to combine the probabilities $p_{\mu\nu}$ into the symmetric matrix $U = (p_{\mu\nu})_{\mu,\nu=1}^q$ as in Fig.~\ref{fig:modulargraph}~(b). 

In this paper, we focus on random modular graphs with $q=2$ blocks, so $\mathscr{B} = \big(B_1, B_2\big)$. We additionally impose the equality \mbox{$p_{12} = p_{21} = p_{\text{out}}$}. Besides, if \mbox{$p_{11} = p_{22} = p_{\text{in}}$}, the SBM is called the planted partition model \cite{Condon2001, Young2017}. Two extreme cases are also of special interest: \mbox{$p_{\text{in}} = p_{\text{out}}=p$} is equivalent to the Erd\H{o}s-R\'enyi model (also known as the Gilbert model) $\mathcal{G}(N, p)$ and $p_{\text{in}} = 0$ is the random bipartite graph.

\subsection{Phase dynamics}
\label{subsec:phase_dyn_synchro}

The dynamics of an oscillator is described by a dynamical system which possesses at least one periodic orbit (cycle) for a certain set of parameters. When multiple oscillators interact, their trajectories are perturbed from their periodic state. For small perturbations, the oscillators stay in the basin of attraction of their limit cycles and the variation in the oscillation amplitudes are small. The whole dynamics then becomes mainly controlled by a system of ordinary differential equations that only involve the phases of the oscillations \cite{Pikovsky2003, Izhikevich2007, Pietras2019}. 
For a network with adjacency matrix $A = (\,A_{jk}\,)_{j,k=1}^N$, this system typically looks like
\begin{equation}
        \dot{\theta}_j = f(\theta_j) + \omega_j\,g(\theta_j) +  \sum_{k = 1}^{N} A_{jk}\, h(\theta_j, \theta_k),
        \label{eq:phasedyn}
\end{equation}
where $j\in\{1,\ldots, N\}$. Moreover, $\theta_j$ is a real-valued function such that $\theta_j(t)$ gives the phase of oscillator $j$ at time $t \in \mathbb{R}$,  $\dot{\theta}_j = {\mathrm{d}\theta_j}/{\mathrm{d}t}$ is the instantaneous phase velocity of oscillator $j$, $\omega_j$ is a dynamical parameter related to oscillator $j$, $f$ and $g$ are real analytic functions representing the intrinsic dynamics of each oscillator, and $h$ is a real analytic function describing the coupling among the oscillators. The functions $f$, $g$, and $h$ are assumed to be periodic with period $2\pi$. We also assume that their Fourier series are finite, which will be useful in Sec.~\ref{subsec:derivation}.

In 1967, Winfree proposed one of the first models of coupled phase oscillators in which synchronization is possible \cite{Winfree1967}. The coupling function of his model takes the form $h(\theta_j, \theta_k) = h_1(\theta_j)h_2(\theta_k)$, where $h_1(\theta_j)$ encodes the phase response of oscillator $j$ to perturbations $\sum_{k=1}^N A_{jk} h_2(\theta_k)$ caused by its oscillating neighbors $k$ \cite{Gallego2017}. 
Following Ref.~\cite{Ariaratnam2001}, we choose \mbox{$h_1(\theta_j) = - \sin\theta_j$} and \mbox{$h_2(\theta_k) =\sigma \big(1 + \cos\theta_k\big)/N$}, where $\sigma\in \mathbb{R}_+$ is a coupling constant. Thus, in this version of the \emph{Winfree model}, the rate of change of the $j$-th phase is
\begin{equation}
    \dot{\theta}_j = \omega_j - \frac{\sigma}{N}\sin\theta_j\sum_{k = 1}^{N} A_{jk}\, (1+\cos\theta_k).
    \label{winfree}
\end{equation}
where $\omega_j$ is the natural frequency of oscillator $j$, \mbox{$f(\theta_j)=0$}, and \mbox{$g(\theta_j) = 1$}.

Inspired by the work of Winfree, Kuramoto introduced another model of non-linearly coupled phase oscillators in 1975 that has now become a classic for the study of synchronization in populations of oscillators \cite{Kuramoto1975}. In the network version of the \emph{Kuramoto model}, the $j$-th phase evolves according to 
\begin{equation}
    \dot{\theta}_j = \omega_j +  \frac{\sigma}{N}\sum_{k = 1}^{N} A_{jk}\, \sin(\theta_k - \theta_j).
    \label{kuramoto}
\end{equation}
Despite the simple form of its dynamics, the Kuramoto model is quite useful \cite{Moon2006}. Indeed, for weak couplings and a phase coupling function $h$ well approximated by its first harmonics, a large class of phase oscillator dynamics can be described by the Kuramoto model \cite{Pietras2019}. The model is relevant for the study of Josephson junctions \cite{Wiesenfeld1996}, nanoelectromechanical oscillators \cite{Matheny2019}, and exhibits a rich variety of dynamical behaviors when transposed on complex networks \cite{Rodrigues2016}. Adding a global phase lag $\alpha$ between the oscillators of the Kuramoto model lead to the \emph{Kuramoto-Sakaguchi model} \cite{Sakaguchi1986} whose dynamical equations are
\begin{equation}
    \dot{\theta}_j = \omega_j +  \frac{\sigma}{N}\sum_{k = 1}^{N} A_{jk} \sin(\theta_k - \theta_j - \alpha),
    \label{sk}
\end{equation} 
for $j\in\{1,\ldots, N\}$. This model possesses partially synchronized solutions, called chimera states \cite{Abrams2008}, which will be investigated in Sec.~\ref{sec:chimeras}.

\begin{figure}[b] 
 \centering
   \includegraphics[width=1\linewidth, clip=True]{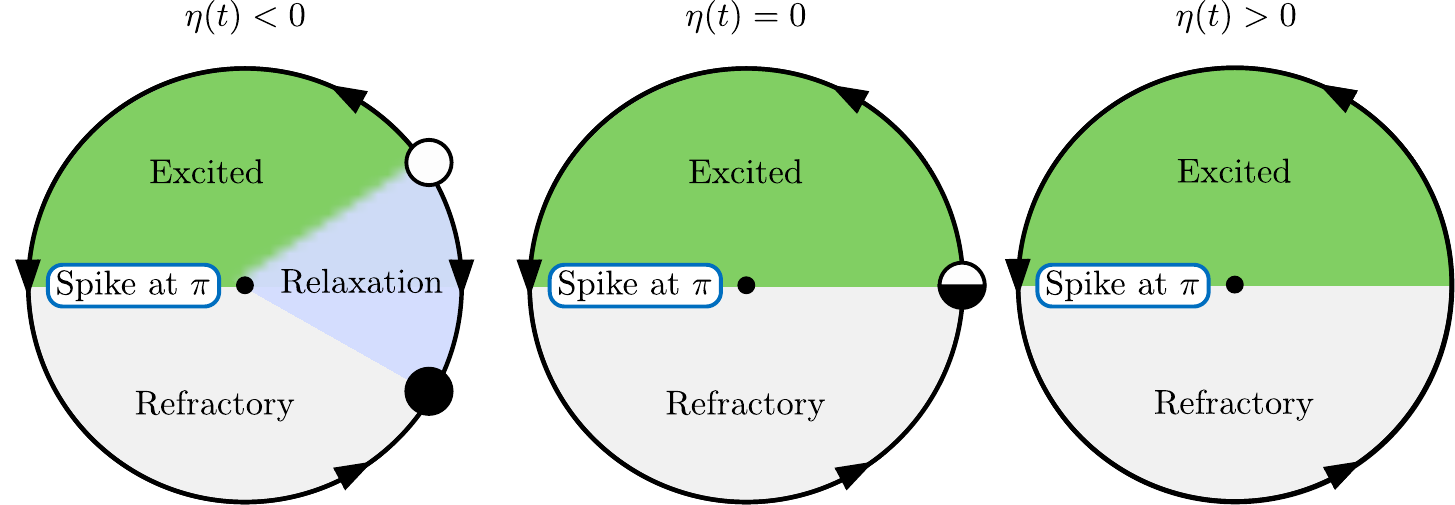}
   \caption{(Color online) Schematization of the dynamical states in the theta model.  The interaction term $\eta(t)$ increases from left to right, producing a SNIC bifurcation. The white dot is an unstable equilibrium point, the black dot is a stable equilibrium point, while the black and white dot is a half-stable equilibrium point. The flow of the phase on the circle is represented by the arrows.}
   \label{fig:SNIC}
 \end{figure}

Partly due to the inherent complexity of neural systems and their nonlinear oscillatory behavior, much research has been conducted to obtain simplified models of neurons that capture their main dynamical properties \cite{Izhikevich2004, Ermentrout2010}. For example, in 1986, the \emph{theta model} (also known as the Ermentrout-Kopell model) was introduced to provide a low-dimensional dynamical description of parabolic bursting while preserving some basic characteristics of higher-dimensional models that belong to the Hodgkin-Huxley family \cite{Ermentrout1986}. In the theta model, the rate of change of the $j$-th oscillator is
\begin{equation}
    \dot{\theta}_j = (1-\cos \theta_j)+(1+\cos \theta_j)\eta_j,
    \label{theta}
\end{equation}
where $\eta_j$ stands for the interaction term defined as
\begin{equation*}
    \eta_j = \omega_j + \frac{\sigma}{N}\sum_{k=1}^{N}A_{jk}(1 - \cos \theta_k).
    \label{current}
\end{equation*}
In the last equation, $\omega_j$ is interpreted as the current injected into neuron $j$.  The above model essentially differs from the quadratic integrate-and-fire model by a change of variables $V_j = \tan(\theta_j/2)$, where $V_j$ is the membrane potential of neuron $j$ \cite{Montbrio2015}. Equation~\eqref{theta} is also the normal form for the saddle-node on invariant circle (SNIC) bifurcation illustrated in Fig.~\ref{fig:SNIC}.

\section{\texorpdfstring{DART: Dynamics Approximate\\$\qquad\qquad\quad$Reduction Technique}{Lg}}
\label{sec:dart}

We now introduce DART that approximately reduces $N$-dimensional dynamical systems, akin to Eq.~\eqref{eq:phasedyn}, to new $n$-dimensional ones with $n\ll N$. For this, we first define $n$ new linearly independent dynamical observables $Z_\mu$ for $\mu \in\{1,\ldots, n\}$. We force their time evolution to obey a closed system of $n$ ordinary differential equations. In turn, this constraint leads us to impose three compatibility conditions, taking the form of three systems of linear algebraic equations, which, in general, cannot be satisfied simultaneously. We nevertheless clarify the circumstances in which a particular condition should be favored and explain  how to get satisfactory solutions, allowing us to successfully apply the method to phase dynamics and to predict their synchronization behavior on modular graphs.  

\subsection{Definition of the observables}
\label{subsec:observables}

Inspired by Koopman's seminal work on ergodic theory (e.g., see \cite{Budisic2012}), we define an observable of a dynamical system as a complex-valued function on its phase space. We restrict the observables to be smooth functions. Center of mass, total kinetic energy, and linear momentum are classical examples of smooth observables. All such observables form a countably infinite-dimensional complex algebra $\mathscr{O}$.  

In that context, the goal of dimension reduction consists in selecting $n\ll N$ observables in $\mathscr{O}$ whose dynamics is determined, at least approximately, by a closed system of $n$ ordinary differential equations. Selecting the right set of observables is a difficult task.  Yet, as argued in \cite{Laurence2019}, this task is greatly simplified if, rather than looking for observables in the whole algebra $\mathscr{O}$,  we limit our search to \emph{linear} observables, which form a $N$-dimensional complex vector space.

However, when studying the synchronization of oscillators, linear observables such as $\sum_{j=1}^Nc_j\theta_j$, where $c_j\in\mathbb{C}$ for each $j$, have limited value.  To get observables that quantify synchronization, we must first make the change of variables
\begin{equation} \theta_j\longmapsto z_j = e^{\mathrm{i}\theta_j},\qquad j\in \{1,\ldots,N\},\label{eq:def_z}
\end{equation}
which maps the phase of oscillator $j$ onto the unit circle $\mathbb{T}$ in the complex plane. This means that we should look for $n$-linear observables of the form 
\begin{equation}
    Z_{\mu} = \sum_{j=1}^N M_{\mu j}z_j, \quad \mu \in \{1, ..., n\},
    \label{mesoobs}
\end{equation}
where $M$ is a $n\times N$ matrix with real elements. In matrix form, 
\[\bm Z = M\bm z, \]
where
\[ \bm Z=\begin{pmatrix} Z_1\\\vdots\\ Z_n\end{pmatrix}\quad \text{and}\quad \bm z =  \begin{pmatrix}z_1\\\vdots\\ z_N\end{pmatrix}.\]
Choosing $n$ linear observables $Z_1\ldots Z_n$ is thus equivalent to choosing a matrix $M$. We call this matrix a ``reduction matrix'' since it reduces $N$ variables $z_j$ into $n$ variables $Z_{\mu}$.

To further restrict the set of linear observables, we impose two additional conditions: 
\begin{itemize}  
    \labitem{A}{itm:condA} The rank of matrix $M$ is $n$;
    \labitem{B}{itm:condB} Each row of $M$ is a probability vector, i.e., 
    \begin{itemize}   
        \labitem{B$_1$}{itm:condB1} $\sum_{j=1}^N M_{\mu j}=1$ for all $\mu$,
        \labitem{B$_2$}{itm:condB2} $M_{\mu j }\geq 0$ for all $\mu$ and $j$.
    \end{itemize}
\end{itemize}
These conditions can be reformulated as follows: 
\begin{itemize}  
    \item[A] The observables $Z_1,\ldots Z_n$ are linearly independent; 
    \item[B] Each observable $Z_\mu$ is a weighted average of the activity variables $z_1,\ldots, z_N$.
\end{itemize}

Condition~\ref{itm:condA} ensures that the dimension reduction is optimal in the sense that there is no redundancy among the observables. 

Condition~\ref{itm:condB} makes each observable easily interpre\-table, namely, it is possible to decompose each linear observable as
\begin{equation}
    Z_{\mu} = R_{\mu}e^{i\Phi_{\mu}},\quad R_\mu = |Z_\mu|,\quad \Phi_\mu=\text{arg}(Z_{\mu}).
    \label{eq:observables}
\end{equation}  
This second condition then directly implies that the inequality $R_\mu\leq 1$ holds and $R_\mu$ reaches its maximal value when, and only when, all phases $\theta_j$ are equal modulo $2\pi$. In other words, thanks to Condition~\ref{itm:condB}, each $R_\mu$ can be interpreted as a synchronization measure. Although recent works \cite{Schroder2017, Gallego2017} suggest that other observables may provide better quantitative measures of phase coherence, we use $R_\mu$ 
because of its simple properties and because it is easily obtained from the linear observable $Z_{\mu}$.

Let us illustrate Conditions~\ref{itm:condA}--\ref{itm:condB} with an example. First, suppose that for each $\mu\in \{1,\ldots, n\}$, $n_\mu$ nodes are selected to form the subset $O_{\mu} \subset \mathcal{V}$. Second, assume that $(O_{\mu})_{\mu=1}^n$ is a partition of $\mathcal{V}$. Third, set
\begin{equation}
M_{\mu j}=\frac{1}{n_\mu}\times \begin{cases}1,&j\in O_\mu,\\
                                             0,&j\notin O_\mu. \end{cases}
\label{eq:semi-ortho}
\end{equation}
Then, the matrix $M$ satisfies both Conditions~\ref{itm:condA}--\ref{itm:condB} with the corresponding observables
\begin{equation*}
    Z_{\mu} = \frac{1}{n_\mu}\sum_{j \in O_{\mu}}z_j,\quad  \mu\in \{1,\ldots, n\}.
    \label{typicalobservable}
\end{equation*}

In the last example, the reduction matrix $M$ of Eq.~\eqref{eq:semi-ortho} is such that if node $j$ contributes to the  observable $Z_{\mu}$, meaning $M_{\mu j}\neq 0$, then the same node does not contribute to any other observable $Z_{\nu}$ with $\mu \neq \nu$. This last property together with Condition~\ref{itm:condB} imply the following:
\begin{itemize} 

    \vspace{0.5em}
    
    \labitem{A'}{itm:condA'} $M$ is row-orthogonal, i.e., $MM^\top = I_{n\times n}$.
    
    \vspace{0.5em}
    
\end{itemize}
This is a stronger version of Condition~\ref{itm:condA}. Indeed, the orthogonality between each row of $M$ implies that the rows are linearly independent, so the rank is equal to $n$. Condition~\ref{itm:condA'} therefore implies Condition~\ref{itm:condA}.

Although Condition~\ref{itm:condA'} will not be strictly imposed in the paper, it will always be considered as desirable. The reason is quite simple: together, Conditions~\ref{itm:condA'} and \ref{itm:condB} induce a partition of the set of nodes into $n$ disjoint blocks, as in Sec.~\ref{subsec:modular_graph}. In other words, when the reduction matrix $M$ satisfies both Conditions \ref{itm:condA'} and \ref{itm:condB}, then $M$ also provides a modular structure and each $Z_\mu$ is interpretable as a weighted average inside the $\mu$-th block. This module, however, is not necessarily related to that of the network, since information has not yet been extracted from the adjacency matrix $A$.

In the following sections, we will only use linear observables satisfying at least Conditions~\ref{itm:condA}--\ref{itm:condB}. As a consequence, any row probability vector $\bm \ell = (\ell_\mu)_{\mu=1}^n$ ensures that 
\begin{equation}\label{eq:global_weights}
    \bm m = (m_j)_{j=1}^N =\bm \ell M,
\end{equation}
is a probability vector. Throughout the paper, we choose $\ell_{\mu}$ for all $\mu$ as the number of non-zero elements in row $\mu$ of $M$ divided by the total number of non-zero elements in $M$. The observables $Z_1,\ldots Z_n$ can therefore be linearly combined to produce a single weighted average of the activity variables $z_1,\ldots, z_N$: 
\begin{equation}
    Z =  \sum_{\mu=1}^n \ell_\mu Z_{\mu} = \sum_{j=1}^N m_j z_j.
    \label{eq:macroscopic_order_parameter}
\end{equation}
The observable $R = |Z|$ is then bounded as $0\leq R\leq 1$ and serves as a global measure of synchronization (i.e., order parameter). Additionally, we will often use the time-averaged measure of synchronization
\begin{equation}
    \langle R \rangle_{t} = \frac{1}{t_1 - t_0}\int_{t_0}^{t_1} R(\bm \theta(t))\,\text{d}t,
    \label{eq:avg_R}
\end{equation}
where we choose sufficiently large time values, $t_0$ and \mbox{$t_1 > t_0$}, such that $R(t)$ oscillates around a certain mean or reaches a stable solution. We will sometimes denote by $\langle R \rangle$ the average synchronization measure over multiple quantities (e.g., over random initial conditions or  various network realizations).

\subsection{Derivation of the reduced dynamical system}
\label{subsec:derivation}
We now obtain a reduced system of $n$ differential equations that describes the dynamics of the linear observables $Z_1,\ldots, Z_n$.
We summarize the main result in Table~\ref{tab:main_result} but the reader is invited to look at the derivation to better understand the quantities in play.

First, the $N$-dimensional system that we are going to reduce is Eq.~\eqref{eq:phasedyn} under the change of variables of Eq.~\eqref{eq:def_z}:
\begin{multline}
       \quad \dot{z}_j = F(z_j, \bar{z}_j) + \omega_j G(z_j, \bar{z}_j) \\
        +  \sum_{k = 1}^{N} A_{jk} H(z_j, \bar{z}_j, z_k, \bar{z}_k),\quad
        \label{eq:complex_complete_dyn}
\end{multline}
for $j\in\{1,\ldots, N\}$.  This system defines a dynamics on the complex torus $\mathbb{T}^N$. The functions $F$, $G$, and $H$ are directly related to the functions $f$, $g$ and $h$. They therefore inherit some of their properties. 

In particular, $F$ and $G$ are holomorphic functions with domain $\mathbb{C}^2$ and codomain $\mathbb{C}$ while $H$ is a holomorphic function from $\mathbb{C}^4$ to $\mathbb{C}$. For example, with the Kuramoto dynamics,
\begin{align*}
    h(\theta_j, \theta_k) &= \frac{\sigma}{N}\sin(\theta_k - \theta_j),\\
    H(z_j, \bar{z}_j, z_k, \bar{z}_k) &= \frac{\sigma}{2N}\left[z_k - z_j^2\bar{z}_k\right].
\end{align*}

Second, we use Eq.~\eqref{eq:complex_complete_dyn} to compute the time derivative of the observable $Z_\mu$ defined in Eq.~\eqref{mesoobs}:
\begin{multline}\label{eq:derivativeZ}
     \dot{Z}_{\mu} = \sum_{j=1}^N M_{\mu j}\, F(z_j, \bar{z}_j) + \sum_{j=1}^N M_{\mu j}\,\omega_j\, G(z_j, \bar{z}_j) \\
     + \sum_{j,k =1}^N M_{\mu j} A_{jk}H(z_j, \bar{z}_j, z_k, \bar{z}_k) 
\end{multline}
for all $\mu \in \{1,\ldots, n\}$.
 
Third, we take advantage of the fact that $F$, $G$, and $H$ are holomorphic functions to apply Taylor's theorem:
\begin{align*}
    F&(z_j, \bar{z}_j) = F(\beta_\mu, {\beta}'_\mu) + (z_j -\beta_{\mu})F_1 + (\bar{z}_j -{\beta}'_{\mu})F_2 + r^F_{\mu j}\\
    G&(z_j, \bar{z}_j) = G(\gamma_\mu, {\gamma}'_\mu) + (z_j -\gamma_{\mu})G_1 + (\bar{z}_j -{\gamma}'_{\mu})G_2 +r^G_{\mu j}\\
    H&(z_j, \bar{z}_j, z_k, \bar{z}_k) = H(\delta_\mu, {\delta}'_\mu, \epsilon_\mu, {\epsilon}'_{\mu}) \\&\quad\qquad\qquad\qquad\,\,\,\,+ (z_j -\delta_{\mu})H_1 + (\bar{z}_j -{\delta}'_{\mu})H_2\\ &\quad\qquad\qquad\qquad\,\,\,\,+ (z_k -\epsilon_{\mu})H_3 + (\bar{z}_k -{\epsilon}'_{\mu})H_4+r^H_{\mu j k}
\end{align*}
where $r^F_{\mu j}$, $r^G_{\mu j}$, $r^H_{\mu j k}$ are second order Lagrange remainders and $\beta_\mu, \gamma_\mu, \delta_\mu, \epsilon_\mu, \beta'_\mu, \gamma'_\mu, \delta'_\mu, \epsilon'_\mu$ are arbitrary complex numbers around which we apply the expansions. Also, $F_1$ and $F_2$ are the derivatives of $F$ with respect to the first and second arguments ($z_j$ and $\bar{z}_j$) respectively and they are evaluated at $(\beta_\mu, {\beta}'_\mu)$. The same applies to $G_1$, $G_2$, $H_1$, $H_2$, $H_3$, and $H_4$. The substitution of these Taylor expansions into  Eq.~\eqref{eq:derivativeZ} then leads to the equation
\begin{multline}
    \dot{Z}_{\mu} = F(\beta_\mu,  {\beta}'_\mu) + \Omega_{\mu} G(\gamma_\mu,  {\gamma}'_\mu) \\+ \kappa_{\mu}H(\delta_\mu, {\delta}'_\mu, \epsilon_\mu,  {\epsilon}'_\mu)+\Upsilon_{\mu}+\Xi_{\mu}, \phantom{\int}
    \label{eq:eq:general_reduced_equation_remainder}
\end{multline}
where $\Upsilon_{\mu}$ is a homogeneous polynomial of degree one in the variables $z_j-\beta_\mu$, $ \,\, \,\bar{z}_j-{\beta}'_\mu$, $\,\, \,z_j-\gamma_\mu$, $\,\, \,\bar{z}_j-{\gamma}'_\mu$, $\,$and so forth, \mbox{$\Xi_{\mu} = \sum_j (r_{\mu j}^F + r_{\mu j}^G) + \sum_{j,k} r_{\mu jk}^H$}, and we have defined the \mbox{parameters}
\begin{equation*}
\begin{aligned}[c]
\Omega_{\mu} = \sum_{j=1}^N M_{\mu j} \omega_j,
\end{aligned}
\quad\quad
\begin{aligned}[c]
\kappa_{\mu} = \sum_{j = 1}^N M_{\mu j} k_j,
\end{aligned}
\end{equation*}
with the in-degree of node $j$,
\begin{equation*}
k_j =\sum_{k=1}^N A_{jk}.
\end{equation*}

By introducing the ``dynamical parameter matrix'' \mbox{$W = \text{diag}(\omega_1,...,\omega_N)$} and the ``degree matrix'' \mbox{$K = \text{diag}(k_1,...,k_N)$}, the parameters $\Omega_{\mu}$ and $\kappa_{\mu}$ can be written in matrix form as in Eqs.~(\ref{eq:Omega}-\ref{eq:kappa}). These parameters represent weighted averages of the dynamical parameters $\omega_j$ and the in-degrees $k_j$, respectively.

Finally, we close the system of differential equations $\dot{Z}_{\mu}$, with $\mu \in\{1,\ldots,n\}$. We thus need to convert each equation of the form of Eq.~\eqref{eq:eq:general_reduced_equation_remainder} into an equation that involves only the linear observables $Z_1,\ldots,Z_n$ and their complex conjugates, without explicit reference to the variables $z_1,\ldots,z_N$ and their complex conjugates.  To do so, we impose three new conditions.  

\begin{enumerate}
\labitem{I}{itm:condI} Exact cancellation of the first-order terms: each polynomial $\Upsilon_{\mu}$ is identically zero;
\labitem{II}{itm:condII} Linearity of the dynamical variables: all the variables $\beta_\mu$, $\gamma_\mu$, $\delta_\mu$, $\epsilon_\mu$, $\beta'_\mu$, $\gamma'_\mu$, $\delta'_\mu$, $\epsilon'_\mu$ are linear combinations, with real coefficients, of the observables ${Z}_{1},\ldots, {Z}_{n}$ or their complex conjugates;
\labitem{III}{itm:condIII} Approximation at second order: each term $\Xi_{\mu}$ is neglected.
\end{enumerate}
Condition~\ref{itm:condI} readily provides formulae expressing $\beta_\mu$, $\gamma_\mu$, $\delta_\mu$, and $\epsilon_\mu$ as linear combinations of the $z_j$'s:
\begin{equation*}
\begin{aligned}[c]
\beta_\mu &= \sum_{j=1}^{N} M_{\mu j} z_j = Z_{\mu},\\
\delta_\mu &= \kappa_{\mu}^{-1}\sum_{j = 1}^{N} M_{\mu j} k_{j}z_j,
\end{aligned}
\quad\quad
\begin{aligned}[c]
\gamma_\mu &= \Omega_{\mu}^{-1}\sum_{j=1}^{N} M_{\mu j}\omega_j z_j,\\
\epsilon_\mu &=\kappa_{\mu}^{-1}\sum_{j,k = 1}^{N} M_{\mu j} A_{jk}z_k.
\end{aligned}
\end{equation*}
The equations for $\beta'_\mu$, $\gamma'_\mu$, $\delta'_\mu$, and $\epsilon'_\mu$ have the same form as the previous equations, but with the complex conjugate of $z_j$ and $z_k$. This implies that $\beta'_\mu$, $\gamma'_\mu$, $\delta'_\mu$, and $\epsilon'_\mu$ are the complex conjugate of $\beta_\mu$, $\gamma_\mu$, $\delta_\mu$, and $\epsilon_\mu$, respectively. 

Condition~\ref{itm:condII} requires to rewrite $\beta_\mu$, $\gamma_\mu$, $\delta_\mu$, and $\epsilon_\mu$ as linear combinations of the observables $Z_1,\ldots,Z_n$. This condition is directly satisfied for $\beta_\mu$, but not for the others. Satisfying Condition~\ref{itm:condII} for $\gamma_\mu$, $\delta_\mu$, and $\epsilon_\mu$ is the challenging step in our calculation. Yet, it can be done if 
\begin{align*}
    \sum_{j=1}^{N} M_{\mu j}\omega_jz_j &= \sum_{j = 1}^{N}\sum_{\nu=1}^{n} \mathcal{W}_{\mu \nu} M_{\nu j}z_j = \sum_{\nu=1}^{n} \mathcal{W}_{\mu \nu} Z_{\nu}, 
    \\
    \sum_{j = 1}^{N} M_{\mu j} k_{j}z_j &= \sum_{j = 1}^{N}\sum_{\nu=1}^{n} \mathcal{K}_{\mu \nu} M_{\nu j}z_j = \sum_{\nu=1}^{n} \mathcal{K}_{\mu \nu} Z_{\nu},
    \\
    \sum_{j,k = 1}^{N} M_{\mu j} A_{jk}z_k &= \sum_{k = 1}^{N}\sum_{\nu=1}^{n} \mathcal{A}_{\mu \nu} M_{\nu k} z_k = \sum_{\nu=1}^{n} \mathcal{A}_{\mu \nu} Z_{\nu},
\end{align*}
where we have introduced three unknown $n\times n$ matrices $\mathcal{W}$, $\mathcal{K}$, and $\mathcal{A}$. We observe that if the \emph{compatibility equations}~(\ref{freq_eq}-\ref{spec_eq}) are all satisfied, then the previous equations are satisfied along with Condition~\ref{itm:condII}. The consistency of these compatibility equations is a subtle issue that will be addressed in the next subsection. 

Condition~\ref{itm:condIII} guarantees that no further restriction is imposed on $\gamma_\mu$, $\delta_\mu$, and $\epsilon_\mu$.

Let us now state the main result of this paper, presented in Table~\ref{tab:main_result}. If the three compatibility equations are satisfied, then the linear observables $Z_1,\ldots, Z_n$ obey Eqs.~\eqref{eq:general_reduced_equation}, where the symbol~$\approx$ means ``equality up to second order corrections'' and where the variables $\gamma_\mu$, $\delta_\mu$, $\epsilon_\mu$ are the components of the vectors in Eqs.~(\ref{freq_constraint}-\ref{spec_constraint}). 

\onecolumngrid

\begin{table*}[b]
\caption{\label{tab:main_result} \textbf{Reduced dynamics obtained from DART.}}
\begin{ruledtabular}

\begin{tabular}{ c c c }

&  \multicolumn{1}{c}{\textbf{Differential equations}} & \vspace{-0.2cm}\\

\multicolumn{3}{c}{\parbox{15cm}{\begin{equation}  \dot{Z}_{\mu} \approx F(Z_\mu, \bar{Z}_\mu) + \Omega_{\mu}\, G(\gamma_\mu, \bar{\gamma}_\mu) + \kappa_{\mu}\,H(\delta_\mu, \bar{\delta}_\mu, \epsilon_\mu, \bar{\epsilon}_\mu) \label{eq:general_reduced_equation},\qquad \mu \in \{1, ..., n\} \end{equation}}}

\vspace{0.1cm}
\\

\textbf{Arguments of} $G$ \textbf{and }$H$  & \textbf{Parameters} & \textbf{Compatibility equations}
\vspace{-0.5cm}
\\

\parbox[t]{4cm}{\begin{align}
                             \gv{\gamma} &=  \mathcal{D}_{\Omega}\mathcal{W}\,\bm{Z}\label{freq_constraint}\\
                             \gv{\delta} &=  \mathcal{D}_{\kappa}\mathcal{K}\,\bm{Z} \label{deg_constraint}\\
                             \gv{\epsilon}  &= \mathcal{D}_{\kappa}\mathcal{A}\,\bm{Z} \label{spec_constraint}
              \end{align}} &
\parbox[t]{6cm}{\begin{align}    
                              \bm{\Omega} &= MW \bm{1}_N^\top\label{eq:Omega}\\ 
                              \bm{\kappa} &= MK \bm{1}_N^\top\label{eq:kappa}\\
                              \mathcal{D}_{\Omega} &= \text{diag}\left(\Omega_1^{-1}, ..., \Omega_n^{-1}\right)\\
                              \mathcal{D}_{\kappa} &= \text{diag}\left(\kappa_1^{-1}, ..., \kappa_n^{-1}\right)
              \end{align}} &
                           
\parbox[t]{4cm}{\begin{align}
                             \mathcal{W}M &= MW  \label{freq_eq}\\
                             \mathcal{K}M &= MK  \label{deg_eq}\\
                             \mathcal{A}M &= MA \label{spec_eq}\\
                             &\nonumber
              \end{align}} 

\end{tabular}
\end{ruledtabular}
\end{table*}

\vspace{1cm}

\twocolumngrid

\onecolumngrid

\begin{table}[t]
\caption{\label{tab:table2}
The reduced dynamics of the Winfree, Kuramoto, and theta model.}
\begin{ruledtabular}
\begin{tabular}{l c c}
Model & Reduced phase dynamics \\
\colrule
Winfree & \parbox{14cm}{\begin{align} \dot{Z}_{\mu} &= i\sum_{\nu=1}^n \mathcal{W}_{\mu \nu}Z_{\nu}  + \frac{\sigma\kappa_{\mu}}{2N} + \frac{\sigma}{4N}\sum_{\nu=1}^n \mathcal{A}_{\mu\nu}(Z_{\nu} + \Bar{Z}_{\nu}) -\frac{\sigma}{2N\kappa_{\mu}}\sum_{\xi, \tau = 1}^{n} \mathcal{K}_{\mu \xi}\mathcal{K}_{\mu \tau} Z_{\xi}Z_{\tau}\nonumber\\ &\qquad- \frac{\sigma}{4N\kappa_{\mu}^2}\sum_{\nu, \xi, \tau = 1}^{n} \mathcal{A}_{\mu \nu}\mathcal{K}_{\mu \xi}\mathcal{K}_{\mu \tau} Z_{\xi}Z_{\tau}(Z_{\nu} + \Bar{Z}_{\nu}).
\label{redwinfree}\end{align}} \\
\colrule
Kuramoto & \parbox{14cm}{\begin{align}\dot{Z}_{\mu} &= i\sum_{\nu=1}^n \mathcal{W}_{\mu \nu}Z_{\nu} +\frac{\sigma}{2N}\sum_{\nu = 1}^{n} \mathcal{A}_{\mu\nu} Z_{\nu}-\frac{\sigma}{2N\kappa_{\mu}^2}\sum_{\nu, \xi, \tau = 1}^{n} \mathcal{A}_{\mu\nu}\mathcal{K}_{\mu \xi}\mathcal{K}_{\mu \tau} Z_{\xi}Z_{\tau}\Bar{Z}_{\nu}\quad \qquad
\label{redkuramoto} \end{align}} \\
\colrule
theta &  \parbox{14cm}{\begin{align}\dot{Z}_{\mu} &= -\frac{i}{2}(Z_{\mu} - 1)^2 + \frac{i\Omega_{\mu}}{2}\left(\Omega_{\mu}^{-1}\sum_{\nu=1}^n \mathcal{W}_{\mu\nu}Z_{\nu}+1\right)^2 \nonumber \\&\qquad+\frac{i\sigma}{4N}\left(\kappa_{\mu}^{-1}\sum_{\nu = 1}^{n} \mathcal{K}_{\mu\nu}Z_{\nu} + 1\right)^2\left[2\kappa_{\mu} -\sum_{\nu = 1}^{n} \mathcal{A}_{\mu\nu}(Z_{\nu} + \bar{Z}_{\nu})\right] \qquad\qquad\qquad
    \label{redtheta}\end{align}}\\
\end{tabular}
\end{ruledtabular}
\end{table}

\twocolumngrid

We stress that the result of Table~\ref{tab:main_result} remains valid for real dynamical systems. This means that if the variables $z_j$ in Eqs.~\eqref{eq:complex_complete_dyn} satisfy $\bar{z}_j=z_j$, then the linear observables $Z_\mu$ also evolve according to Eqs.~\eqref{eq:general_reduced_equation}, but with the additional conditions $\bar{Z}_{\mu} = Z_{\mu}$, $\bar{\gamma}_{\mu} =\gamma_{\mu}$, $\bar{\delta}_{\mu} =\delta_{\mu}$, and $\bar{\epsilon}_\mu =\epsilon_\mu$. Actually, in the real case, DART can be carried out a step further and be applied to dynamical systems with more than one dynamical equation per
node. This is discussed in Appendix~\ref{appendix:DART_DN}.

In general, the reduced system cannot be totally consistent with the complete set of differential equations~\eqref{eq:phasedyn}. The reduced system should thus be interpreted as an approximation whose quality strongly depends upon the choice of the $n\times N$ reduction matrix $M$. This will be discussed in depth in the next subsection.

\begin{figure}[t]  
 \centering
   \includegraphics[width=0.8\linewidth, clip=true]{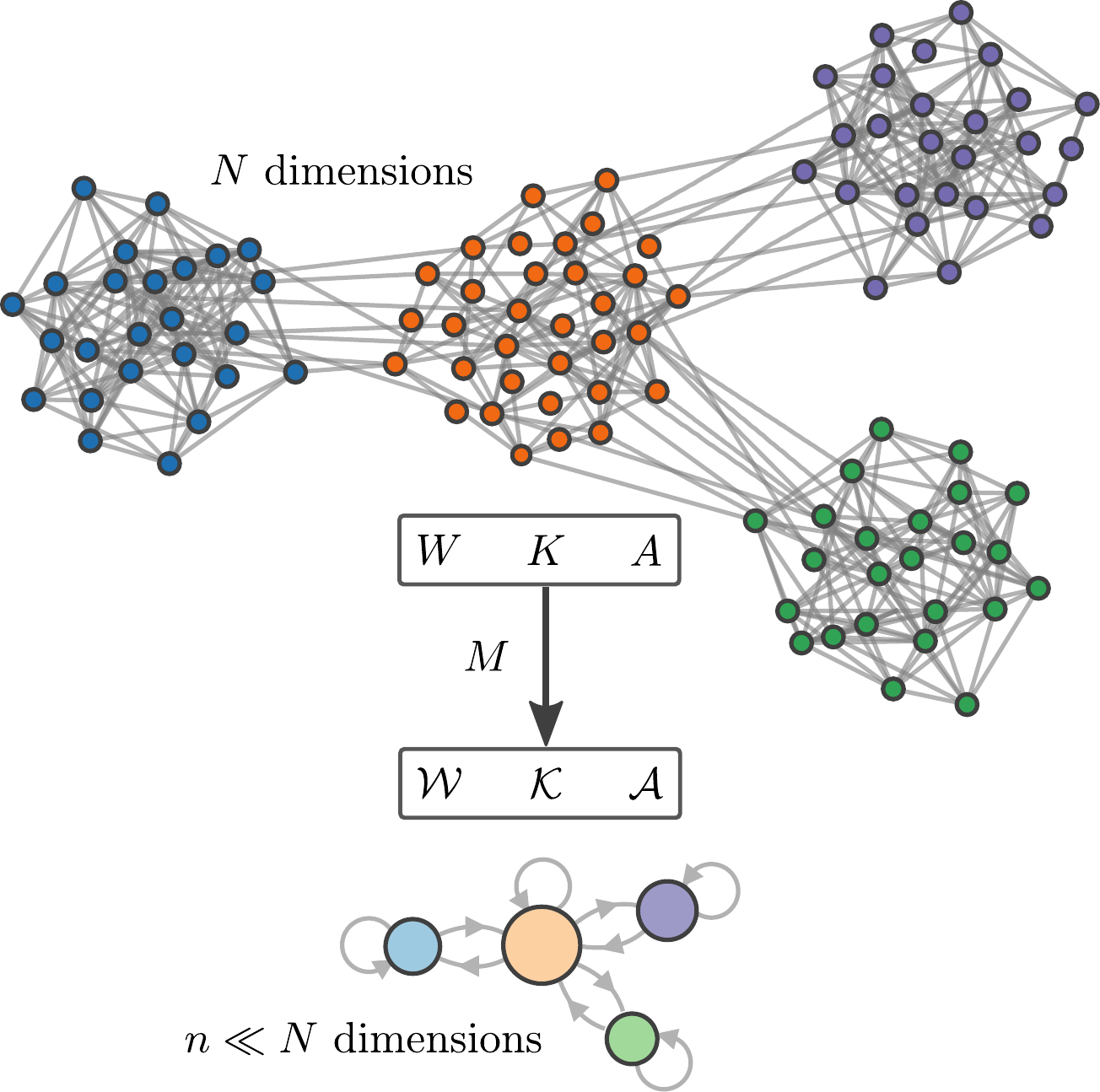}
   \caption{(Color online) Schematization of DART for a modular graph. The graph of adjacency matrix $A$ with $N$ nodes represents the structure of the complete dynamics, while the small graphs of adjacency matrix $\mathcal{A}$ with $n$ nodes illustrate the structure of the reduced dynamics. The $N\times N$ matrices of dynamical parameters $W$ and of degrees $K$ are also reduced with the reduction matrix $M$ to $n \times n$ matrices $\mathcal{W}$ and $\mathcal{K}$ respectively.}
   \label{fig:SDR}
 \end{figure}

Although the matrices $\mathcal{W}$, $\mathcal{K}$, and $\mathcal{A}$ seem to have been introduced for the sole purpose of mathematical convenience, they actually have a simple interpretation. As illustrated at the bottom of Fig.~\ref{fig:SDR}, $\mathcal{A}$ is the reduced adjacency matrix, that is, the matrix that regulates the interactions in the reduced system. Indeed, $\mathcal{A}$ is related to the graph with four nodes in the figure. Similarly, $\mathcal{K}$ and $\mathcal{W}$ respectively describe the in-degrees and the dynamical parameters of the reduced system. Figure~\ref{fig:SDR} therefore gives a basic intuition of DART.

To get a better idea of what a reduced system looks like, we apply DART to specific cases, namely, the Winfree, Kuramoto, and theta dynamics. The reduced dynamics are presented in \mbox{Table~\ref{tab:table2}}. The related phase and modulus dynamics are given in Table~\ref{tab:table_reduced_dynamics} of Appendix~\ref{appendix:reduced_dynamics}.

\subsection{Construction of the reduction matrix}
\label{subsec:weights}

The problem is to determine whether we can construct a reduction matrix $M$ that ensures the consistency of the compatibility equations. In this subsection, we prove that there is an infinite number of reduction matrices that lead to exact solutions for at least one compatibility equation while providing approximate solutions to the other equations. We then propose two procedures that aim at cleverly selecting a reduction matrix that minimizes the errors associated to the approximate solutions.

\subsubsection{Existence of the reduction matrix \texorpdfstring{$M$}{Lg} and~its~factorization}

In Appendix~\ref{appendix:existence}, we establish that for any $N\times N$ matrix $T$, if $M$ is a $n\times N$ matrix of rank $n$, with $n<N$, then there exists at most one solution to the matrix equation
\begin{equation}
    \mathcal{T}M = MT, 
    \label{eq:mat}
\end{equation}
with unknown reduced matrix $\mathcal{T}$ of size $n\times n$. If the solution exists, then it is equal to 
\begin{equation}
\mathcal{T} = MTM^+,
\label{eq:unique_sol}
\end{equation}
where $^+$ denotes Moore-Penrose pseudo-inversion.

Moreover, if the solution does not exist, Eq.~\eqref{eq:unique_sol} provides the best approximate solution in the sense that it minimizes the mean squared error (MSE) between $MT$ and $\mathcal{T}M$.

To tackle the problem of the existence of a solution to Eq.~\eqref{eq:mat}, we focus on the case where $T$ is real and symmetric, as are the matrices $W$, $K$, and $A$ in the compatibility equations. In Appendix~\ref{appendix:existence}, we prove that for
\begin{enumerate}
    \labitem{(1)}{itm:cond1} a factorization $M = CV$ where
    \labitem{(2)}{itm:cond2} $C$ is a real non-singular $n\times n$ matrix and
    \labitem{(3)}{itm:cond3} $V$ is a $n\times N$ real matrix composed of $n$ real \mbox{orthonormal row} eigenvectors of $T$,
\end{enumerate} 
Eq.~\eqref{eq:mat} has the unique solution 
\begin{equation}
   \mathcal{T} = C\Lambda C^{-1}, 
    \label{eq:solutionC}
\end{equation}
where $\Lambda$ is the diagonal matrix whose $\mu$-th element on the diagonal is equal to the eigenvalue $\lambda_\mu$ corresponding to the $\mu$-th eigenvector in $V$. 

Conditions~\ref{itm:cond1}--\ref{itm:cond3} are not particularly restrictive. They simply ensure that the rows of $M$ form a (not necessarily orthogonal) basis of a $n$-dimensional subspace of~$\mathbb{R}^N$. This subspace is spanned by the row eigenvectors of~$T$ used to build the \textit{eigenvector matrix} $V$. 

We call $C$ the \textit{coefficient matrix}, as its main role is to combine the eigenvectors of $T$ in a suitable manner such that Conditions~\ref{itm:condA} and \ref{itm:condB} are respected. We note that, due to the non-singularity of $C$ and the orthonormality of the rows of $V$, the matrix $M=CV$ automatically complies with Condition A.

These results lead to sufficient criteria for the consistency of the compatibility equations. Indeed, if
\begin{enumerate} 
\labitem{\textit{i}}{itm:condi} the matrices $W$, $K$, and $A$ share the row eigenvectors $\bm{v}_\mu$, with $\mu\in\{1,\ldots, n\}$,
\labitem{\textit{ii}}{itm:condii} $M=CV$, where $C$ is a $n\times n$ non-singular matrix and $V$ is the $n\times N$ matrix whose $\mu$-th row is $\bm{v}_\mu$,
\end{enumerate}
then the compatibility equations are all consistent and DART is exact to first-order.

In general, however, the matrices $W$, $K$, and $A$ do not share eigenvectors. This makes difficult or even impossible to determine whether or not there is a reduction matrix $M$ whose corresponding $n$-dimensional reduced system is exact to first-order. Nevertheless, the sufficient criteria \ref{itm:condi}-\ref{itm:condii} suggest how to use eigenvectors to get a reduced system that is almost exact to first-order.

\subsubsection{Selection of one target matrix among \texorpdfstring{$\{W, K, A\}$}{Lg}}

Even if the criterion \ref{itm:condi} is generally not satisfied, we can still define $V$ with the eigenvectors of one matrix $T$ among $\{W, K, A\}$, called the \textit{target matrix}, and solve exactly its compatibility equation while solving the others approximately. The intermediate steps to achieve this are detailed in Procedure~\ref{proc:1}.

The procedure uncovers a \textit{trichotomy} in DART: we must choose either $W$, $K$, or $A$ as a target. To link the chosen target $T$ to its reduction matrix, we denote $M$ as $M_T$, $C$ as $C_T$, and $V$ as $V_T$.

\begin{algorithm}[H]
\begin{flushleft}
    \caption{Reduction with a single target matrix}
    \label{proc:1}
    \begin{tabular}{ll} 
        \textbf{Input:}& $N\times N$ matrices $W$, $K$, and $A$\hfill\\
        & positive integer $n<N$\\
        \textbf{Output:} & $n\times n$ matrices $\mathcal{W}$, $\mathcal{K}$, and $\mathcal{A}$\\
    \end{tabular}
    \begin{algorithmic}[1]
        
        \State \label{state:proc1_1} Select a target matrix $T$ from the set $\{W,K,A\}$.
        
        \State \label{state:proc1_2} Use $n$ orthonormal row eigenvectors of $T$ to form the $n\times N$ matrix $V_T$.
        
        \State \label{state:proc1_3} Find a non-singular $n\times n$ matrix $C$ such that             
        \begin{equation*} M_T=C_TV_T 
        \end{equation*}
        satisfies Condition~\ref{itm:condB} [see Appendix~\ref{appendix:calculation}].
        
        \State \label{state:proc1_4} Solve exactly or approximately the compatibility equations using the formulae
            \begin{equation}\label{eq:threesolutions}
                \mathcal{W} = M_TWM_T^+,\quad 
                \mathcal{K} = M_TKM_T^+,
                \quad \mathcal{A} = M_TAM_T^+.
            \end{equation}

    \end{algorithmic}
\end{flushleft}
\end{algorithm}

Note that finding the coefficient matrix $C_T$ in Step~\ref{state:proc1_3} is generally not straightforward. In simple cases, where all elements of $V_T$ are nonnegative, $C_T$ can take the form of a diagonal matrix. When $V_T$ has negative values, the coefficient matrix is more difficult to obtain and we must turn to more sophisticated factorization methods, including semi-nonnegative matrix factorization (SNMF) and orthogonal nonnegative matrix factorization (ONMF). The technical details of the calculation of $C_T$ and $V_T$ are given in \mbox{Appendix~\ref{appendix:calculation}}. 

Let us discuss about Procedure~\ref{proc:1} when the adjacency matrix $A$ is chosen as the target matrix in Step~\ref{state:proc1_1}. 
There are a number of reasons to look at this choice in more details.  
One recalls that $A$ encodes the interactions among the oscillators, so the first-order errors induced by the inconsistency in the equation $\mathcal{A}M_A = M_A A$ may lead to poor estimates for the evolution of the linear observables $Z_{\mu}$, unless the interactions are negligible.  Moreover, a recent work \cite{Laurence2019} has revealed that the dominant eigenvector of $A$ provides an observable that is better estimated through the reduced dynamics compared to the corresponding observable based on the degrees, and thus based on $K$.  

For $T=A$, $V_A$ is a $n\times N$ matrix where the rows are orthonormalized eigenvectors of $A$. For \textit{any} $n\times n$ non-singular matrix $C_A$, setting $M_A=C_AV_A$ makes the matrix equation $\mathcal{A}M_A = M_AA$ consistent. The unique solution to the matrix equation is then
\begin{equation}
    \mathcal{A}=M_A A M_A^+= C_A\Lambda_A C_A^{-1},
    \label{eq:reducedA}
\end{equation} 
where $\Lambda_A$ is the diagonal matrix that contains the eigenvalues associated with the eigenvectors in $V_A$. Therefore, the elements of the reduced adjacency matrix $\mathcal{A}$ are combinations of eigenvalues of $A$. In other words, the reduced graph is weighted by combinations of eigenvalues. Moreover, any eigenvalue of $\mathcal{A}$ is also an eigenvalue of $A$ and this result does not depend on the matrix $C_A$. Hence, in addition to its structural importance, this choice reveals something quite useful about an extension of the method to more than one target matrices.

\subsubsection{Selection of two or three target matrices}

In Procedure~\ref{proc:1}, there is a lot of freedom to choose the non-singular coefficient matrices $C_W$, $C_K$, and $C_A$. Yet, we have not exploited this freedom. Even if the procedure allows to solve one of the compatibility equations exactly, the resulting reduction matrix $M_T$ could generate considerable errors in the other ones. We should therefore seek for a procedure that leverages the freedom on the coefficient matrix to minimize these errors. 

For instance, let us consider \mbox{$T = A$} and \mbox{$M_A = C_A V_A$}. Then, \mbox{$\mathcal{A} = M_A A M_A^{+}$} is exact. We now want to find a coefficient matrix $C_A$ that minimizes the MSE between $M_A$ and \mbox{$M_W=C_WV_W$}, which ensures the consistency of the equation \mbox{$\mathcal{W}M_W = M_WW$}. 
In such a situation, we say that $W$ is the second target matrix. 
The solution to the minimization problem is
\begin{equation}
    C_A = M_W V_A^+=C_W V_W V_A^+,
    \label{C_from_target_weight}
\end{equation}
which implies that the best reduction matrix is
\begin{equation}
    M_A = M_W V_A^+ V_A = C_W V_W V_A^\top V_A.
    \label{M_from_target_weight}
\end{equation}
We note that $V_A^+=V_A^\top$ since the rows of $V_A$ are orthonormal. The matrix $C_W$ is not yet determined, but this can be done by imposing Condition~\ref{itm:condB}.

In another context, one could want to minimize the error related to the degree matrix $K$ instead of $W$. The second target matrix would then become $K$ and the MSE would be minimized between $M_A=C_AV_A$ and the matrix $M_K=C_KV_K$ leading to the equations
\begin{align}
        C_A &= M_K V_A^+,
    \label{C_from_target_deg}\\
        M_A &= M_K V_A^+ V_A=C_K V_K V_A^\top V_A,
    \label{M_from_target_deg}
\end{align}
where $C_K$ is a non-singular matrix chosen so that Condition~\ref{itm:condB} is satisfied.

We have thus succeeded in targeting two matrices, first $A$, then $W$ or $K$. We will therefore use the notation \mbox{$A \to W$}, \mbox{$A \to K$}, or more generally \mbox{$T_1 \to T_2$}, to denote the choice of target $T_1$ followed by $T_2$. The first target is reached exactly in the sense that the equation $\mathcal{A}M_A = M_AA$ is consistent, while the second cannot be reached exactly in general, but nevertheless allows the approximate resolution of the equation \mbox{$\mathcal{W}M_A = M_AW$} or \mbox{$\mathcal{K}M_A = M_AK$}. Procedure~\ref{proc:2} is a formalized version of this method with multiple targets in which the first and the second target matrices are arbitrary. The same procedure is also applicable to the case of three target matrices \text{\footnote{For $u = 3$, one can also set \mbox{$M_3 = C_{T_3}V_{T_3}V_{T_2}^+V_{T_2}V_{T_1}^+V_{T_1}$}, but Eq.~\eqref{eq:M_allcases}  has been favored based on its better performance in numerical experiments.}}. 

Procedure~\ref{proc:2} includes more constraints on the coefficient matrix than Procedure~\ref{proc:1} in the hope of satisfying more accurately the compatibility equations \footnote{One may wonder if solving the compact compatibility equation $\mathcal{Q}M = MQ$, where $Q = aW + bK + cA$, helps to satisfy the compatibility equations. If the compatibility equations are satisfied, then $\mathcal{Q} = a\mathcal{W} + b \mathcal{K} + c \mathcal{A}$. 
Otherwise, we find $|| MQ - \mathcal{Q} M || \leq |a| \,|| MW - \mathcal{W} M || + |b| \,|| MK - \mathcal{K} M || + |c| \,|| MA - \mathcal{A} M ||$ using the triangle inequality. However, recall that to minimize the first-order errors in DART, we must minimize $|| MW - \mathcal{W} M ||$, $|| MK - \mathcal{K} M ||$, and $|| MA - \mathcal{A} M ||$. Since $|| MQ - \mathcal{Q} M ||$ is a lower bound on these errors, minimizing the error between $MQ$ and $\mathcal{Q} M$ is not helpful.}. Indeed, Steps~\ref{state:proc2_3} and \ref{state:proc2_4} can be seen as a successive imposition of constraints on the coefficient matrix, from constraints to satisfy the different compatibility equations to constraints to fulfill Condition~\ref{itm:condB}.

Step~\ref{state:proc2_4} is not trivially satisfied when $V_A$ is involved. Indeed, only the first dominant eigenvector possesses solely real positive entries which is guaranteed by the Perron-Frobenius Theorem \cite[Theorem 38]{VanMieghem2011}. To satisfy Condition~\ref{itm:condB2}, we once again use SNMF. We also use ONMF to ensure compliance of Condition~\ref{itm:condA} (exactly) and Condition~\ref{itm:condA'} (approximately). The details are given in \mbox{Appendix \ref{appendix:calculation_proc2}}.

\begin{algorithm}[H]
  \caption{Reduction with $u$ target matrices}
  \label{proc:2}
   \begin{tabular}{ll} 
   \textbf{Input:}& $N\times N$ matrices $W$, $K$, and $A$\\
  & positive integer $n<N$\\
  & $u\in \{1,2,3\}$\\
  \textbf{Output:} & $n\times n$ matrices $\mathcal{W}$, $\mathcal{K}$, and $\mathcal{A}$\\
   \end{tabular}\\
   \begin{algorithmic}[1]
        \State \label{state:proc2_1} For $k\in\{1,\ldots, u\}$,  select a target $T_k$ from the set  \begin{equation*}S_k=\begin{cases}\{W,K,A\},& k=1,\\
        \{W,K,A\}\setminus\{T_1\}, & k=2, \\ 
         \{W,K,A\}\setminus\{T_1,T_2\}, & k=3.
        \end{cases}
        \end{equation*}  
        
    \State \label{state:proc2_2} For $k\in\{1,\ldots, u\}$,  use $n$ orthonormal row eigenvectors of $T_k$ to form the $n\times N$ matrix $V_{T_k}$.
    
    \State \label{state:proc2_3} Set 
    \begin{equation}
    M_u=
    \begin{cases}C_{T_1}V_{T_1},& u=1,\\
     C_{T_2}V_{T_2}V_{T_1}^+V_{T_1} , & u=2, \\ 
     C_{T_3}V_{T_3}V_{T_1}^+V_{T_1}V_{T_2}^+V_{T_2}V_{T_1}^+V_{T_1}, & u=3. \\ 
    \end{cases}
    \label{eq:M_allcases}
    \end{equation}
    For arbitrary $C_{T_u}$, $M_u$ must satisfy Condition~\ref{itm:condA}. Therefore, if $u = 2$ and $V_{T_2}V_{T_1}^+$ is singular, return to Step~\ref{state:proc2_2} and choose different eigenvectors. If $u = 3$ and $V_{T_3}V_{T_1}^+V_{T_1}V_{T_2}^+V_{T_2}V_{T_1}^+$ is singular, return to Step~\ref{state:proc2_2} and choose different eigenvectors. 
    
    \State \label{state:proc2_4} Find a non-singular $n\times n$ matrix $C_{T_u}$ by imposing Condition~\ref{itm:condB} [see Appendix~\ref{appendix:calculation_proc2}]. 
    
    \State \label{state:proc2_5} Solve exactly or approximately the compatibility equations using the formulae
    \begin{equation} 
    \mathcal{W} = M_uWM_u^+,\quad 
    \mathcal{K} = M_uKM_u^+,
    \quad \mathcal{A} = M_uAM_u^+.
    \end{equation}
    
    \end{algorithmic}
\end{algorithm}

\vspace{-0.5cm}
 
\subsubsection{Choice of eigenvectors for the adjacency matrix }

There is yet another aspect that needs to be clarified for the case where the adjacency matrix is selected as a target matrix, that is, the choice of eigenvectors. Although any eigenvector could be chosen in principle, many reasons speak in favor of prioritizing the dominant eigenvectors, which are the eigenvectors of $A$ whose corresponding eigenvalues are away from the bulk of the spectrum. It is known, for instance, that the largest eigenvalue of $A$ plays a crucial role for predicting the stability \cite{May2001}, the dynamic range \cite{Larremore2011}, and the critical values of dynamical parameters \cite{Restrepo2005, VanMieghem2012, Castellano2017a} in dynamical systems on networks. Moreover, from the graph structure perspective, the eigenvector with the largest eigenvalue has proven to be a useful centrality measure \cite{VanMieghem2011, Newman2018} while the dominant eigenvectors are key objects in spectral methods for community detection \cite{Lei2015}, graph clustering \cite{Schaeffer2007}, and graph partitioning \cite{Fiedler1975, Barnes1982, Barnes1984, Powers1988}. Finally, Eq.~\eqref{spec_constraint} and Eq.~\eqref{eq:reducedA} reveal that the dominant eigenvalues have the strongest impact on the reduced dynamics.

\begin{figure}[tb] 
 \centering
   \includegraphics[width=1\linewidth, clip=true]{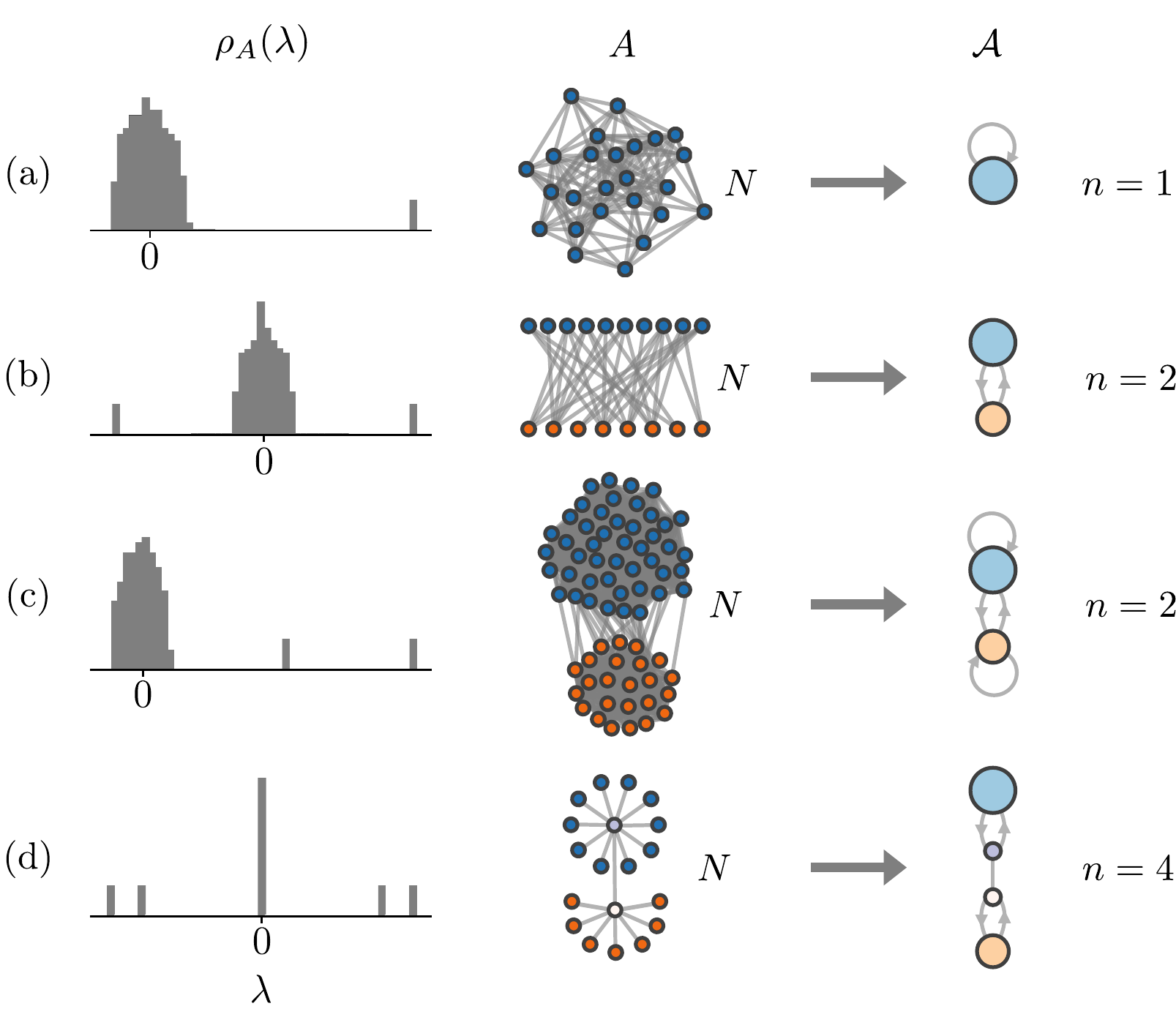}
   
   \vspace{-0.4cm}
   
\caption{(Color online) Graph schematization of DART for (a) the Erd\H{o}s-R\'enyi model, (b) the random bipartite graph, (c) the SBM, and (d) the two-star graph. An example of spectral density $\rho_A(\lambda)$ is given for each graph. Note that the dominant eigenvalues bins have been enlarged for the sake of  visualization.}
   \label{fig:SDR_spectrum}
 \end{figure}

We should therefore use the dominant eigenvectors of $A$ when constructing the matrix $V_A$. But how many of them should we consider?  A rule of thumb is to choose $n$ as the number of eigenvalues that are away from the bulk of the spectrum, which is the set that contains the eigenvalues $\lambda_i$ of $A$ such that $\lambda_i/\lambda_{\text{D}}\approx 0$, where $\lambda_{\text{D}}$ is the largest eigenvalue (see Fig.~\ref{fig:SDR_spectrum}). In random graphs with one community ($q = 1$), such as the Erd\H{o}s-R\'enyi model, only one eigenvalue is separated from the bulk. We then simply choose $C_A = 1$ and $M_A = V_A = \mathbf{v}_D$, the dominant eigenvector of $A$, to get a reduced dynamics of dimension $n = 1$ as in Refs.~\cite{Barlev2011, Laurence2019}. For a graph with two communities ($q = 2$), there are usually two dominant eigenvectors, suggesting to make a reduction of dimension two ($n = q = 2$). When the communities are densely connected among themselves, it is not possible to apply twice the one-dimensional reduction method, as used in Ref.~\cite{Laurence2019}. We then need to combine the eigenvectors using a second target matrix, $W$ for example.

\subsection{Errors in DART}
\label{subsec:errors}

 \begin{figure}[b] 
 \centering
  \includegraphics[width=1\linewidth, clip=true]{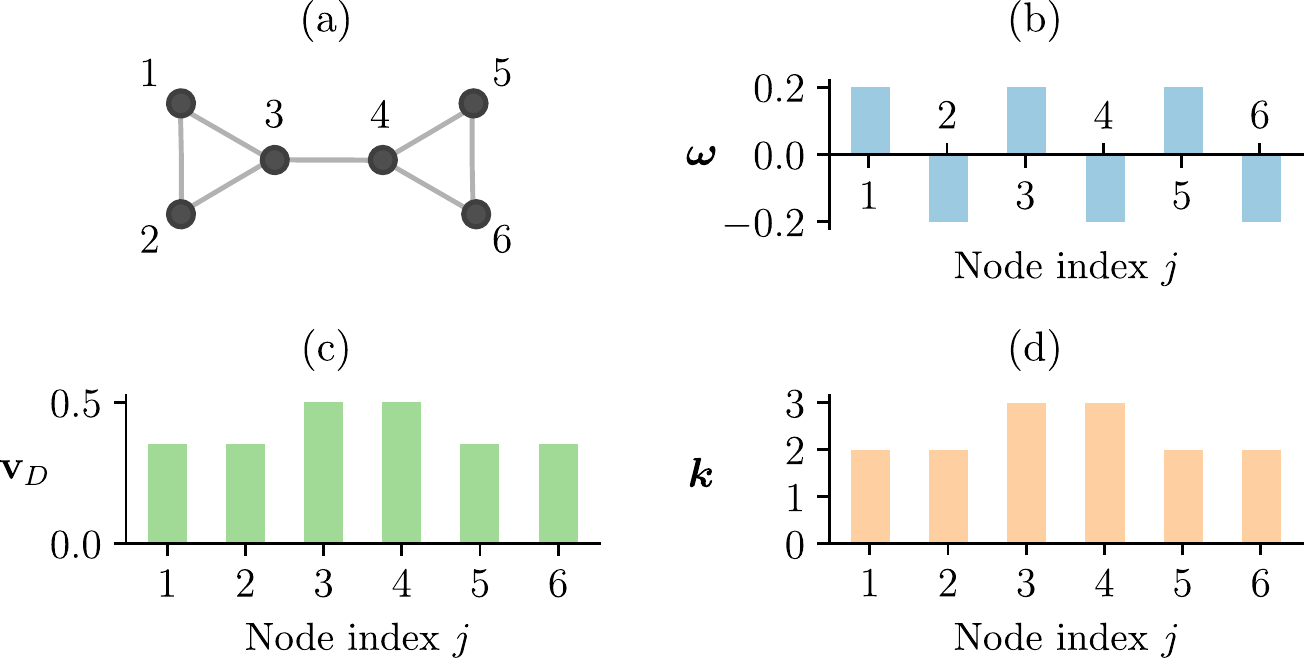}
  
  \vspace{-0.35cm}
  
  \caption{(Color online) Setup for the dimension reduction of the Kuramoto model on the two-triangle graph. (a)~Two-triangle graph where the nodes are labeled from 1 to 6. (b)~Frequency sequence $\bm{\omega}$. (c) Dominant eigenvector $\mathbf{v}_D$ of the adjacency matrix. (d) Degree sequence $\bm{k}$.
  }
   \label{fig:small_graphs_reduction_setup}
 \end{figure}

\begin{figure*}[t] 
 \centering
  \includegraphics[width=0.9\linewidth, clip=true]{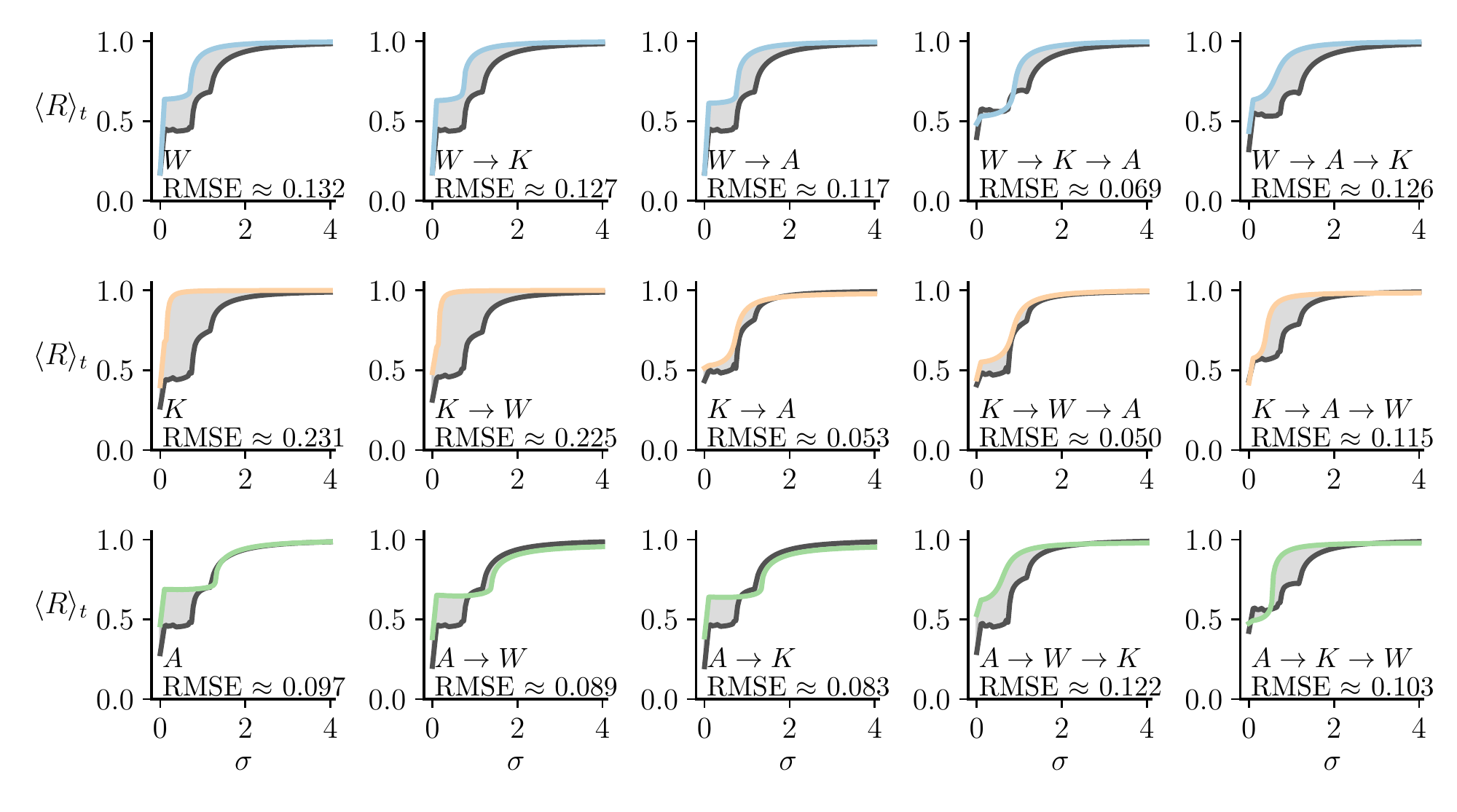}
  
  \vspace{-0.6cm}
  
  \caption{(Color online) 
  Time-averaged global synchronization observable $\langle R \rangle_t$ for the complete Kuramoto dynamics with dimension $N = 6$ (dark gray lines) and its reduced dynamics~\eqref{redkuramoto} with dimension $n = 2$ (colored lines) vs.\ the coupling constant $\sigma$ for the 15 choices of target matrices denoted as $T_1 \rightarrow T_2 \rightarrow T_3$ in the plots. The RMSE is the root-mean-squared error between the complete dynamics curve and the reduced dynamics curve. The gray area between the synchronization curves is shown to illustrate qualitatively the error made by the reduced dynamics. The initial conditions are the same for each plot and are drawn from a uniform distribution $\mathcal{U}(0, 2\pi)$.}
   \label{fig:small_graphs_reduction}
 \end{figure*}

In the previous subsections, we have made approximations to get the reduced dynamics~\eqref{eq:general_reduced_equation}. Until now, we have not considered the impact of these approximations on the accuracy of the reduced dynamics synchronization curves compared to those of the complete dynamics~\eqref{eq:phasedyn}. In this subsection, we first specify the sources of errors and then expose their effects on the reduced dynamics synchronization curves.

The errors in DART come from the Taylor expansion made in Eq.~\eqref{eq:eq:general_reduced_equation_remainder}. They are of two types: first-order terms [Condition~\ref{itm:condI}] and higher-order ($> 1$) terms [Condition~\ref{itm:condIII}]. In this paper, we focus on the first-order errors which are directly related to satisfying the compatibility equations. The sources of first-order errors depend on the choice of: 
\begin{enumerate}
    \item Target matrices;
    \item Dimension $n$ of the reduced dynamics;
    \item Eigenvectors;
    \item Methods to calculate of $C_{T_u}$. 
\end{enumerate}
 
The third and fourth sources of errors are discussed in Appendices~\ref{appendix:calculation} and \ref{appendix:calculation_proc2}. To illustrate the impact of the first and second sources of errors, let us consider the simple setup presented in Fig.~\ref{fig:small_graphs_reduction_setup}. We reduce the dimension of the Kuramoto dynamics on a two-triangle graph ($N=6$) to the dynamics~\eqref{redkuramoto} with $n=2$. We choose this small graph and natural frequencies that are uncorrelated to the structure (ex. the degrees) to accentuate the errors between the synchronization curves of the complete and reduced dynamics. By doing so, the matrices $W$, $K$ and $A$ will have very different eigenvectors which makes it more difficult to satisfy the compatibility equations.
 
We choose the orthogonal eigenvector matrices,
\begin{align}
    V_W &= \begin{pmatrix} \frac{1}{\sqrt{3}}&0&\frac{1}{\sqrt{3}}&0&\frac{1}{\sqrt{3}}&0\\0&\frac{1}{\sqrt{3}}&0&\frac{1}{\sqrt{3}}&0&\frac{1}{\sqrt{3}}\end{pmatrix}\nonumber,\\
    V_K &= \begin{pmatrix} \frac{1}{\sqrt{3}}&\frac{1}{\sqrt{3}}&0&0&\frac{1}{\sqrt{3}}&0\\0&0&\frac{1}{\sqrt{2}}&\frac{1}{\sqrt{2}}&0&0\end{pmatrix}\label{eq:vep_matrices},\\
    V_A &= \begin{pmatrix} \frac{\sqrt{2}}{4} & \frac{\sqrt{2}}{4} & \frac{1}{2}  &   \frac{1}{2}   &  \frac{\sqrt{2}}{4} & \frac{\sqrt{2}}{4}\\[0.1cm]\frac{-1}{c} & \frac{-1}{c} & \frac{1 - \sqrt{3}}{c} & \frac{\sqrt{3}-1}{c} & \frac{1}{c} &  \frac{1}{c}\end{pmatrix}\nonumber,
\end{align}
where $c = [2(\sqrt{3}-1)^2 + 4]^{1/2}$ and the two rows of $V_A$ are the first two dominant eigenvectors of $A$. These eigenvector matrices ensure that $V_{T_2}V_{T_1}^+$ and $V_{T_3}V_{T_1}^+V_{T_1}V_{T_2}^+V_{T_2}V_{T_1}^+$ (in Step~\ref{state:proc2_3} of Procedure \ref{proc:2}) are non-singular regardless of the combination of target matrices \footnote{
For instance, choosing $\bm{v}_1 = (1/2,  1/2, 0, 0, 1/2, 1/2)$ as the first row of $V_K$ would create singular matrices (e.g., $V_KV_A^+$ would be singular). Note that if $V_K$ includes $\bm{v}_1$, the predictions in Fig.~\ref{fig:small_graphs_reduction} when targeting $K$ are not improved and therefore, the conclusion drawn in this section would still be valid.}. 

\subsubsection{Target matrices}

To evaluate the impact of the choice of target matrices on first-order errors, we proceed as follows. With Procedure \ref{proc:2} and the eigenvector matrices in Eqs.~\eqref{eq:vep_matrices}, there are 15 possible choices of target matrices: three single targets ($u = 1$), six combinations of two targets ($u = 2$), and six combinations of three targets ($u = 3$). We thus obtain 15 reduction matrices and 15 reduced dynamics each related to a choice of target matrices. From these results, we find the synchronization curves $\langle R \rangle_t$ vs.\ the coupling constant $\sigma$ for the complete and reduced Kuramoto dynamics which are presented in Fig.~\ref{fig:small_graphs_reduction}. In each plot, the choice of target matrices and the RMSE between the synchronization curves of the complete and reduced dynamics are specified just above the $\sigma$-axis.

As expected, we first observe that choosing $T_1 = W$ allows an accurate prediction of the synchronization observable at $\sigma = 0$ compared to $T_1 = K$ or $T_1 = A$. This is explained by the fact that the compatibility equation $\mathcal{W}M_1 = M_1W$ is perfectly satisfied when $T_1 = W$, but not when $T_1 = K$ or $T_1 = A$.

Interestingly, the reduced dynamics with the one target procedure $T_1 = A$ and the two-target procedures $A \to W$, $A\to K$ accurately predict the form of the complete dynamics synchronization curve for high couplings \footnote{Comparable predictions can also be achieved with the targets $L \to W$, where $L$ is the Laplacian matrix, and the compatibility equations $\mathcal{W}M = MW$, $\mathcal{L}M = ML$. For $n=2$, we can build the Laplacian eigenvector matrix $V_L$ with the eigenvectors $\bm{v}_1$ and $\bm{v}_2$ corresponding to the two lowest eigenvalues: $\lambda_1 = 0$ and the Fiedler eigenvalue $\lambda_2$ \cite{Fiedler1973}. The uniformity of $\bm{v}_1$ helps to get a positive reduction matrix, while $\bm{v}_2$ is useful for graph partitioning \cite{Hall1970} and community detection \cite{Newman2006}.}. We can explain this result by the fact that, for higher couplings, the structure becomes increasingly important. Therefore, observables defined from structural quantities (like the eigenvectors of the adjacency matrix) should be favored in this coupling regime.

However, if we choose $T_1 = K$, the predictions of the reduced dynamics are inaccurate even when the dynamics is strongly influenced by the structure ($\sigma > 1$ for instance). From our numerical experiments, choosing $K$ as a first target is often a bad choice. One key reason for this is that the degrees only give a local information on the nodes whereas the eigenvectors of the adjacency matrix contains global information on the network \cite[Chapter 3]{VanMieghem2011}. Each eigenvector of $A$ depends on all the nodes in contrast to the eigenvectors of $K$ in general. Yet, using $A$ as a last target appears to be helpful for the predictions and it is not excluded that mixing global and local informations on the structure gives accurate results.

Another aspect to consider is the number of target matrices $u$. Adding targets can improve or worsen the predictions of the reduced dynamics. Indeed, by looking at the first line of Fig.~\ref{fig:small_graphs_reduction} where the first target is $W$, we get a better RMSE between the synchronization curves for the target combinations $W\to K \to A$ than the one target procedure with $T = W$. The result is the opposite in the third line of Fig.~\ref{fig:small_graphs_reduction} where the first target is $A$ : the one-target procedure with $T = A$ leads to more accurate results than the three-target procedures. Hence, the repercussions of choosing more or less target matrices is very subtle. 

One should nonetheless be careful with the interpretation of the RMSE: a smaller RMSE does not mean that we have better captured the underlying features of the complete dynamics. In our method, there are multiple possible sources of errors and they could compensate each other. For instance, this could be the case for the target combinations $K\to W \to A$.

\subsubsection{Reduced dynamics dimension}

Choosing an appropriate reduced dynamics dimension~$n$ is an inherent difficulty not only for DART, but also for all dimension-reduction methods. For instance, when the dynamics mainly relies on the structure, DART becomes a problem of graph partitioning for which determining the ideal number of groups $n$ is a well-known problem with no obvious answer \cite[Sec. IV.A]{Fortunato2016}. When no graph structure is introduced into the dynamics, even the most classical methods for deriving systems of reduced dimension (e.g., reduced order models) often fail to provide clear bounds on $n$ \cite[Sec. I]{Kramer2019}.

For these reasons, we have not looked for a systematic way to choose $n$. We have nevertheless gathered some theoretical and numerical evidence suggesting, at least for the phase dynamics considered in the paper, that the reduced dynamics can monotonically improves as $n$ increases.  

First, in Appendix~\ref{appendix:n=N}, we prove that if $n = N$, the reduction matrix $M$ can be written as the identity matrix and a reduced dynamics obtained with DART is equivalent to its complete dynamics if and only if the coupling function $H$ is of the form
\begin{equation}
    H(u, v, w, z) = E(u, v)w + Q(u, v) z + S(u,v),
\end{equation}
where $u,v,w,z \in \mathbb{C}$ and $E,Q,S$ are any complex-valued holomorphic functions. Interestingly, the coupling functions for the Winfree, Kuramoto, and theta models all satisfy this criterion, which means that the reduced dynamics in Eqs.~(\ref{redwinfree}, \ref{redkuramoto}, \ref{redtheta}) are equivalent to the complete dynamics in Eqs.~(\ref{winfree}, \ref{kuramoto}, \ref{theta}) when $n = N$. 

\begin{figure}[t] 
        \centering    
        \includegraphics[width=1\linewidth, clip=true]{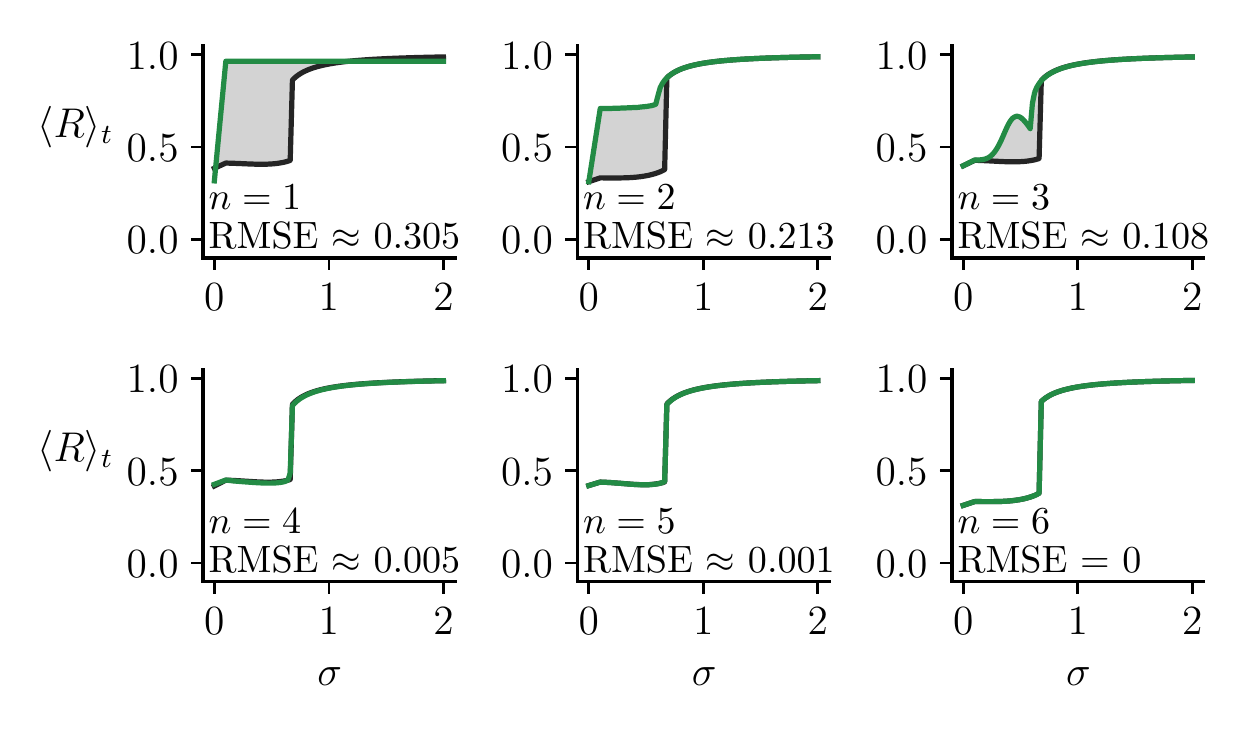}
        
        \vspace{-0.7cm}
        
        \caption{(Color online) Impact of $n$ on the quality of the reduced dynamics for the Kuramoto dynamics on the two-triangle graph. The targets are $A\to W$, the frequency matrix is $W = \text{diag}(0.1, 0.1, -0.2, -0.2, 0.1, 0.1)$, and the initial conditions are the same for each plot and are drawn from a uniform distribution $\mathcal{U}(0, 2\pi)$.}
        \label{fig:different_n}
\end{figure}

Second, using the Kuramoto dynamics on the two-triangle graph with $W = \text{diag}(0.1, 0.1,$ -$0.2,$ -$0.2, 0.1, 0.1)$, we observe a monotonic improvement of DART's predictions with the targets $A \to W$ when increasing the dimension of the reduced dynamics [Figure~\ref{fig:different_n}]. At $n = N = 6$, as expected, the prediction is perfect.  For other choices of targets and other dynamics, it is far from guaranteed that this monotonic improvement will occur, but in this example, ``the more eigenvectors, the better" \cite{Alpert1995}.

It is clear that the choice of target matrices and the dimension $n$ have a crucial impact on the predictions of the reduced dynamics. It is non-trivial to set $n$ wisely and to decide which targets to choose and in which order. Depending on the situation, one should find the appropriate targets to minimize the first-order errors. While the choice of the second and third target matrices (if any) depends on the situation, Figs.~\ref{fig:small_graphs_reduction}-\ref{fig:different_n} and our many numerical experiments suggest selecting $A$ as the first target.

 \begin{figure*}[t] 
 \centering
   \includegraphics[width=0.9\linewidth, clip=true]{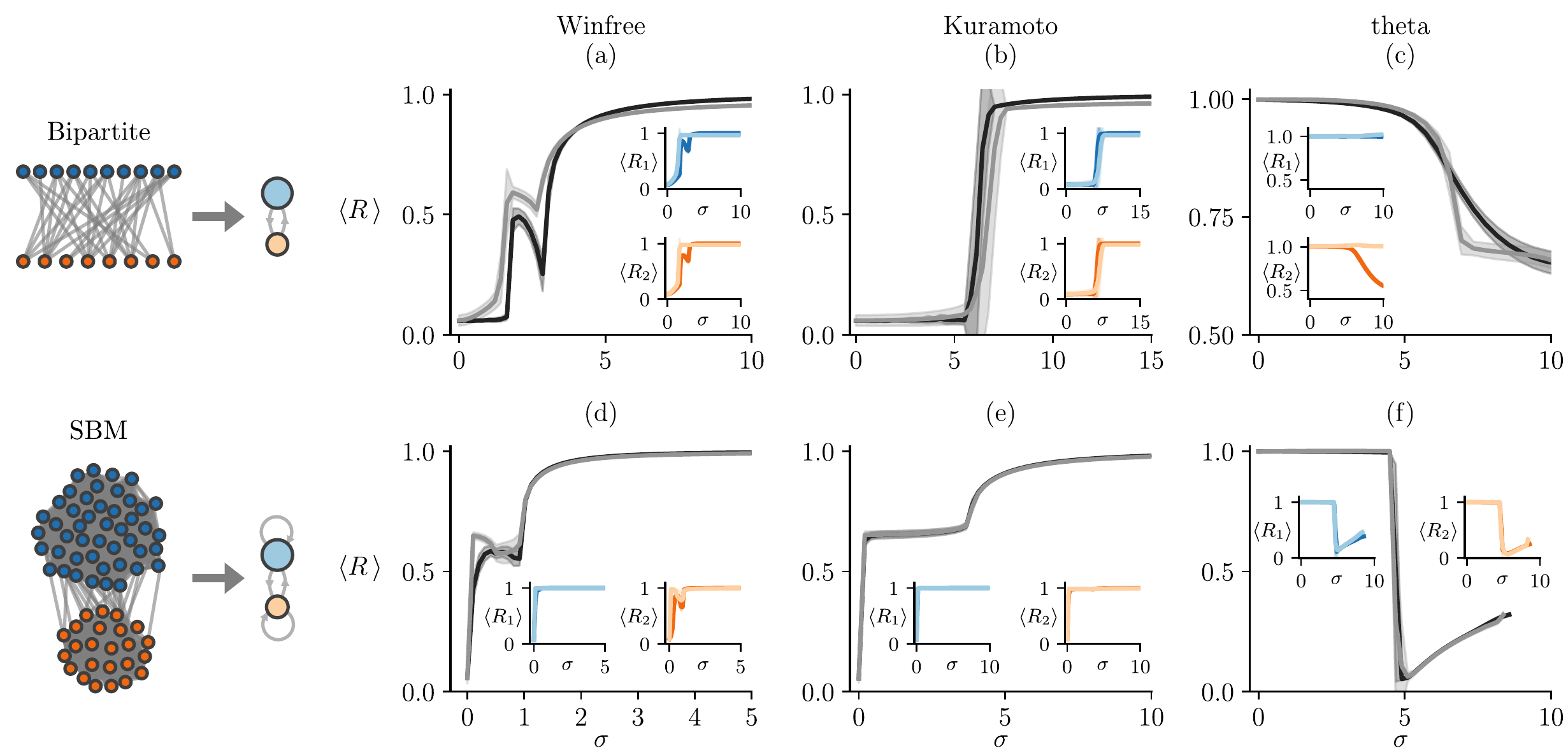}
\caption{(Color online) Comparison of the synchronization curves between three complete phase dynamics (black lines) and their reduced dynamics in Table~\ref{tab:table2} (gray lines) on random modular graphs. (Insets) Mesoscopic synchronization curves of the complete phase dynamics (dark blue and orange lines) and the reduced dynamics (light blue and orange lines). (a, b, d, e) The natural frequencies of the first community are drawn from the normal distribution $\mathcal{N}(0.3, 0.0001)$ and the natural frequencies of the second community are chosen to ensure that $\sum_{j=1}^N \omega_j = 0$. 
(c, f) The currents of the first community are drawn from the normal distribution $\mathcal{N}(-1.1, 0.0001)$ and the currents of the second community are drawn from $\mathcal{N}(-0.9, 0.0001)$. Common parameters to all subplots: $N_1 = 150$, $N_2 = 100$. For the bipartite, $p_{11} = p_{22} = 0$ and $p_{12} = p_{21} = 0.2$. For the SBM, $p_{11} = 0.7$, $p_{22} = 0.5$ and $p_{12} = p_{21} = 0.2$. The observables are averaged over the second half of the temporal series (25000 time steps), 50 graphs of the ensembles, 50 parameter matrices $W$, and 50 initial conditions randomly chosen from the standard normal distribution. The shaded region around each line is the standard deviation of the time-averaged synchronization observable.}
   \label{fig:all_trans}
 \end{figure*}

\subsection{Application to phase dynamics on~modular~graphs}
\label{sec:compmesobs}
 
So far, we have used DART only for a small graph and the Kuramoto dynamics. In this subsection, we show that our formalism is effective when the dimension of the system $N$ gets larger and can be successfully applied to other phase dynamics.

Let us first consider modular oscillator networks with two communities $B_1$ and $B_2$ of size $N_1$ and $N_2$ respectively. We assume that these modular networks are drawn from the SBM and that the dynamical parameters of each community are drawn from normal distributions. We denote these normal distributions $\mathcal{N}(\omega_{\mu}, v_{\mu})$, where $\omega_{\mu}$ and $v_{\mu}$ are respectively the mean and the variance of the dynamical parameters in the community $B_{\mu}$, and we consider that $v_{\mu}$ is small compared to $\omega_{\mu}$. With this setup and DART, we investigate the possibility of predicting the macroscopic and mesoscopic phase synchronization curves for the Winfree, Kuramoto, and theta models [Sec.~\ref{subsec:phase_dyn_synchro}].

For the reasons mentioned in Sec.~\ref{subsec:weights}, we select \mbox{$T_1 = A$}
and define $V_A$ with the two corresponding dominant eigenvectors. We also set $T_2 = W$ since the dynamical parameters are known to be important in the prediction of the synchronization critical value \cite{Acebron2005}.

Selecting $T_2 = W$ when there are $N$ different frequencies raises a problem for choosing the eigenvector matrix $V_W$. The reason is that there are $N$ eigenvectors that cannot be combined to form two eigenvectors with as many non-zero elements as possible (see Appendix~\ref{appendix:calculation}).

We get around this difficulty by choosing approximate eigenvectors of $W$ to build an approximate eigenvector matrix $V_W$. By approximate (left) eigenvectors of $W$, we mean row vectors $\bm{v}$ such that $\bm{v}W \approx \omega \bm{v}$ where $\omega$ is an approximate eigenvalue (frequency). Let $\bm{\omega}_{\mu}$ be the row vector of length $N_{\mu}$ that contains the dynamical parameters of the community $B_{\mu}$. Because the parameter variance in each community is low, the following eigenvector matrix yields good approximate eigenvalues of $W$: 
\begin{equation*}
   V_W = \left(
\begin{array}{c|c}
\frac{\bm{\omega}_1}{\sqrt{\sum_{j=1}^{N_1} (\bm{\omega}_1)_j }} &\bm{0}_{N_2} \\[0.4cm]
\hline
\bm{0}_{N_1} &\frac{\bm{\omega}_2}{\sqrt{\sum_{j=1}^{N_2} (\bm{\omega}_2)_j }}
\end{array}
\right),
\end{equation*}
where $\bm{0}_{N_\mu}$ is a null row vector of length $N_{\mu}$. 

The reduction matrix is obtained using Eq.~\eqref{M_from_target_weight} where $C_W$ is determined as explained in Appendix~\ref{appendix:calculation_proc2}. Note that since the adjacency matrix $A$ and the dynamical parameters $W$ are random matrices, the eigenvector matrices $V_A$ and $V_W$ are random matrices. As a consequence, the reduction matrix $M$ becomes a random matrix as well and the synchronization observables, random variables.

By computing $L$ reduction matrices from $L$ graph and dynamical parameter realizations, we solve approximately the compatibility equations and get $L$ reduced dynamics for the Winfree, Kuramoto, and theta model.

Figure~\ref{fig:all_trans} shows successful predictions of the average synchronization observable $\langle R \rangle$ for the three phase dynamics on $L = 50$ realizations of random modular graphs, dynamical parameters, and initial conditions. 

Double phase transitions occur both in the complete and reduced dynamics in Fig.~\ref{fig:all_trans}~(a), (d), and (e). The first synchronization transition corresponds to the emergence of synchronization at the level of the communities, as seen in the insets, while the second transition is explained by the increase of synchronization between the communities.

The reduced dynamics also predicts abrupt transitions at the macroscopic and mesoscopic levels for the Kuramoto model on bipartite networks \footnote{
When the oscillators in the two layers of the bipartite graph have opposite natural frequencies [which is almost the case in Fig.~\ref{fig:all_trans} (b)], they are called Janus oscillators \cite{Nicolaou2019a, Peron2020}. These oscillators exhibit an impressive diversity of oscillatory phenomena and DART could be useful to get further analytical insights.} [Fig.~\ref{fig:all_trans}~(b) and its insets]. The transitions at the macroscopic and mesoscopic levels occur at the same coupling value $\sigma$ contrarily to the Winfree model on the random bipartite graph. 

Figure \ref{fig:all_trans} (c) and (f) show that desynchronization cur\-ves are captured by the solutions of the reduced theta model. For the complete dynamics in Fig.~\ref{fig:all_trans}~(c), the phase trajectories first reach the closest equilibrium points on the unit circle and then oscillate when $\sigma$ increases. However, this behavior is not observed in the reduced dynamics which explains the result in the bottom inset (second community) of Fig.~\ref{fig:all_trans}~(c). In Fig.~\ref{fig:all_trans}~(f), the transition from oscillation death to an excited state (SNIC bifurcation, see Sec.~\ref{subsec:phase_dyn_synchro} and Fig.~\ref{fig:SNIC}) is well predicted by the reduced dynamics at the macroscopic and mesoscopic level.

Let us now investigate the impact of specific graph realizations on the performance of DART in Fig.~\ref{fig:all_trans}. Actually, we want to quantify how the RMSE between the complete and reduced synchronization curves varies according to the graph realizations used in Fig.~\ref{fig:all_trans}. To distinguish the different random graph realizations, we use the distance $d$ between their adjacency matrix $A$ and the mean adjacency matrix $\langle A \rangle$ given by 
\begin{equation}
    d = \frac{\sqrt{\sum_{j,k = 1}^N (A_{jk} - \langle A \rangle_{jk})^2}}{\sqrt{\sum_{j,k = 1}^N \text{max}(\langle A \rangle_{jk},\, 1- \langle A \rangle_{jk})^2}},
    \label{eq:distA}
\end{equation}
which is a normalized Frobenius norm. Moreover, by computing the probability distribution $P(d)$ in the random graph ensemble \footnote{Note that $P(d)$ can be computed numerically or analytically from the probability distribution $P(A)$ of adjacency matrices given in Eq.~\eqref{eq:probA}.}, we can identify which adjacency matrices among the chosen realizations are the most typical in the random graph ensemble.

\begin{figure}[b] 
 \centering
   \includegraphics[width=1\linewidth, clip=true]{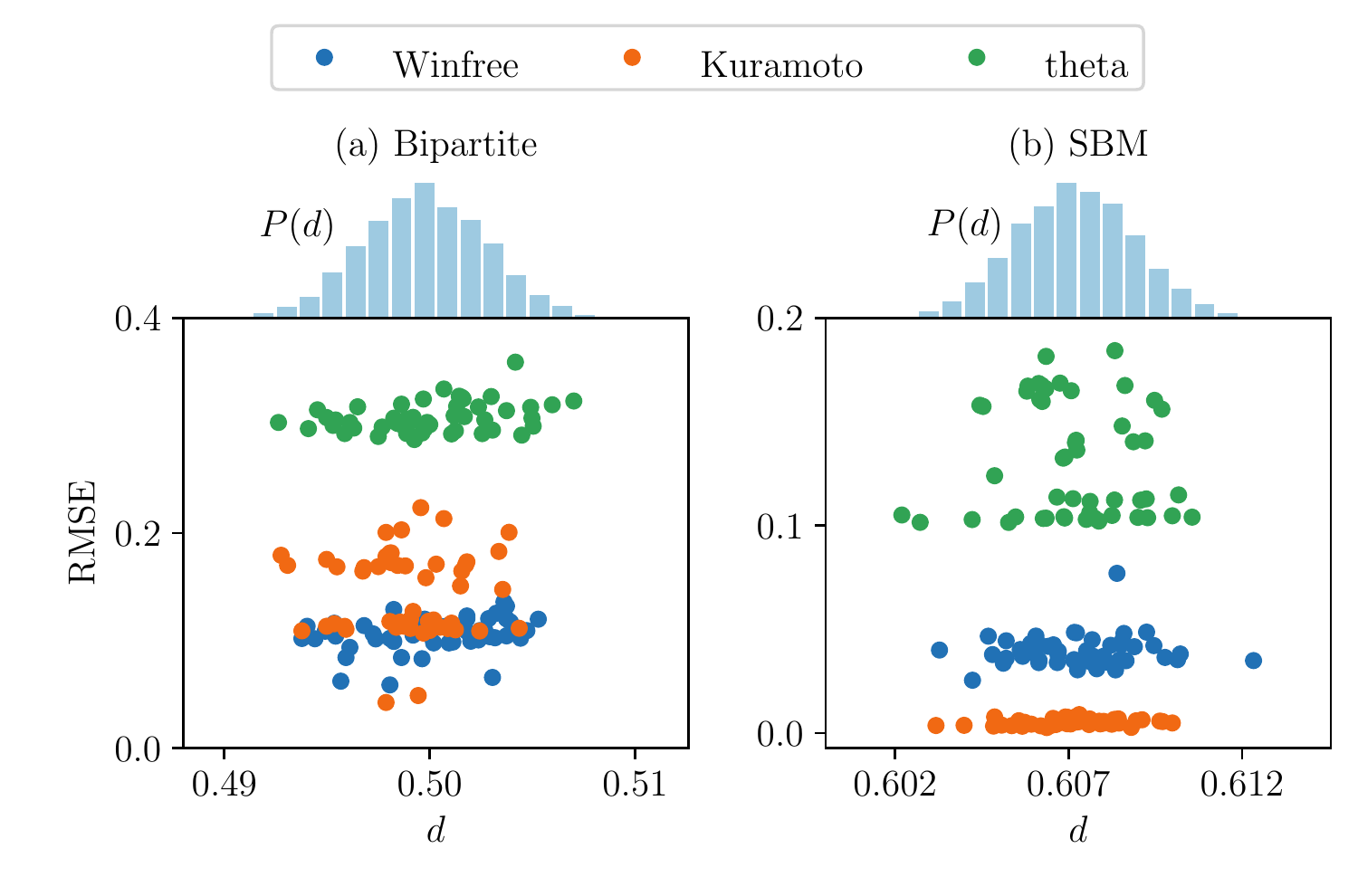}
   
   \vspace{-0.7cm}
   
\caption{(Color online) RMSE error between the complete and reduced dynamics synchronization curves of Fig.~\ref{fig:all_trans} vs.\ the distance $d$ [Eq.~\eqref{eq:distA}] for (a) the random bipartite ensemble and (b) the SBM. The histograms on top provide the distribution of the distance $d$ in each random graph ensemble.}
   \label{fig:rmse_frobenius}
\end{figure}
 
Figure~\ref{fig:rmse_frobenius} shows that the realizations, drawn from (a) the bipartite ensemble and (b) the SBM, cover a wide range of possible distances in the probability distribution $P(d)$. The figure reveals that the RMSE error between the complete and reduced dynamics synchronization curves does not vary significantly with the \mbox{distance~$d$}. 

This last observation is crucial. It suggests that phase dynamics on real modular networks, which differ considerably from the mean of a random graph, can be reduced using the dominant eigenvectors of its adjacency matrix.

\section{Chimeras and explosive synchronization}
\label{sec:chimeras}

Dimension reductions are useful in predicting synchronization regimes, even the more exotic ones \cite{Abrams2008, Kotwal2017, Chen2017, Xu2018b}. In this section, we use the formalism proposed in Sec.~\ref{sec:dart} to get analytical and numerical insights about the emergence of chimeras and explosive synchronization. Interestingly, we find that the reduced dynamics obtained with DART are similar to those deduced from the Ott-Antonsen Ansatz \cite{Abrams2008, Chen2017}. Yet, we provide a new perspective on the existence of chimera states for homogeneous and heterogeneous modular graphs. In addition to the observation of a new kind of chimera state on the two-star graph, we find that the asymmetry of the community sizes can make the difference between getting a chimera state or not. We finally recover results from various papers on explosive synchronization with the potential to push further in certain directions.

\subsection{Preliminary remarks on chimeras}
\label{subsec:chimeras}

By using the Ott-Antonsen Ansatz, it was shown in Ref.~\cite{Abrams2008} that a graph with two communities of identical phase-lagged oscillators can reach a dynamical state where one community is perfectly synchronized ($R_{\mu} = 1$ for any $\mu$ in $\{1, 2\}$) while the other is partially synchronized ($R_{\nu} < 1$ for $\nu$ in $\{1, 2\}\setminus \mu$). This spectacular state of coherence and incoherence, first observed in Ref.~\cite{Kuramoto2002}, was called a \textit{chimera state} or, more succinctly, a \textit{chimera}~\cite{Abrams2004}.   

The phenomenon was later observed experimentally for chemical \cite{Totz2018, Tinsley2012}, mechanical \cite{Wojewoda2016, Kapitaniak2014, Martens2013}, electro-optical \cite{Hagerstrom2012}, and nanoelectromechanical systems \cite{Matheny2019}. There is also evidence about the existence of chimeras in neural systems \cite{Bansal2019,Calim2018,Andrzejak2016,Hizanidis2016} and even in ecological systems \cite{Saha2019}. Theoretically, they were observed under various forms \cite{Abrams2008,Kemeth2016}, at multiple scales (e.g., $N \to \infty$ \cite{Kotwal2017}, $N = 3$ \cite{Maistrenko2017}), and with multiple kinds of coupling \cite{bera2017} (local \cite{Sethia2014,Laing2015} and non-local \cite{Abrams2004,Yeldesbay2014}).

The term ``chimera state" has been used in diverse ways in the literature. Some efforts have been made to clarify the notion \cite{Ashwin2015, Kemeth2016, Boccaletti2018}, but there is still no universally accepted definition. Our definition is based on Ref.~\cite{Abrams2004} in which a chimera state is defined as ``an array [(a graph)] of identical oscillators [that] splits into two domains: one coherent and phase locked, the other incoherent and desynchronized''. Careful attention should be paid to the word ``identical''. By identical, we mean that all oscillators have the same natural frequency $\omega_j = \omega$ for all $j\in\{1,...,N\}$. However, we allow the oscillators to have different neighborhoods (degrees). The oscillators are thus dynamically, but not structurally, identical.

\subsection{Chimeras in modular graphs}
\label{subsec:chimeras_modular_graphs}

We focus on the existence of chimeras in the Kuramoto-Sakaguchi dynamics on modular graphs with asymmetric modules and identical natural frequencies.

We first apply DART to get a simple set of differential equations. We then use the latter to find the equations governing the equilibrium points of the dynamics. Finally, we analyze those points to identify the regions in the space defining the graph structure where different types of chimeras can be found.

\vspace{-1em}
\subsubsection{Reduced dynamics}

We study a specific case of the SBM, namely, the planted partition model with two blocks (see Sec.~\ref{subsec:modular_graph}). The mean adjacency matrix is \footnote{Note that {$\langle A \rangle_{\text{SBM}}$} is the mean adjacency matrix of a SBM for which self-loops are allowed. Indeed, a node has a probability $p_{\text{in}}$ to have a link with itself.}
\begin{equation*}
A = \langle A \rangle_{\text{SBM}} =
\left(
\begin{array}{c|c}
p_{\text{in}}\bm{1}_{N_1}^\top\bm{1}_{N_1} & p_{\text{out}} \bm{1}_{N_1}^\top\bm{1}_{N_2}  \\
\hline
p_{\text{out}} \bm{1}_{N_2}^\top\bm{1}_{N_1} & p_{\text{in}} \bm{1}_{N_2}^\top\bm{1}_{N_2}
\end{array}
\right),
\end{equation*} 
where $\bm{1}_{N_\mu}$ is a row vector of 1's of length $N_{\mu}$. The degree matrix is 
\begin{equation*}
K = \left(\begin{array}{c|c}K_{11}I_{N_1\times N_1} & 0_{N_1\times N_2}  \\\hline0_{N_2\times N_1} & K_{22}I_{N_1\times N_1}\end{array}\right),
\end{equation*}
where $I_{N_{\mu}\times N_{\mu}}$ is an identity matrix of dimension $N_{\mu}\times N_{\mu}$,  $0_{N_{\mu}\times N_{\nu}}$ is a null matrix of dimension $N_{\mu}\times N_{\nu}$ and 
\begin{align*}
    K_{11} &= N_1 p_{\text{in}} + N_2 p_{\text{out}}, \\
    K_{22} &= N_1 p_{\text{out}} + N_2 p_{\text{in}}.
\end{align*}
Because the oscillators are identical,
\begin{equation*}
    W = \omega\,I_{N\times N}.
\end{equation*}
To find the reduced dynamics of the Kuramoto-Sakaguchi model on the mean SBM, let us follow Procedure \ref{proc:2} with $n = 2$ and $u = 2$.

\begin{enumerate}[label=\arabic*:]
    \item The compatibility equation~\eqref{freq_eq} is already satisfied since $W$ is proportional to the identity matrix, so there is no need to consider choosing $W$ as a target. We therefore select the targets $T_1 = A$ and $T_2 = K$.
    
    \item We compute the eigenvectors of $\langle A \rangle_{\text{SBM}}$ and form the matrix 
\begin{equation*}
   V_A = \left(
\begin{array}{c|c}
\frac{\bm{1}_{N_1}}{\sqrt{N_1 + N_2\ell_{+}^2}} &\frac{\ell_{+}\,\bm{1}_{N_2}}{\sqrt{N_1 + N_2\ell_{+}^2}} \\[0.3cm]
\hline
\frac{\bm{1}_{N_1}}{\sqrt{N_1 - N_2\ell_{-}^2}} &\frac{\ell_{-}\,\bm{1}_{N_2}}{\sqrt{N_1 + N_2\ell_{-}^2}}
\end{array}
\right),
\end{equation*}
where
\begin{equation*}
    \ell_{\pm} = \frac{\lambda_{\pm} - N_1 p_{\text{in}}}{N_2 p_{\text{out}}}
\end{equation*}
while
\begin{equation*}
   \qquad \lambda_{\pm} =\frac{Np_{\text{in}}\pm \sqrt{(N_2-N_1)^2p_{\text{in}}^2 + 4N_1N_2p_{\text{out}}^2}}{2}
\end{equation*}
 are the dominant eigenvalues of $\langle A \rangle_{\text{SBM}}$. The pseudo-inverse of $V_A$ is found by using the formula \mbox{$V_A^+ = V_A^\top (V_AV_A^\top)^{-1}$} (not shown here). Moreover, we choose two orthogonal eigenvectors of $K$ by inspection  to get
\begin{equation*}
   V_K = \left(
\begin{array}{c|c}
\frac{1}{\sqrt{N_1}}\bm{1}_{N_1} & \bm{0}_{N_2}  \\[0.1cm]
\hline
 \bm{0}_{N_1} & \frac{1}{\sqrt{N_2}}\bm{1}_{N_2}
\end{array}
\right).
\end{equation*}

\item We set $M = C_K V_K V_A^+ V_A$.
\item We choose 
\begin{equation*}
    C_K  = \begin{pmatrix}
    \frac{1}{\sqrt{N_1}} & 0 \\0 & \frac{1}{\sqrt{N_2}}
    \end{pmatrix},
\end{equation*}
which ensures that the resulting reduction matrix
\begin{equation*}
M_A = \left(
\begin{array}{c|c}
\frac{1}{N_1}\bm{1}_{N_1} &  \bm{0}_{N_2}  \\[0.1cm]
\hline
 \bm{0}_{N_1} & \frac{1}{N_2}\bm{1}_{N_2}
\end{array}
\right),
\end{equation*}
satisfies Condition~\ref{itm:condB}.
\item We find that  $(M^+)_{j\mu} = \delta_{\mu\,s(j)}$ which allows us to solve exactly the compatibility equations with $\,\,\mathcal{W} = \text{diag}(\omega, \omega)$, $\,\,\mathcal{K} = \text{diag}(K_{11}, K_{22})$, $\,\,$and
\begin{equation*}
\mathcal{A} = \begin{pmatrix}N_1p_{\text{in}} & N_2 p_{\text{out}}\\N_1 p_{\text{out}}& N_2 p_{\text{in}}\end{pmatrix}.
\end{equation*}

\end{enumerate}
The reduced dynamics is therefore exact to first order and is given by
\begin{align*} 
\dot{R}_{\mu} &=   \sigma\left(\frac{1-R_{\mu}^2}{2N}\right)\sum_{\nu = 1}^2  \mathcal{A}_{\mu\nu}R_{\nu}\cos(\Phi_{\nu} -\Phi_{\mu}-\alpha),\\
\dot{\Phi}_{\mu} &= \sigma\left(\frac{1+R_{\mu}^2}{2NR_{\mu}}\right)\sum_{\nu = 1}^2  \mathcal{A}_{\mu\nu}R_{\nu}\sin(\Phi_{\nu} -\Phi_{\mu}-\alpha),
\end{align*}
where, without loss of generality,  we have set the system in the center of mass by taking $\omega = 0$ \footnote{
Indeed, this choice gives the same differential equations as if we had made the substitution  $\theta_j = \phi_j + \omega t$, where $\phi_j$ is the phase variable in the center of mass referential.}. 

\subsubsection{Equilibrium points related to chimeras}

We want to identify structural and dynamical conditions that are necessary for the existence of chimeras.   For this, we first set $R_1 = 1$ and $R_2 = R_I < 1$ as in Ref.~\cite{Abrams2008}. Then, we further simplify the reduced dynamics by introducing the phase difference $\Phi = \Phi_1 - \Phi_2$, by merging the phase equations together, and by introducing a characteristic time 
\begin{equation}
\tau = \sigma t/N.
\label{eq:characteristic_time}
\end{equation}
The resulting differential equations are
\begin{align}
    R_I' &= \left(\frac{1-R_I^2}{2}\right) \left[N_2 p_{\text{in}} R_I \cos\alpha +  N_1 p_{\text{out}} \cos(\Phi - \alpha)\right],\label{rired}\\
    \Phi' &=   \left(\frac{1+R_I^2}{2R_I}\right)\left[N_2 p_{\text{in}} R_I \sin\alpha - N_1 p_{\text{out}} \sin(\Phi - \alpha)\right]\nonumber\\ &\quad - N_2 p_{\text{out}} R_I \sin(\Phi+\alpha) -  N_1 p_{\text{in}} R_I \sin\alpha, \label{phired}
\end{align}
where the prime indicates the derivative with respect to $\tau$. These equations are similar to those obtained in Ref.~\cite{Abrams2008} by following the Ott-Antonsen method,  except that we now have an explicit dependence over the network parameters $N_1, N_2, p_{\text{in}}$ and $p_{\text{out}}$.
 
We have proved that if the reduced system is in a chimera state, then Eqs.~\eqref{rired} and \eqref{phired} are satisfied. We now require the chimeras to be equilibrium points of the dynamics.  This is equivalent to imposing  $R_I' = 0$ and $\Phi' = 0$. A few basic manipulations allow us to conclude that the chimeras are equilibrium points only if the following equations are satisfied:  
\begin{align}
    p_{\text{in}} &= \frac{-f p_{\text{out}}\cos(\Phi - \alpha)}{ R_I(\Phi) \cos \alpha},\label{pin}\\ R_I(\Phi) &= \sqrt{\frac{2f\sin\alpha\cos(\Phi - \alpha) - \sin\Phi}{2f^{-1}\sin(\Phi + \alpha)\cos\alpha + \sin\Phi}}\label{ri},
\end{align}
where $f = N_1/N_2$ is the block asymmetry parameter. Note that we have redefined $R_I$ as a function~of~$\Phi$. 

\vspace{-1em}
\subsubsection{Chimeras in the density space}

We aim to clarify the impact of the modular structure on the existence of chimeras. It is already known that the difference between the intra-community coupling strength and the extra community coupling strength plays a critical role in the emergence of chimeras \cite{Abrams2008, Kotwal2017}.  We therefore introduce a new parameter $\Delta$ that captures this difference in coupling strength. Following Ref.~\cite{Young2017}, we make the change of variables
\begin{align*}
    \Delta &= p_{\text{in}} - p_{\text{out}},\\
    \rho &= \beta p_{\text{in}} + (1-\beta) p_{\text{out}},
\end{align*}
where 
\begin{equation*}
    \beta = \frac{N_1^2+N_2^2}{N^2}
\end{equation*}
is the sum of the maximum possible number of links within each community divided by the maximum number of possible links in the network. Note that $\rho$ is the average density in the SBM. The coordinates $(\rho, \Delta)$ form the \textit{density space} that characterizes all the possible graphs in the planted partition model. 

Let us go back to Eq.~\eqref{pin}. Using the new coordinates $(\rho, \Delta)$, the equation becomes
\begin{equation}
\Delta = \left[\frac{\:R_I(\Phi)\cos\alpha + f \cos(\Phi - \alpha)}{\beta f \cos(\Phi-\alpha) - (1-\beta)R_I(\Phi)\cos\alpha}\right]\rho\,\,\,\label{delta}.
\end{equation}
 
 \begin{figure}[tb]  
 \centering
   \includegraphics[width=1\linewidth]{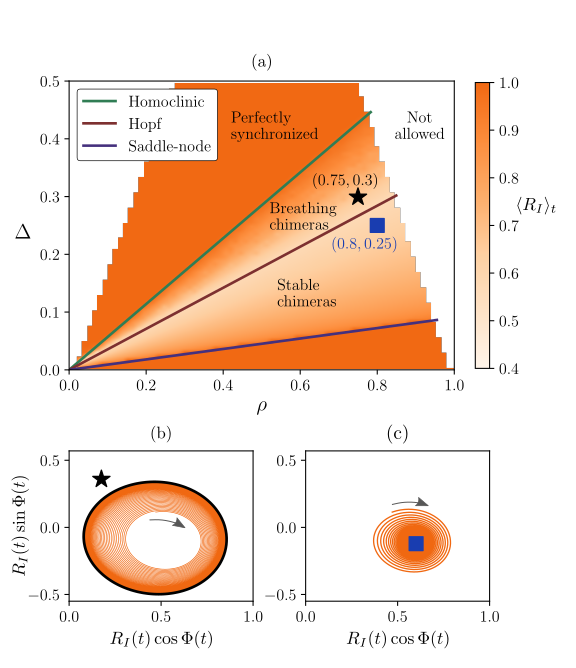}
   \vspace{-0.5cm}
   \caption{(Color online) Chimera state regions in the Kuramoto-Sakaguchi dynamics on the mean SBM in the density space. (a) Each point represents the value of $\langle R_I \rangle_t$, the time average of the phase synchronization observable of the incoherent community obtained with the integration of the complete dynamics. The initial conditions were taken at random from a uniform distribution [see Appendix~\ref{appendix:initial_conditions_chimeras} for more details]. Parameters: $f=1.5$, $\alpha=1.45$, $N=500$. (b) Breathing chimera at $(\rho, \Delta) = (0.75,0.3)$ obtained from the reduced dynamics~(\ref{rired}-\ref{phired}). (c) Stable chimera at $(\rho, \Delta) = (0.8,0.25)$ obtained from the reduced dynamics. The trajectories in the bottom figures have the same initial condition $(\:R_{I}(0)\:,\: \Phi(0) \:) = (0.48,0.24)$.
   }
   \label{fig:chimap}
 \end{figure}
 
We now identify the bifurcations of the reduced dynamics in the density space for fixed $\alpha$ and $f$~\footnote{Our analysis is different from previous studies where the bifurcation diagram for the parameters $\Delta$ and $\alpha$ is investigated by imposing that the sum of the in and out coupling values is equal one \cite{Abrams2008, Kotwal2017, Martens2010}. In our approach, this choice is equivalent to fixing the density at $\rho = 0.5$.}. Equation~\eqref{delta} already reveals the form of the bifurcation curves: they are straight lines in the density space whose slopes depend on the value of the equilibrium point  $\Phi^*$.

To obtain these different equilibrium points $\Phi^*$, we analyze
the Jacobian matrix of the differential equations~(\ref{rired}-\ref{phired}). We find a Hopf bifurcation (a stable chimera/equilibrium point loses stability and trajectories converge to a stable breathing chimera/limit cycle) by setting the trace of the Jacobian matrix equal to zero, i.e., 
\begin{equation}
    0 = f\cos(\Phi - \alpha) + \left[R_I(\Phi)\right]^2 \cos(\Phi+\alpha).
    \label{tr0}
\end{equation}
We also find a saddle-node bifurcation (a stable chimera is created or destroyed) by setting the determinant equal to zero \cite{Abrams2008}, i.e.,
\begin{align}\label{det0}
     0 = 1 - \left[R_I(\Phi)\right]^2&\Big[2\tan\alpha \sin(\Phi - \alpha)\cos(\Phi - \alpha) \\  &\,-\cos(2\alpha - 2\Phi) - 2f^{-1}\cos(2\Phi) \Big].\nonumber
\end{align}

Equations~(\ref{tr0}-\ref{det0}) allow us to compute the appropriate zeros for $\Phi$ by using standard root-finding algorithms. 
Substituting these zeros into Eq.~\eqref{ri} and then into Eq.~\eqref{delta} gives the slopes of the straight lines corresponding to the Hopf and the saddle-node bifurcations. 

The results are displayed in Fig.~\ref{fig:chimap}~(a) where we have integrated the $N$-dimensional Kuramoto-Sakaguchi dynamics for different values in the density space. We only show the assortative region ($p_{\text{in}} > p_{\text{out}} \Rightarrow \Delta > 0$) because the dissortative region ($p_{\text{in}} < p_{\text{out}} \Rightarrow \Delta < 0$) contains no chimeras. The white regions are ``Not allowed" because they are regions where $p_{\text{in}} \text{ or } p_{\text{out}} \notin [0, 1]$. We also illustrate the incoherent trajectories $R_Ie^{i\Phi}$ for two different kinds of chimeras: a breathing chimera (the incoherent trajectory reach the stable limit cycle in black) [Fig~\ref{fig:chimap}~(b)] and a stable chimera (the incoherent trajectory reach the stable equilibrium point in blue) [Fig~\ref{fig:chimap}~(c)].

More importantly, Fig.~\ref{fig:chimap}~(a) shows the agreement of the predicted Hopf (solid brown line) and saddle-node (solid purple line) bifurcations with the bifurcations in the complete dynamics (heatmap). The homoclinic bifurcation (solid green line) was obtained numerically and is shown to better see the complete chimera region, which is located between the homoclinic bifurcation and the saddle-node bifurcation.

\subsection{Periphery chimeras in two-star graphs}
\label{subsec:periphery_chimeras}

Two-star graphs are modular (see Sec.~\ref{subsec:modular_graph}).  Yet, their structural properties differ considerably from those of the planted partition model studied in the previous subsection.  We use DART to show that these structural differences also cause significant dynamical differences. In particular, we find that for two-star graphs, the Kuramoto-Sakaguchi dynamics with identical frequencies can lead to the emergence of a new kind of chimera.

\vspace{-1em}
\subsubsection{Reduced dynamics}

We consider a two-star graph divided into the modules $B_1,...,B_4$ defined in Sec.~\ref{subsec:modular_graph}. The
adjacency matrix is 
\begin{equation*}
A =
\left(
\begin{array}{c|c|c|c}
   & \bm{1}_{N_{p_1}} & 1 &  \\[0.1cm]
\hline
\bm{1}_{N_{p_1}}^\top &     &   &  \\[0.1cm]
\hline
1       &      &      & \bm{1}_{N_{p_2}}\\[0.1cm]
\hline
        &     & \bm{1}_{N_{p_2}}^\top   & 
\end{array}
\right),
\end{equation*} 
where we have separated the modules with lines and where the empty blocks are filled with zeros. The degree matrix is therefore 
\begin{equation*}
    K = \text{diag}(N_{p_1}+1, \underbrace{1,...,1}_{N_{p_1} \text{ times}}, N_{p_2}+1, \underbrace{1,...,1}_{N_{p_2} \text{ times}}),
\end{equation*}
and the frequency matrix is $W = \omega I_{N\times N}$.

Let us now go through all the steps of Procedure~\ref{proc:2} with $n = q = 4$ and $u = 2$.  

\begin{enumerate}[label=\arabic*:]
    \item We select the targets $T_1 = A$ and $T_2 = K$.
    
    \item We analytically find the eigenvector matrix $V_A$ and the eigenvector matrix of the degree matrix,
\begin{equation*}
   V_K = \left(
\begin{array}{c|c|c|c}
1 &   &   &  \\
\hline
  &  \frac{1}{\sqrt{N_{p_1}}}\bm{1}_{N_{p_1}}  &   &   \\
\hline
  &  & 1 &   \\
\hline
  &   &   & \frac{1}{\sqrt{N_{p_2}}}\bm{1}_{N_{p_2}} 
\end{array}
\right).
\end{equation*}
\item We set $M = C_K V_K V_A^+ V_A$.
\item We choose
\begin{equation*}
    C_K  = \begin{pmatrix}
    1 & 0 & 0 & 0 \\0 & \frac{1}{\sqrt{N_{p_1}}} & 0 & 0\\0 & 0 & 1 & 0\\0 & 0 & 0 & \frac{1}{\sqrt{N_{p_2}}}
    \end{pmatrix},
\end{equation*}
which gives the reduced matrix
\begin{equation*}
    M_A = \left(
\begin{array}{c|c|c|c}
1 &   &   &  \\
\hline
  &  \frac{1}{N_{p_1}}\bm{1}_{N_{p_1}}  &   &   \\[0.2cm]
\hline
  &  & 1 &   \\[0.2cm]
\hline
  &   &   & \frac{1}{N_{p_2}} \bm{1}_{N_{p_2}}
\end{array}
\right).
\end{equation*}
The latter satisfies Conditions~\ref{itm:condA}, \ref{itm:condB}, and \ref{itm:condA'}.
\item We solve exactly the  compatibility equations with 
\begin{align*}
\mathcal{W} &= \text{diag}(\omega,\omega,\omega,\omega),\\
\mathcal{K} &= \text{diag}(N_{p_1}+1, 1, N_{p_2}+1, 1),\\
    \mathcal{A} &= 
    \begin{pmatrix}
    0 & N_{p_1} & 1 & 0 \\
    1 & 0 & 0 & 0 \\
    1 & 0 & 0 & N_{p_2}\\
    0 & 0 & 1 & 0 \end{pmatrix}.
\end{align*}
\end{enumerate}

We find that the reduced equations for the radial and phase variables are
\begin{align*}
    \dot{R}_{\mu} &=  \sigma\left(\frac{1-R_{\mu}^2}{2N}\right)\sum_{\nu=1}^{4}\mathcal{A}_{\mu\nu}R_{\nu}\cos{(\Phi_{\nu} - \Phi_{\mu} - \alpha)},\\
    \dot{\Phi}_{\mu} &=\omega+  \sigma\left(\frac{1+R_{\mu}^2}{2NR_{\mu}}\right)\sum_{\nu=1}^{4}\mathcal{A}_{\mu\nu}R_{\nu}\sin{(\Phi_{\nu} - \Phi_{\mu} - \alpha)}.
\end{align*}

The above equations  have the same form as the reduced equations of the Kuramoto-Sakaguchi dynamics on the mean SBM. This observation can be generalized for other types of networks and dynamics.  Indeed, as shown in Table \ref{tab:table4} of Appendix~\ref{appendix:reduced_dynamics}, there is always a simpler form of the reduced dynamics when  $\mathcal{W}_{\mu\nu} = \Omega_{\mu}\delta_{\mu\nu}$ and $\mathcal{K}_{\mu\nu} = \kappa_{\mu}\delta_{\mu\nu}$. 

It is worth pointing out that the cores have their own observables and they are already maximally synchronized (with themselves), which is not true for the peripheries. Thus, extra care should be taken for two-star graphs to avoid any confusion between the cores and the peripheries. With this in mind, 
 we slightly change the notation as follows:
\begin{align*}
(R_1, R_2, R_3, R_4) &= (R_{c_1},R_{p_1}, R_{c_2}, R_{p_2}),\\
(\Phi_1, \Phi_2, \Phi_3, \Phi_4) &= (\Phi_{c_1},\Phi_{p_1},\Phi_{c_2},\Phi_{p_2}),
\end{align*}
where $c_{\mu}$ and $p_{\mu}$  stand for ``core $\mu$'' and ``periphery $\mu$'', respectively.
The trivial conditions now become \mbox{$R_{c_1}=1$} and \mbox{$R_{c_2}=1$}, which imply  \mbox{$\dot R_{c_1}=0$} and \mbox{$\dot R_{c_2}=0$}. Thus, among the eight dynamical equations, only six are nontrivial.

We reduce again the number of equations by introducing variables that describe the phase differences between the communities, i.e., \mbox{$\Phi_1 = \Phi_{p_1} - \Phi_{c_1}$}, \mbox{$\Phi_2 =\Phi_{p_2} - \Phi_{c_2}$}, and \mbox{$\Phi_{12} = \Phi_{c_2} - \Phi_{c_1}$}. 
We end up with five equations to describe the Kuramoto-Sakaguchi dynamics on the two-star graph. Two equations are related to the synchronization observables in the peripheries:
\begin{equation}
\label{eq:R_sk_red_ts}
\begin{split}
    R_{p_1}' &=  \left(\frac{1-R_{p_1}^2}{2}\right) \cos(\Phi_1 + \alpha), \\
    R_{p_2}' &=  \left(\frac{1-R_{p_2}^2}{2}\right) \cos(\Phi_2 + \alpha),
\end{split}
\end{equation}
where we have used the characteristic time defined in Eq.~\eqref{eq:characteristic_time}. The other three equations describe the evolution of the phase observables:
\begin{align}
    \Phi_{1}' = &- \left(\frac{1+R_{p_1}^2}{2R_{p_1}}\right) \sin(\Phi_1 + \alpha)\nonumber \\& \phantom{\frac{1}{I}}- N_{p_1}R_{p_1}\sin(\Phi_1 -\alpha) -  \sin(\Phi_{12}  -\alpha),\nonumber\\
    \Phi_{2}' = &-\left(\frac{1+R_{p_2}^2}{2R_{p_2}}\right) \sin(\Phi_2 + \alpha)\label{eq:Phi_sk_red_ts} \\& \phantom{\frac{1}{I}}- N_{p_2}R_{p_2}\sin(\Phi_2 -\alpha)  +  \sin(\Phi_{12} +\alpha),\nonumber\\ 
    \Phi_{12}' = & \phantom{\frac{1}{I}}N_{p_2}R_{p_2}\sin(\Phi_2 -\alpha) -  \sin(\Phi_{12} +\alpha) \nonumber \\& \phantom{\frac{1}{I}}- N_{p_1}R_{p_1}\sin(\Phi_1 -\alpha)- \sin(\Phi_{12} -\alpha)\nonumber.
\end{align}

\begin{figure}[tb] 
  \centering
  \includegraphics[width=1\linewidth]{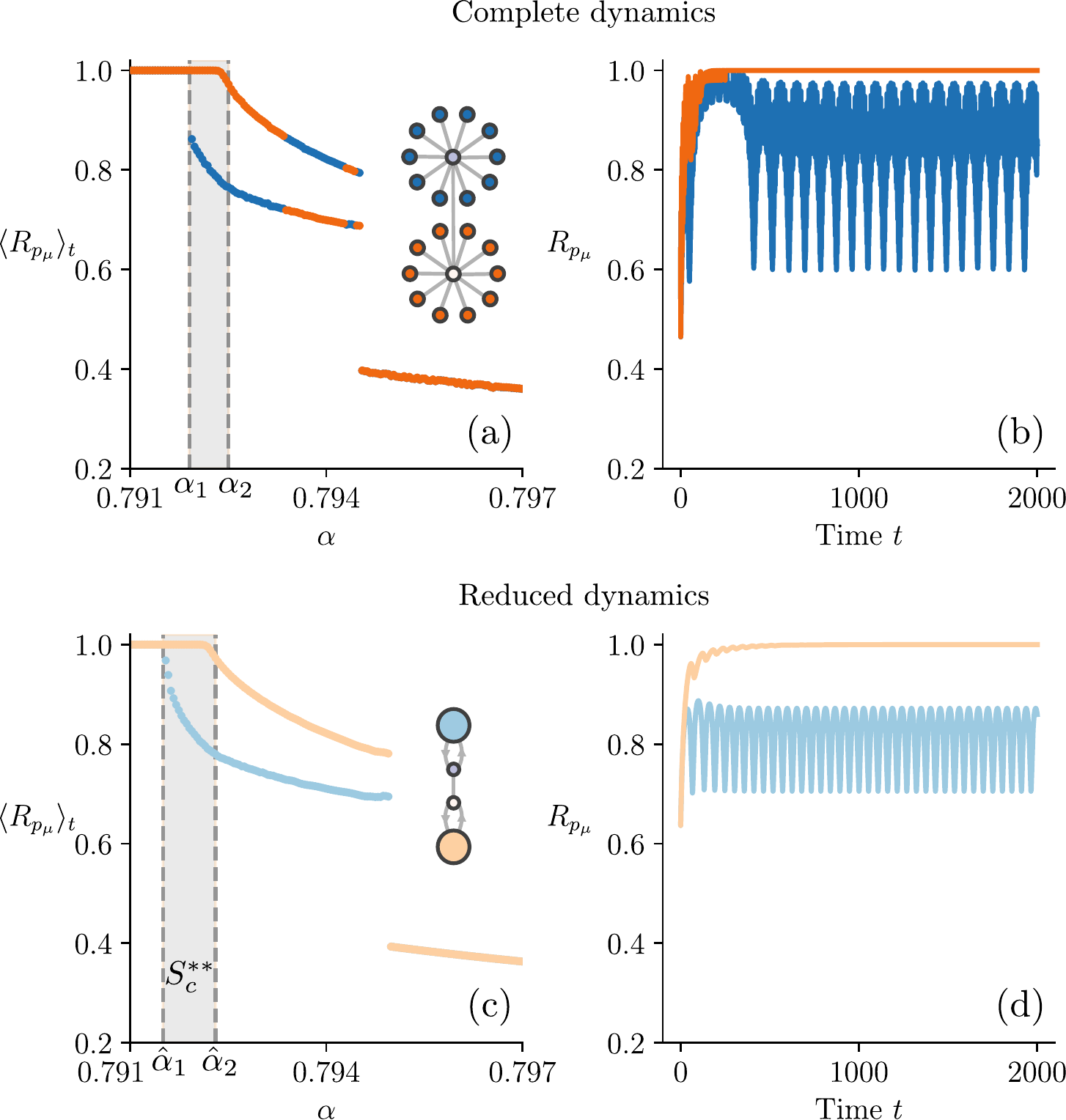}
  \vspace{-0.6cm}
  \caption{(Color online) (Left) Effect of the phase lag $\alpha$ on the time-averaged synchronization observable in periphery $p_\mu$ of a two-star graph for (a) the complete Kuramoto-Sakaguchi dynamics and (c) its reduced version (\ref{eq:R_sk_red_ts}-\ref{eq:Phi_sk_red_ts}). Desynchronization of the peripheries does not happen at the same values of $\alpha$. The region where chimeras exist is represented by the darker region between the two vertical dashed lines at $\alpha_1$ and $\alpha_2$ ($\hat{\alpha}_1$ and $\hat{\alpha}_2$).  (Right) An example of a periphery chimera for (b) the complete and (d) the reduced dynamics, both with $\alpha = 0.792$. Global parameters: $\tau = 5t/N$, $N_{p_1} = N_{p_2} = 100$, $N = 202$. Initially, the $N$ phases are equidistantly distributed between 0 and $2\pi$. The color varies from one periphery to the other. 
  }
  \label{fig:twostarsdesynchro}
\end{figure}

\begin{figure}[tb] 
  \centering
  \includegraphics[width=1\linewidth]{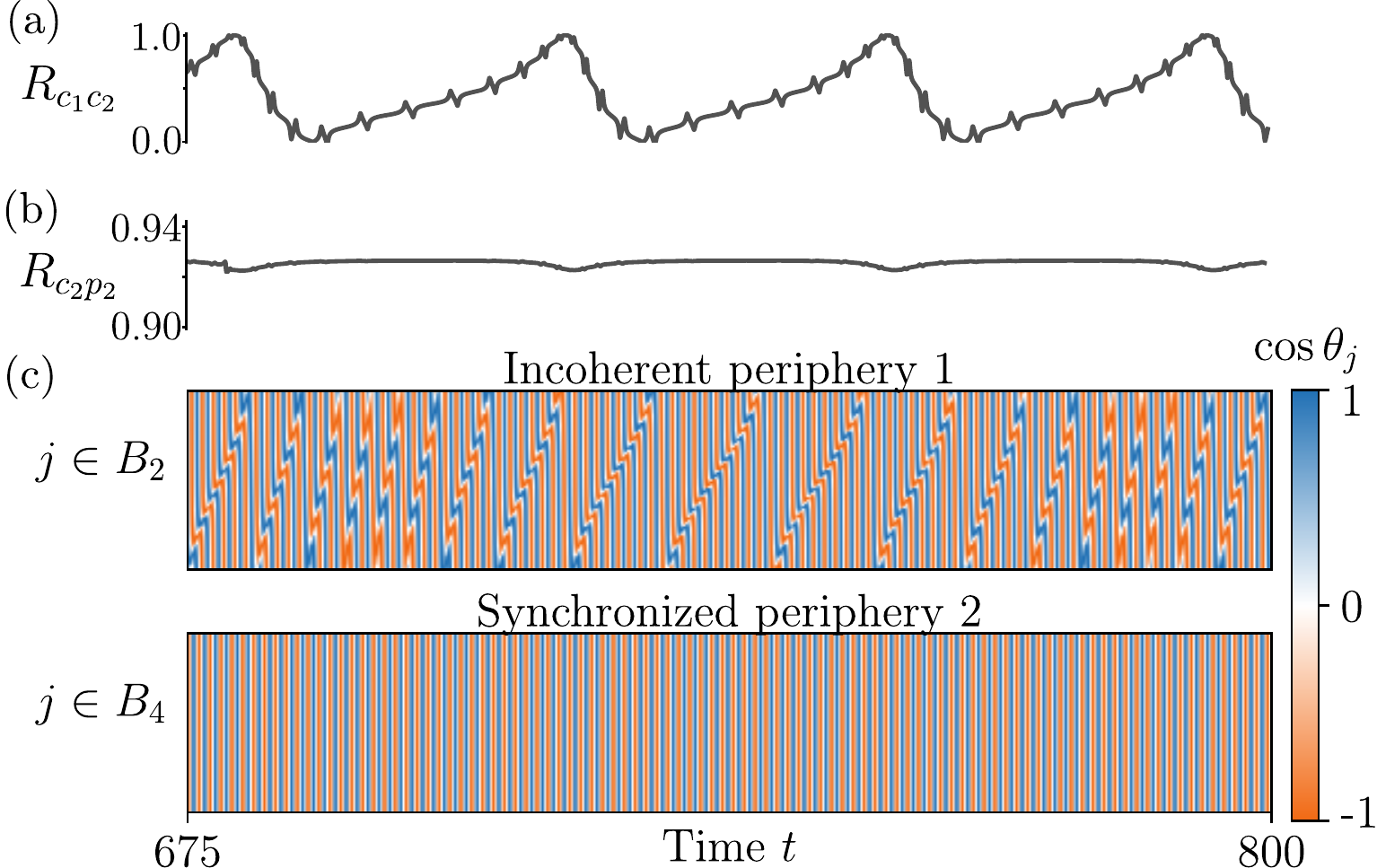}
  
  \vspace{-0.3cm}
  
  \caption{(Color online) Periphery chimera in the complete Kuramoto-Sakaguchi dynamics. (a) Synchronization between the two cores vs. time. (b) Synchronization between the second core and its periphery vs. time. (c) Cosinus of the phases $\theta_j$ for the peripheries vs. time. Parameters: $\alpha = 0.792$, $\tau = 5t/N$, $N_{p_1} = N_{p_2} = 100$, $N = 202$. Initially, the $N$ phases are equidistantly distributed between 0 and $2\pi$.}
  \label{fig:periphery_chimera}
\end{figure}

\subsubsection{Searching for chimeras}

According to our previous analysis, there is a chimera state in the two-star graph if either $R_{p_1}=1$ and $ R_{p_2}<1$ or $ R_{p_1}<1$ and $ R_{p_2}=1$. Yet, imposing these conditions yields four coupled differential equations for which the dynamical analysis is much more complicated. We will therefore rely on numerical analysis. 

We fix the parameters $\sigma$, $N_{p_1}$, and $N_{p_2}$. This allows us to study how the phase lag $\alpha$ affects the synchronization in each periphery. As shown in Fig.~\ref{fig:twostarsdesynchro} (a) and (c), for each periphery, there is a critical value [dashed vertical line] above which desynchronization begins. For the complete (reduced) dynamics, we have denoted these critical values $\alpha_{1}$ and $\alpha_{2}$ ($\hat{\alpha}_{1}$ and $\hat{\alpha}_{2}$). 

Notably, we observe a clear separation between the critical values $\alpha_{1}$ and $\alpha_{2}$ ($\hat{\alpha}_{1}$ and $\hat{\alpha}_{2}$). Let us stress that this separation happens despite the complete similarity of the peripheries: same number of nodes and same natural frequencies. Therefore, there is a region [dark region between the two vertical dashed lines in Fig.~\ref{fig:twostarsdesynchro}~(a) and (c)] where a periphery is coherent (the one that has the higher critical value) while the other is incoherent. We call any state that belongs to that region a periphery chimera. Examples of periphery chimeras are given in Fig.~\ref{fig:twostarsdesynchro}~(b) and (d). In these examples, the periphery chimeras are breathing~\cite{Abrams2008}. 

To characterize periphery chimeras more completely, it is important to know if the two cores are synchronized and if the perfectly synchronized periphery is also synchronized with its core. As portrayed in Figure~\ref{fig:periphery_chimera} (a), the two cores are not synchronized: $R_{c_1 c_1} = |e^{i\theta_1} + e^{i\theta_{N_{p_1}+1}}|/2$ oscillates between 0 and 1. Figure~\ref{fig:periphery_chimera} (b) also shows that the second core is not perfectly synchronized (and even not perfectly phase-locked) with its perfectly synchronized periphery. Indeed, $R_{c_2p_2} = |e^{i\theta_1} + e^{i\theta_{j \in B_4}}|/2$ does not reach 1 and oscillates. We finally observe an interesting periodic pattern for the incoherent periphery in Fig.~\ref{fig:periphery_chimera} (c).

Because star graphs are building blocks of complex networks, our results suggest that periphery chimeras could exist within very small regions of the parameter space describing the phase dynamics of a complex network. Moreover, with the recent experimental success in observing exotic synchronization phenomena \cite{Totz2018, Tinsley2012, Wojewoda2016, Kapitaniak2014, Martens2013, Hagerstrom2012, Matheny2019}, there are reasons to believe that periphery chimeras can be detected experimentally using nanoelectromechanical oscillators \cite{Matheny2019} for instance.

\subsection{Chimera region size}
We investigate the impact of the asymmetry between the communities on the size of the chimera region for the reduced dynamics of the Kuramoto-Sakaguchi model on the mean SBM (\ref{rired}-\ref{phired}) and the two-star graph (\ref{eq:R_sk_red_ts}-\ref{eq:Phi_sk_red_ts}).

First, in the case of the mean SBM, the size of the chimera region in the density space is 
\begin{equation}
    S_c = \frac{1}{2}\left|\rho_2\Delta_3 - \Delta_2\rho_3\right|,
    \label{size_sbm}
\end{equation}
where the coordinates $(\rho_2, \Delta_2)$ and $(\rho_3, \Delta_3)$ are the points of the intersection between the straight lines related to the bifurcations (homoclinic, saddle-node) and the limit points of the allowed region in the density space. Using Eq.~\eqref{det0}, Eq.~\eqref{delta}, and the equation $\Delta = (1-\rho)/(1 - \beta)$ that corresponds to the right limit of the density space, we find that the size of the stable chimera region is given by the equations
\begin{align*}
    S_c^{\,\mathrm{stable}} & = \frac{1}{2} \left| \frac{a_H - a_{S}}{[(1-\beta)a_H +1][(1-\beta)a_{S} + 1]} \right|,\\
    a_X &= \frac{[R_I\cos\alpha + f \cos(\Phi_X - \alpha)]}{\beta f \cos(\Phi_X-\alpha) - (1-\beta)R_I\cos\alpha},
\end{align*}
where $X\in\{H,S\}$ indicates whether the bifurcation is homoclinic ($H$) or saddle-node ($S$).  The symbol $a_X$ denotes the slope of the bifurcation $X$ while $\Phi_X$ is the zero found numerically from Eqs.~(\ref{tr0}-\ref{det0}). 

\begin{figure}[tb] 
 \centering
   \includegraphics[width=1\linewidth]{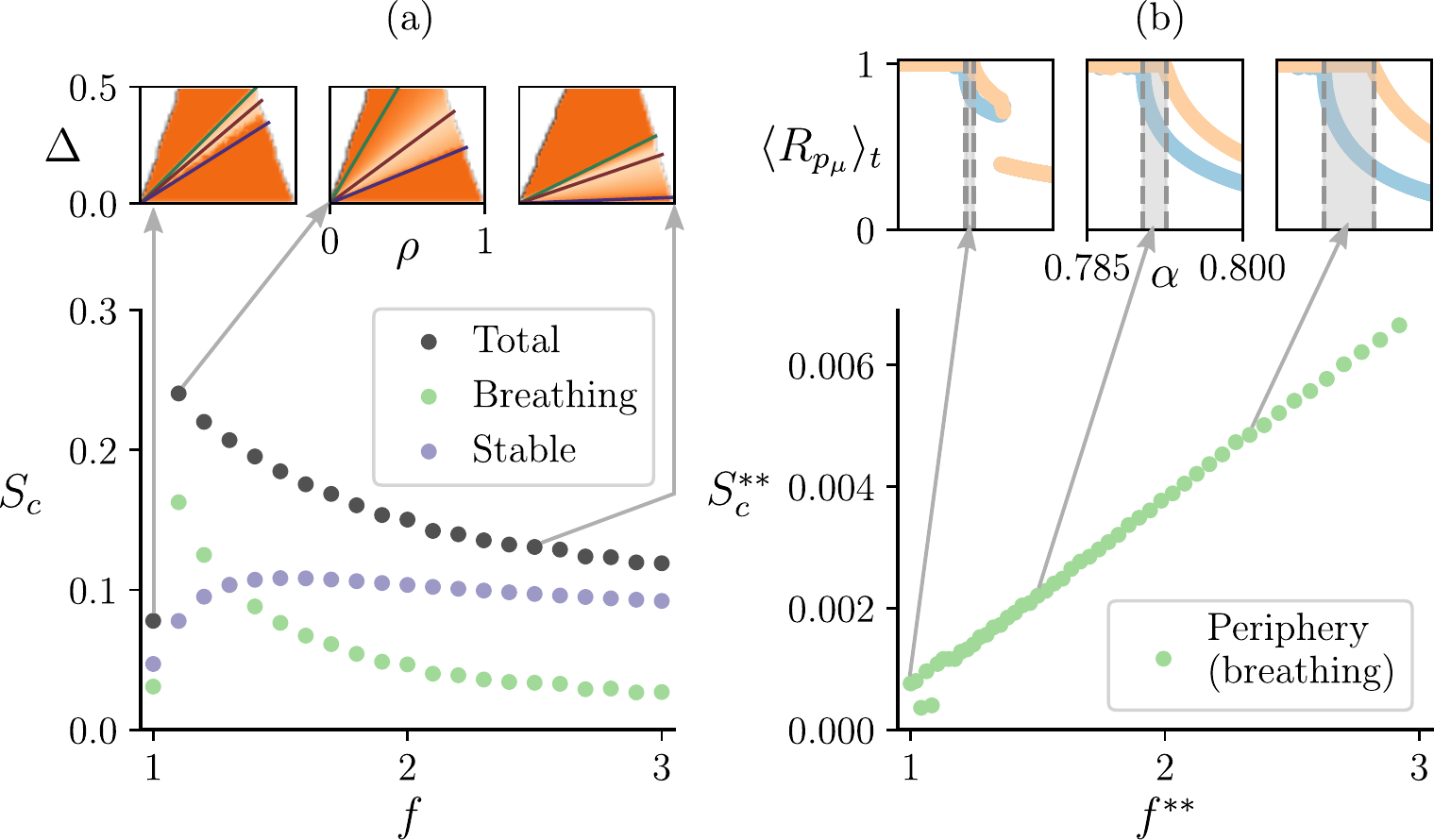}
   \vspace{-0.8cm}
   \caption{(Color online) (a) Size $S_c$ of the chimera regions [Eq.~\eqref{size_sbm}] in the density space for the reduced Kuramoto-Sakaguchi dynamics~(\ref{rired}-\ref{phired}) on the mean SBM vs.\ the block asymmetry parameter $f = N_1/N_2$ for $\alpha = 1.45$. At the top of the plot, chimera regions are shown for $f = 1, \,1.1$, and $2.5$. (b) Size $S_c^{**}$ of the periphery chimera region in terms of $\alpha$ [Eq.~\eqref{size_ts}] for the reduced Kuramoto-Sakaguchi dynamics~(\ref{eq:R_sk_red_ts}-\ref{eq:Phi_sk_red_ts}) on the two-star graph vs.\ the block asymmetry parameter $f^{**} = N_{p_1}/N_{p_2}$ for $\sigma = 5$. At the top of the plot, chimera regions are shown for $f^{**} = 1, \, 1.5$, and $2.3$.}
   \label{fig:sizevsf}
 \end{figure}

The results are illustrated in Fig.~\ref{fig:sizevsf} (a) where we observe that the size of the stable and breathing chimera regions evolves non-linearly as a function of $f$. As a consequence, the total size of the chimera region quickly goes from its minimal value at $f=1.0$ to its peak at $f \approx 1.1$, before decreasing rapidly as $f$ increases. A small structural change in the reduced equations can therefore have a considerable impact on the resulting chimera region. 

Second, in the case of the Kuramoto-Sakaguchi on the two-star graph, we investigate the impact of the asymmetry on the size of the region with periphery chimeras, which is defined along the $\alpha$-axis. For most values of $\alpha$, like the one in Fig.~\ref{fig:twostarsdesynchro}~(c), the size of the periphery chimera region is approximated as 
\begin{equation}
    S_c^{**} \approx |\hat{\alpha}_2 - \hat{\alpha}_1|,
    \label{size_ts}
\end{equation}
where we recall that $\hat{\alpha}_\mu$ is the critical value above which there is desynchronization of the periphery $p_\mu$ in the reduced dynamics. In Fig.~\ref{fig:sizevsf}~(b), we observe that the relationship between the size $S_c^{**}$ and $N_{p_1}/N_{p_2}$ is approximately linear.

Figure~\ref{fig:sizevsf} thus shows that the size of the chimera regions, which can be expressed in terms of specific structural parameters [ex.\ $\rho$ and $\Delta$ in Eq.~\eqref{size_sbm}] or the dynamical parameters [ex.\ $\alpha$ in Eq.~\eqref{size_ts}], is closely related to another structural parameter [i.e., $f$]. When some structural or dynamical parameters are fixed, the asymmetry between the community sizes can therefore dictate whether or not a chimera emerges.

\subsection{Explosive synchronization in star graphs}

Explosive synchronization is characterized by discontinuous (first-order) phase transitions.  It is used to describe the occurrence of sudden catastrophes, such as epileptic seizures or electric breakdown \cite{DSouza2019,Boccaletti2016,Vlasov2015}. A simple but instructive example of a system that can evolve towards explosive synchronization is the Kuramoto model on the star graph with degree-frequency correlation \cite{Gomez-Gardenes2011,Zou2014,Gao2016synchro,Chen2017,Rodrigues2016}. Interestingly, the model's bifurcation diagram for the global synchronization observable exhibits hysteresis \cite{Gomez-Gardenes2011}. 

The goal of this subsection is to use DART to gain analytical insights about the Kuramoto-Sakaguchi model on the star graph. In particular, we want to reproduce some known results on explosive synchronizations. 

\vspace{-1em}
\subsubsection{Reduced dynamics}

The Kuramoto-Sakaguchi model on the two-star graph has already been analyzed in Sec.~\ref{subsec:periphery_chimeras}. Since the star graph is contained in the two-star graph, it is straightforward to adapt our previous calculations to derive the appropriate reduced dynamics. 

First, let us introduce the matrices that define the frequencies and the structure of the complete dynamics of dimension $N = N_p + 1$, i.e.,
\begin{align*}
W &= \text{diag}(\omega_1\:\:,\:\: \underbrace{\omega_2\:\:,\:\:...\:\:,\:\:\omega_2}_{N_{p} \text{ times}}),\\
K &=\text{diag}(N_{p}\:\:,\:\: \underbrace{1\:\:,\:\:...\:\:,\:\:1}_{N_{p} \text{ times}}),\\
A &=\left(\begin{array}{c|c}0 & \bm{1}_{N_{p}}  \\[0.1cm] \hline \bm{1}_{N_{p}}^\top &  0_{N_{p}\times N_{p}} \end{array} \right).
\end{align*} 
Setting $n = q = 2$ and using Procedure \ref{proc:2} with \mbox{$A \to W \to K$} yields
\begin{align*}
\mathcal{W} &= \text{diag}(\omega_1,\omega_2),\\
\mathcal{K} &= \text{diag}(N_{p},1),\\
    \mathcal{A} &= 
    \begin{pmatrix} 
    0 & N_{p}  \\
    1 & 0
\end{pmatrix}.
\end{align*}

The reduced dynamics is therefore given by Eq.~\eqref{redmpkuramoto_K_I} with $\Omega_1 = \omega_1$, $\Omega_2 = \omega_2$, $\sigma/N \mapsto \sigma$ and the reduced adjacency matrix $\mathcal{A}$ defined above. 

Let $R_2 = R_p$ be the phase synchronization observable of the periphery and let $\Phi = \Phi_1 - \Phi_2$ be the difference between the phase of the core and the phase observable of the periphery. Then, the reduced dynamics of the Kuramoto-Sakaguchi model on the star graph is
\begin{align}
    \dot{R}_p &= \sigma \left(\frac{1-R_p^2}{2}\right)\cos(\Phi - \alpha), \label{kuramoto_star_R}\\
    \dot{\Phi} &= \omega_1 - \omega_2 - \sigma (N-1)R_p\sin(\Phi + \alpha)  \label{kuramoto_star_phi}\\&\quad\quad\quad\quad\;\;\:- \sigma\left(\frac{1+ R_p^2}{2R_p}\right)\sin(\Phi - \alpha)\nonumber,
\end{align}
which is also reported in Ref.~\cite{Chen2017} apart from a difference in sign \footnote{The equation of Ref.~\cite{Chen2017} that corresponds to our  Eq.~\eqref{kuramoto_star_R} contains the factor  $\cos(\Phi +\alpha)$, instead of $\cos(\Phi - \alpha)$, which is the correct factor.}.

\vspace{-1em}
\subsubsection{Global synchronization observable at equilibrium}

The observable that measures the global synchronization of the network is the modulus of the complex observable $Z = R e^{i\Phi}$ defined in Eq.~\eqref{eq:macroscopic_order_parameter} with $\bm{\ell} = (1\quad N_{p})$,  that is, 
\begin{equation}
     R = \frac{1}{N}\sqrt{1 + (N-1)^2R_p^2 + 2(N-1)R_p\cos\Phi}.
    \label{R_glob_star}
\end{equation}
To find the equilibrium points of the global synchronization we first notice that 
\begin{equation*}
    \frac{\text{d}(R^2)}{\text{d}t} = 2R\dot{R} = 0 \iff R=0 \text{\:\:or\:\:} \dot{R} = 0. 
\end{equation*}
Hence, the global synchronization is at equilibrium if the derivative of $1 + (N-1)^2R_p^2 + 2(N-1)R_p\cos\Phi$ with respect to time is zero. 

From the above analysis, we deduce that if both $R_p$ and $\Phi$ are at equilibrium, then so is $R$.  Now, the equilibrium solutions for $R_p$ and $\Phi$  are the zeros of  Eqs.~(\ref{kuramoto_star_R}--\ref{kuramoto_star_phi}). Using 
 trigonometric identities (see \footnote{There is a non-trivial part in the procedure where we must find $\Phi$ such that $A\sin\Phi + B\cos\Phi = C$. To solve the equation, we divide both sides by $D = \sqrt{A^2 + B^2}$ and define an angle $\Theta = \arccos(A/D)$.  Hence, $\cos^2\Theta + \sin^2\Theta = 1$. Moreover,  $\cos\Theta\sin\Phi + \sin\Theta\cos\Phi = \sin(\Phi + \Theta) = C/D$.  The last equality then allows  to express $\Phi$ in terms of $A$, $B$, and $C$.}), we readily solve these equations and get 
\begin{align*}
    R_p^* &= 1, \\
    \Phi^* &= \arcsin\left(\frac{\omega_1 - \omega_2}{\sigma r(N, \alpha)}\right) - \arcsin\left(\frac{(N-2)\sin{\alpha}}{r(N, \alpha)}\right),
\end{align*}
where the asterisk indicates equilibrium and  \begin{equation*}
    r(N, \alpha) = \sqrt{N^2 - 4(N-1)\sin^2\alpha}.
\end{equation*}
After substituting these solutions into  Eq.~\eqref{R_glob_star}, we obtain the corresponding equilibrium point for the global synchronization observable: 
\begin{equation}
    R^* = \frac{1}{N}\sqrt{1 + (N-1)^2 + 2(N-1)\cos\Phi^*}.
    \label{top_branch_gen}
\end{equation}
The equilibrium point exists for all values of $\sigma$ that are greater than or equal to the critical coupling
\begin{equation}
    \sigma_c = \frac{\omega_1 - \omega_2}{r(N, \alpha)}.
    \label{critical_sigma_ks}
\end{equation}
Below the critical coupling, a real positive solution for $R^*$ does not exist. At the critical coupling, $R^*$ is equal to
\begin{equation}
    R_c^* = \frac{1}{N}\sqrt{1 + (N-1)^2
    + \frac{s(N,\alpha)}{ r(N, \alpha)}},
    \label{critical_R_ks}
\end{equation}
where 
\begin{equation*}
s(N,\alpha) = 
2(N-1)(N-2)\sin^2\alpha.
\end{equation*}

\begin{figure}[t] 
\includegraphics[width=0.95\linewidth]{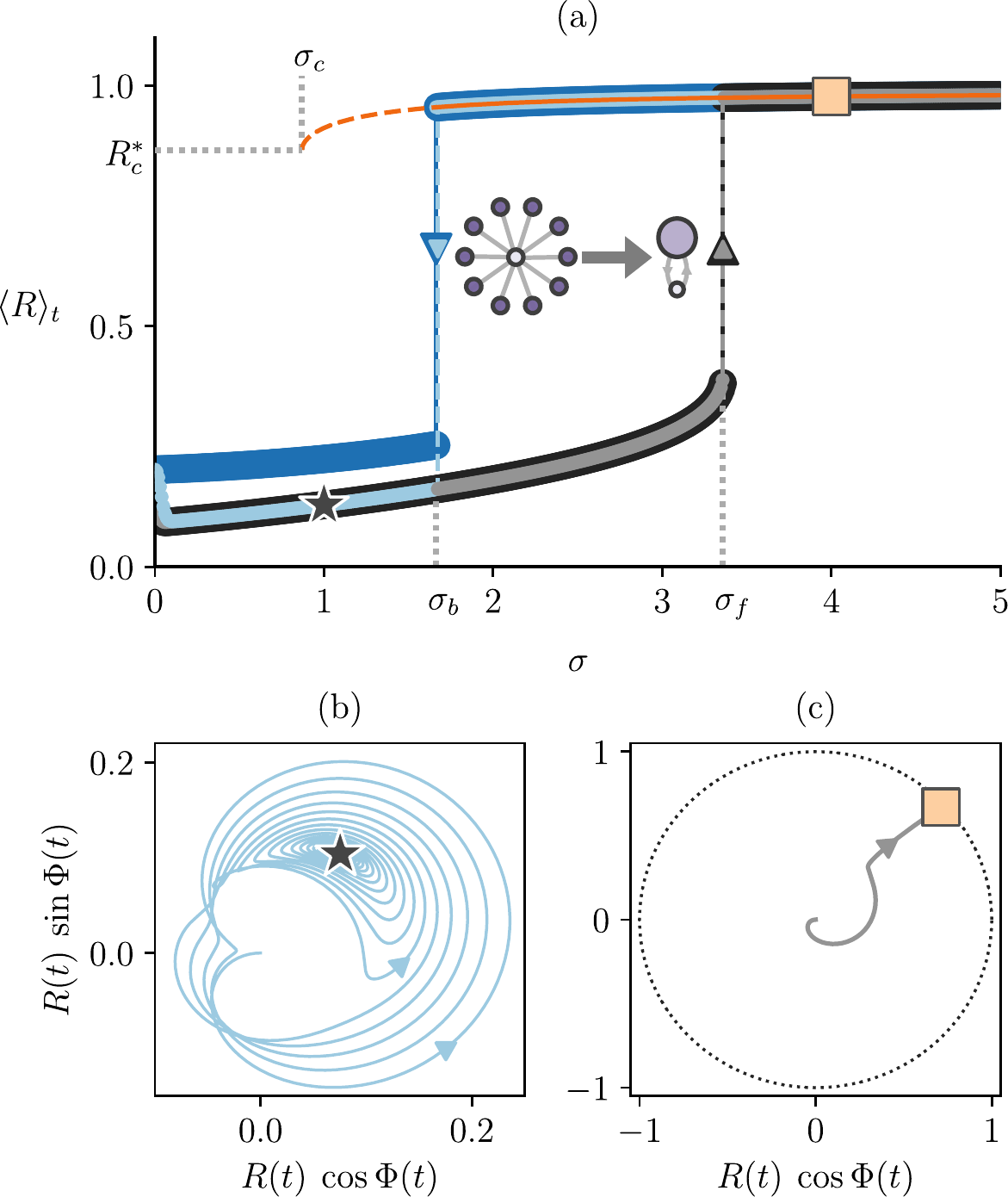}
\vspace{-0.3cm}
  \caption{(Color online) (a) Hysteresis of the time-averaged global synchronization observable $\langle R \rangle_t$ vs.\ the coupling constant $\sigma$ in the Kuramoto-Sakaguchi dynamics on the star graph. The numerical results are shown for the complete dynamics [dark blue  (backward branch) and black (forward branch) markers] and reduced dynamics~(\ref{kuramoto_star_R}-\ref{kuramoto_star_phi}) [light blue (backward branch) and gray (forward branch) markers]. The orange line represents the analytical result \eqref{top_branch_gen}. The dashed lines demarcate the domain  where unstable equilibria are found numerically.
  (b)~Unsynchronized trajectory of the reduced dynamics at $\sigma = 1$. (c)~Synchronized trajectory of the reduced dynamics at $\sigma = 4$. The dotted unit circle is the boundary for the trajectory. Initial conditions: $\theta_j(0) = 2\pi(j-1)/N$. Parameters: $\alpha = -0.2\,\pi$, $N = 11$.
  }
  \label{fig:explosive_synchro}
\end{figure}

The equilibrium points for the global synchronization observable $R$ are illustrated in Fig.~\ref{fig:explosive_synchro} (a). We observe that the equilibrium points form a hysteresis with two branches: backward branch (blue) and forward branch (gray). The numerical solutions obtained from the complete (reduced) dynamics are shown with darker (lighter) markers. We observe that the solutions of the reduced dynamics agree with those of the complete dynamics, except for the backward branch in the domain $0<\sigma<\sigma_b\approx 1.66 $. An example of synchronized trajectory observed in this domain is illustrated in Fig.~\ref{fig:explosive_synchro}~(b), which contrasts with the synchronized solution shown in Fig.~\ref{fig:explosive_synchro}~(c).

The analytical solution \eqref{top_branch_gen} is indicated by the orange line in Fig.~\ref{fig:explosive_synchro} (a).  This solution clearly belongs to the backward branch of the hysteresis.
For $\alpha = 0$, Eq.~\eqref{top_branch_gen} yields the same result as the backward branch reported in Ref.~\cite{Zou2014}. For $\alpha = 0$, $\omega_1 = N_{p}$, and $\omega_2 = 1$, the critical coupling becomes
\begin{equation*}
    \sigma_c = \frac{N_{p} - 1}{N_{p} + 1},
\end{equation*}
which corresponds to value of the top branch of the hysteresis found in Ref.~\cite{Gomez-Gardenes2011}.  The substitution of this result into Eq.~\eqref{top_branch_gen}
leads to the critical value reported in Ref.~\cite{Zou2014}, i.e., 
\begin{equation*}
    R^*_c = \frac{\sqrt{N_{p}^2 +1}\,\,}{N_{p} + 1}.
\end{equation*}
The forward branch critical synchronization value \cite{Chen2017}
\begin{equation*}
    \sigma_f = \frac{\omega_1 - \omega_2}{\sqrt{2(N-1)\cos(2\alpha) + 1}},
\end{equation*}
predicts accurately the numerical results obtained from the complete and the reduced dynamics. We also observe that the backward branch critical desynchronization value $\sigma_b$ occurs before the critical value $\sigma_c$ of the equilibrium point $R^*$ in Eq.~\eqref{top_branch_gen}. This is explained by the fact that the equilibrium point is not generally stable for all $\sigma > \sigma_c$ and for $\alpha \neq 0$ \cite{Chen2017}. However, the coupling value $\sigma_b$ we obtain is not equal to the value
\begin{equation*}
    \frac{\omega_1 - \omega_2}{\sqrt{(N-1)\cos(2\alpha) + 1}},
\end{equation*}
reported in Refs.\ \cite{Chen2017} and \cite{Huang2016}.

To summarize, using DART with $A\to W \to K$, we successively obtain multiple analytical results on explosive synchronization. At first glance, the use of the three targets procedure may seem exaggerated to derive the reduced dynamics~(\ref{kuramoto_star_R}-\ref{kuramoto_star_phi}), since any one-target procedure would give the same reduced dynamics in this case. However, any heterogeneity in the frequencies or the structure  (e.g., connecting nodes in the periphery) would break the possibility to satisfy the three compatibility equations with a one-target procedure. We conjecture that using DART with three target matrices will come in handy for these kinds of perturbed systems and even pave the way for new interesting analytical results on explosive synchronization.


\section{Conclusion}
\label{sec:conclusion}

We have introduced a Dynamics Approximate Reduction Technique (DART) to obtain a low-dimensional dynamical system from a high-dimensional dynamical system on finite networks. DART generalizes previous approaches \cite{Gao2016, Jiang2018, Laurence2019} for modular graphs with strong interactions between the modules, dynamically non-identical nodes, and complex observables. 

Our approach has uncovered a threefold problem: to close the reduced dynamics to first order, one needs to solve three compatibility equations involving a dynamical parameter matrix $W$, the degree matrix $K$, the adjacency matrix $A$, and the reduction matrix $M$. 

The form of these compatibility equations has revealed a spectral aspect of the dimension-reduction problem. If one finds a common eigenvector basis for the matrices $W$, $K$, and $A$, then the compatibility equations are satisfied and the reduced dynamics is exact to the first order. However, these matrices do not share eigenvectors in general. 

To tackle this problem, we have introduced two procedures designed to solve one compatibility equation exactly and the two others approximately. The key for achieving this task is to perform linear transformations that project eigenvectors of one of the three matrices as close as possible to eigenvectors of the other matrices. 

Yet, it is not trivial to choose the eigenvectors defining the observables of the reduced dynamics. The threefold character of DART has demonstrated that the choice of observables is subtle and not universal for dynamics with non-identical nodes: the choice depends on the dynamical and structural setup. Despite this fact, the various numerical experiments performed with the Kuramoto model on a small graph suggest that choosing the eigenvectors of the adjacency matrix, i.e., satisfying the corresponding compatibility equation exactly, is advantageous in most cases to predict the synchronization transitions curves, and should be preferred.

Using the eigenvectors of $A$ and $W$, we have derived the reduced dynamics for the Winfree, Kuramoto, and theta models on random modular networks. The global phase synchronization curves obtained with the reduced dynamics are in agreement with those of the complete dynamics and exhibit, in particular, double phase transitions and SNIC bifurcations. 

We have also analyzed the Kuramoto-Sakaguchi dynamics on different modular graphs with DART to get insights on exotic synchronization states. For the mean SBM, we have first investigated the impact of the network density and community size asymmetry on the existence of chimeras. We have detected new chimera regions in the SBM density space whose size varies in a surprising nonlinear way according to the community size asymmetry. In particular, we have found that the size of the chimera regions peaks when asymmetry is small. For the two-star graph, we have shown the existence of periphery chimera states, i.e., a dynamical regime in which the periphery of one star is perfectly synchronized while the periphery of the other star is partially synchronized. This type of chimera lives within a very narrow range of phase-lag values and can breathe. Finally, for the star graph, we have used DART to recover multiple analytical results from the literature on explosive synchronization. 

Interestingly, despite the apparent differences in the methods, the reduced dynamics obtained from our formalism possess similarities with the ones obtained with the Ott-Antonsen approach. DART, however, is not restricted to phase dynamics: it is suited for real dynamics such as the Wilson-Cowan and Lotka-Volterra models as well as models in which each node has its own system of differential equations [see Appendix~\ref{appendix:DART_DN}]. DART is also specifically designed for finite networks and potentially leads to new revealing perturbative analysis. Thus, it seems worth exploring more in depth the mathematical relationships between the two methods. Moreover, a thorough investigation of the errors (e.g., according to the system size $N$) caused by DART is still missing. Finally, new algorithms should be developed to solve the compatibility equations in an optimized way.

We believe that DART will contribute to solve harder problems where the networks have more heterogeneous properties and the dynamics are driven by more complex functions. Using our approach for dynamics on real networks should provide new analytical insights about critical phenomena in a wide range of fields from epidemiology~\cite{Pan2020} to neurosciences~\cite{Laurence2019} and ecology~\cite{Jiang2018}.

\section*{Acknowledgments}
We are grateful to J.-G.~Young for having recommended to investigate chimeras and for having shared codes related to the SBM. We thank X.~Roy-Pomerleau for his early contribution to the study of chaos in chimeras and for useful discussions. We are also grateful to E.~Laurence, C.~Murphy, and M.~Vegu\'e for helpful comments about dimension reduction, chimeras, and the manuscript, respectively. 

This work was supported by the Fonds de recherche du Qu\'ebec -- Nature et technologies (V.T., P.D.), the Natural Sciences and Engineering Research Council of Canada (V.T., G.S.O, L.J.D., P.D.), and the Sentinel North program of Universit\'e Laval, funded by the Canada First Research Excellence Fund (V.T., G.S.O, L.J.D., P.D.). 

\appendix

\section{\texorpdfstring{$DN$}{Lg}-dimensional real dynamics}
\label{appendix:DART_DN}

In this appendix, we use DART to reduce the dimension of dynamics on networks of $N$ nodes, as before, but for which the state of node $j \in \{1,...,N\}$ is now governed by a $D$-dimensional dynamics of the form
    \begin{equation}\label{eq:complete_dynamics_D}
        \dot{\bm{x}}_j = \omega_j\bm{F}(\bm{x}_j) + \sum_{k=1}^N  A_{jk}\bm{G}(\bm{x}_j, \bm{x}_k), 
    \end{equation}
where $\bm{x}_j = (x_j^{(d)})_{d=1}^D: \mathbb{R} \to \mathbb{R}^D$ is a real-valued function of time representing the state of node $j$, $\bm{F}:\mathbb{R}^D \to \mathbb{R}^{D}$ is the intrinsic dynamics function, and $\bm{G}:\mathbb{R}^D \times \mathbb{R}^D \to \mathbb{R}^{D}$ is the function describing the coupling between the nodes.
    
To get a reduced system, we first need to introduce linear
observables, namely $\bm{X}_{\mu} = (X^{(d)}_\mu)_{d=1}^D$ with
    \begin{equation}\label{eq:observables_real}
        {X}_{\mu}^{(d)} = \sum_{j=1}^N M_{\mu j}{x}_j^{(d)}, \quad \mu\in\{1,\ldots, n\}, 
    \end{equation}
and $M$ being a reduction matrix respecting Conditions~\ref{itm:condA}-\ref{itm:condB} of Sec.~\ref{subsec:observables}. Then, using the same steps as in Sec.~\ref{subsec:derivation}, we obtain the $nD$-dimensional reduced dynamics
\smallskip
\begin{align}\label{eq:reduced_dynamics_D}
        \dot{\bm{X}}_{\mu} &\approx \Omega_{\mu}\bm{F}(\bm{\beta}_{\mu}) + \kappa_{\mu} \bm{G}(\bm{\gamma}_{\mu}, \bm{\delta}_{\mu}), \quad \mu \in \{1,...,n\},
\end{align}
\smallskip
together with the compatibility equations $MW = \mathcal{W}M$, $MK = \mathcal{K}M$, $MA = \mathcal{A}M$, and
\begin{align}
    \bm{\beta}_{\mu} &= \Omega_{\mu}^{-1} \sum_{\nu=1}^N \mathcal{W}_{\mu \nu} \bm{X}_{\nu},\label{eq:beta_D}\\
    \bm{\gamma}_{\mu} &= \kappa_{\mu}^{-1} \sum_{\nu=1}^N \mathcal{K}_{\mu \nu} \bm{X}_{\nu},\label{eq:gamma_D}\\
    \bm{\delta}_{\mu} &= \kappa_{\mu}^{-1} \sum_{\nu=1}^N \mathcal{A}_{\mu \nu} \bm{X}_{\nu}\label{eq:delta_D}.
\end{align}

Let us give an example of how DART applies for $N$ Lorenz oscillators with diffusive coupling:
\begin{align*}
    \dot{x}_j &= a(y_j - x_j) + \textstyle{\sigma \sum_{k=1}^N A_{jk}(x_k - x_j)},\\
    \dot{y}_j &= bx_j - y_j - x_j z_j,\\
    \dot{z}_j &= x_jy_j - c z_j,
    \end{align*}
where $a>0$ is the Prandtl number, $b>0$ is the Rayleigh number, and $c>0$ is sometimes called the aspect ratio \cite{Strogatz2018a}. We set $a=10$, $b=28$, $c=8/3$ as it is typically done to study the oscillators on the Lorenz attractor.  With Eqs.(\ref{eq:reduced_dynamics_D}-\ref{eq:delta_D}) and the linear observables 
\begin{align*}
    \mathcal{X}_{\mu} = \sum_{j=1}^n M_{\mu j} x_j,\quad
    \mathcal{Y}_{\mu} = \sum_{j=1}^n M_{\mu j} y_j, \quad
    \mathcal{Z}_{\nu} = \sum_{j=1}^n M_{\mu j} z_j,
\end{align*}
 we get the following $3n$-dimensional reduced dynamics:
\begin{align*}
    \dot{\mathcal{X}}_{\mu} &= a(\mathcal{Y}_{\mu} - \mathcal{X}_{\mu}) - \textstyle{\sigma\sum_{\nu=1}^n \mathcal{L}_{\mu\nu}\mathcal{X}_{\nu}}, \\
    \dot{\mathcal{Y}}_{\mu} &= b\mathcal{X}_{\mu} - \mathcal{Y}_{\mu} - \mathcal{X}_{\mu}\mathcal{Z}_{\mu},\\
    \dot{\mathcal{Z}}_{\nu} &= \mathcal{X}_{\mu}\mathcal{Y}_{\mu} - c\mathcal{Z}_{\mu},
\end{align*}
where $\mathcal{L} = \mathcal{K} - \mathcal{A}$ is the reduced Laplacian matrix.

We now aim to measure phase synchronization for both the complete \textit{and} reduced dynamics. For the complete dynamics and for the oscillators reaching the Lorenz attractor, the phase of oscillator $j$ can be defined as \cite{Pikovsky2003, Osipov2007}
\begin{equation}
    \theta_j = \arctan\left[\frac{z_j - z_0}{\sqrt{x_j^2 + y_j^2} - u_0}\right]
\end{equation}
with $z_0 = 27$ and $u_0 = 10$. The global phase synchronization is then computed using the modulus of Eq.~\eqref{eq:macroscopic_order_parameter}.

Measuring global phase synchronization is not so simple for the reduced dynamics. Indeed, the linear observables $\mathcal{X}_{\mu}$, $\mathcal{Y}_{\mu}$, and $\mathcal{Z}_{\mu}$, do not contain information about the synchronization between the oscillators \textit{inside} the communities. To apply DART to these oscillator dynamics \textit{with the aim of measuring synchronization} \footnote{One could still be interested in the reduced dynamics of linear observables and get good results in some cases.}, nonlinear observables must be introduced. For instance, one could use 
$ \mathcal{V}_{\mu} = \sum_{j=1}^N M_{\mu j} \|\bm{x}_j -  \bm{X}_{\mu}\|^{2}$
as synchronization observables, but this would require a number of lengthy manipulations and would lead to a more complicated reduced system \cite[Annexe B]{Thibeault2020}.

\begin{figure}[ht] 
        \centering    
        \includegraphics[width=1\linewidth, clip=true]{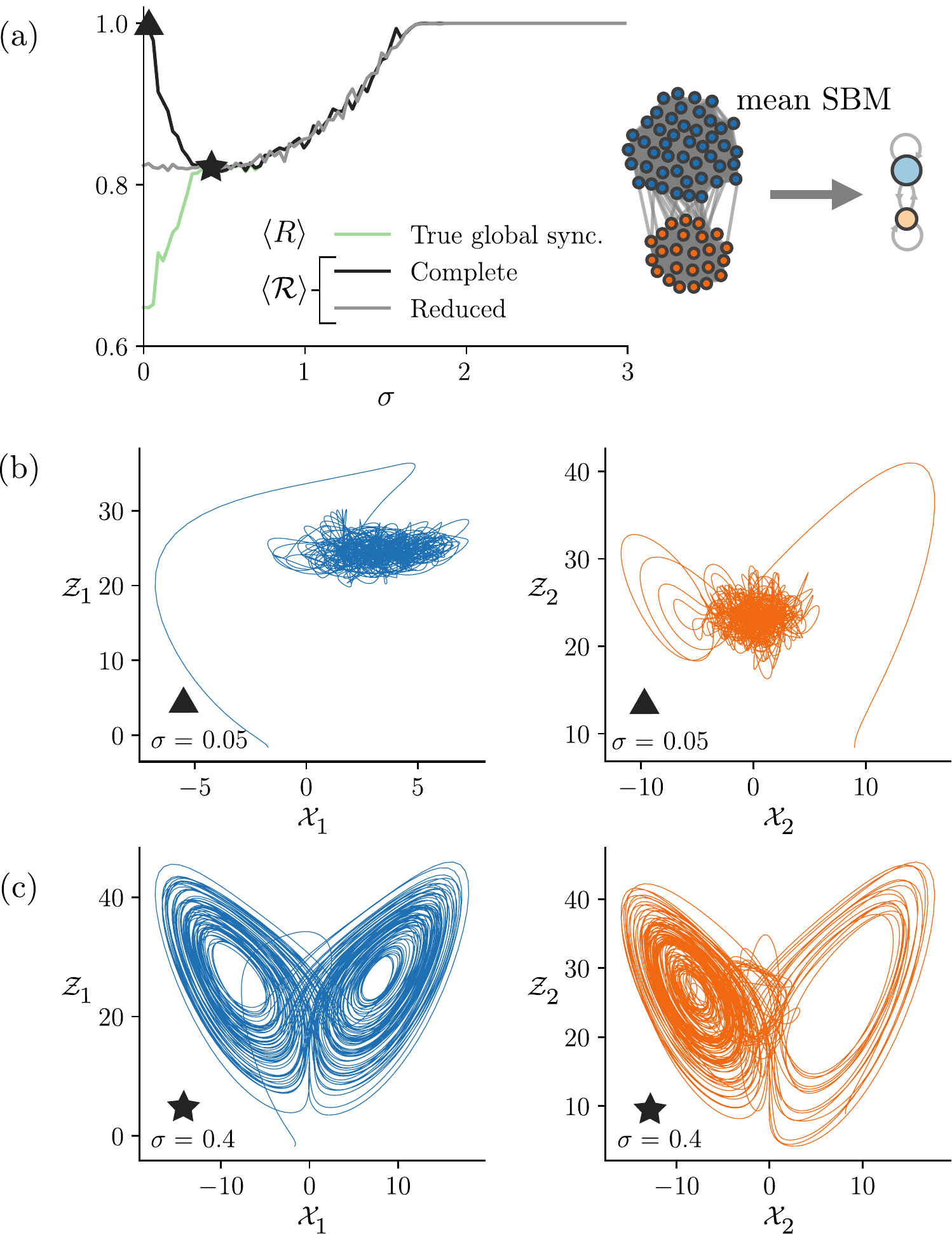}
        \caption{(Color online) Lorenz dynamics on the mean SBM. (a) Global phase synchronization curve (green)  vs. the coupling. The synchronization curves related to the measure \eqref{eq:phase_synchro_between_communities} are also shown for the complete (black) and reduced (grey) dynamics. (b-c) Observable trajectories in each community (blue and orange) in the $\mathcal{X}_{\mu}$-$\mathcal{Z}_{\mu}$ plane. The trajectories do not reach the Lorenz attractor for small coupling in (b), but reach the attractor for stronger coupling in (c). Parameters: $N = 50$, $N_1 = 30$, $N_2 = 20$, $p_{\text{in}} = 0.8$, $p_{\text{out}} = 0.1$, $a=10$, $b=28$, $c=8/3$. $\langle \cdot \rangle$ denotes the average over time and 100 randomly chosen initial conditions.}
        \label{fig:lorenz}
     \end{figure}  
    
Despite these difficulties, there is still a way to use the reduced dynamics of linear observables to get some insights on the synchronization \textit{between} the communities. First, we define the phase of a community $B_{\mu}$ of Lorenz oscillators as
\begin{equation}\label{eq:phase_community}
        \Psi_{\mu} = \arctan\left[\frac{\mathcal{Z}_{\mu} - z_0}{\sqrt{\mathcal{X}_{\mu}^2 + \mathcal{Y}_{\mu}^2} - u_0}\right],
    \end{equation}
which is valid if the mean position of the oscillators in the community (i.e., the observables) reaches the Lorenz attractor. Then, we use the coefficients $\ell_{\mu}$ introduced in Eq.~\eqref{eq:global_weights} and the phases $\Psi_\mu$ to get the following measure of phase synchronization between the communities:
\begin{equation}\label{eq:phase_synchro_between_communities}
    \mathcal{R} = \left|\sum_{\mu=1}^n \ell_{\mu}e^{i\Psi_{\mu}}\right|.
    \end{equation}
    
We are now ready to compare the complete and the reduced Lorenz dynamics on the mean SBM. The agreement between the temporal series of $\bm{X}_{\mu}=(\mathcal{X}_{\mu}, \mathcal{Y}_{\mu},\mathcal{Z}_{\mu})$ in the complete dynamics and the ones of the reduced dynamics is poor, which is expected, because the oscillators are chaotic. However, when looking at the curves describing the global phase synchronization between the communities in Fig.~\ref{fig:lorenz}~(a), the agreement is surprisingly good. The curves are obtained by extracting the phase of each community with Eq.~\eqref{eq:phase_community} and by measuring the synchronization between the communities with Eq.~\eqref{eq:phase_synchro_between_communities}.

For small coupling values, the discrepancies between $\langle R \rangle$ and $\langle \mathcal{R} \rangle$ are caused by the fact that the measure of synchronization is not well defined. In this regime,  Fig.~\ref{fig:lorenz}~(b) shows that the trajectories of the complete dynamics do not reach the Lorenz attractor, preventing us from using the observable of Eq.~\eqref{eq:phase_synchro_between_communities}.  Also, since the reduced dynamics does not carry any information about global phase synchronization, the grey synchronization curve of Fig.~\ref{fig:lorenz}~(a) is expected to be different from the true global phase synchronization defined in Eq.~\eqref{eq:macroscopic_order_parameter}.
 
For larger coupling values, the oscillators in each community of the complete dynamics become synchronized and the trajectories reach a Lorenz attractor as shown in Fig.~\ref{fig:lorenz}~(c). At that point, only the synchronization between the communities can further increase phase synchronization. This is why the synchronization curves for $\langle R \rangle$ and $\langle \mathcal{R} \rangle$ agree with one another.

\section{\texorpdfstring{Phase and modulus equations for\\$\qquad\qquad$the reduced phase dynamics}{Lg}}\label{appendix:reduced_dynamics}

In the context of synchronization, the observable of interest is not $Z_{\mu}$ itself, but rather its modulus $R_{\mu}$ and its phase $\Phi_{\mu}$ as in Eq.~\eqref{eq:observables}. To get the differential equations describing the evolution of $R_{\mu}$ and $\Phi_{\mu}$, one simply uses \mbox{$\dot{R}_{\mu} = (\dot{Z}_{\mu}\Bar{Z}_{\mu} + Z_{\mu}\dot{\Bar{Z}}_{\mu} )/(2R_{\mu})$} and \mbox{$\dot{\Phi}_{\mu} = (\dot{Z}_{\mu}\Bar{Z}_{\mu} - Z_{\mu}\dot{\Bar{Z}}_{\mu} )/(2iR_{\mu}^2)$} which lead to real reduced systems of dimension $2n$. For the Winfree, Kuramoto, and theta models, the reduced dynamics for the moduli and phases are given in Table~\ref{tab:table_reduced_dynamics}. Specific versions of the equations in Table~\ref{tab:table_reduced_dynamics} are shown in Table~\ref{tab:table4}.

\section{\texorpdfstring{Existence and uniqueness of solutions\\$\qquad\quad$to the compatibility equations}{Lg}}\label{appendix:existence}

We prove the claims made in Sec.~\ref{subsec:weights} about the solutions to the compatibility equations. 

Let $M$ and ${T}$ be complex matrices of respective size  $n\times N$ and $N\times N$ where $n < N$.  As shown in Ref.~\cite{Penrose1955}, when $\mathcal{T}$ is an unknown matrix of size $n\times n$, the equation
\begin{equation}
    \mathcal{T}M = MT,
    \label{eq:mat_appendix}
\end{equation} 
has a solution if and only if 
\begin{equation}
    MTM^{+}M = MT,
    \label{penrose}
\end{equation}
where $M^{+}$ denotes the Moore-Penrose pseudo-inverse of $M$. If the latter condition is respected, then any solution of Eq.~\eqref{eq:mat_appendix} takes the form 
\begin{equation*}
    \mathcal{T} = MTM^+ + Y - YMM^{+},
\end{equation*}
where $Y$ is an arbitrary $n\times n$ matrix.  If, moreover, 
\begin{equation*}
    MM^+ = I,
    \label{mpri}
\end{equation*}
then the only possible solution is $\mathcal{T} = MTM^+$. 

The equation $MM^+ = I$ is satisfied if the rank of $M$ is $n$, which is precisely Condition~\ref{itm:condA} that we have imposed on $M$ to guarantee the linear independence of the observables $Z_\mu$. Thus, in the context of dimension reduction, $MM^+ = I$ must be satisfied, which implies that Eq.~\eqref{eq:mat_appendix} has at most one solution, namely, $\mathcal{T} = MTM^+$. Now, the condition
\begin{equation}
    M^+ M = I,
    \label{mpmi}
\end{equation}
is sufficient for satisfying Eq.~\eqref{penrose}. However, Eq.~\eqref{mpmi} is satisfied only if the rank of $M$ is $N$, which contradicts our first assumption that $n<N$.  

We have proved so far that, in the context of dimension reduction, there exists at most one solution to Eq.~\eqref{eq:mat_appendix}, i.e., $\mathcal{T} = MTM^+$.  However, the existence of this solution remains non-trivial. To go further in our analysis, we need to impose an additional restriction to the matrix~$T$. 

Thus, let us suppose that $T$ is real and symmetric,  as are the matrices $W$, $K$, and $A$ in Eqs.~(\ref{freq_eq}-\ref{spec_eq}). Then, $T$ possesses $N$ real orthonormal  eigenvectors of size $1\times N$ (i.e., row vectors).  Let us choose $\ell\geq n$ eigenvectors and form the $\ell\times N$ matrix $V$ whose \mbox{$\mu\text{-th}$} row is the \mbox{$\mu\text{-th}$} eigenvector of $T$.  Then, $VV^\top=I$ and $V^+=V^\top$.  

Additionally, let us suppose that  $M$ can be factorized as $M = C V$, where $C$ is a real $n\times \ell$ matrix of rank $n$, so $n\leq \ell$. Simple manipulations then lead to the following conclusion:  Eq.~\eqref{eq:mat_appendix} is satisfied if and only if 
\begin{equation}
   \mathcal{T}C = C\Lambda,
    \label{eq:TC}
\end{equation}
where $\Lambda$ is the diagonal matrix whose $\mu$-th element on the diagonal is equal to the eigenvalue $\lambda_\mu$ of the $\mu$-th eigenvector in $V$.  In turn, Eq.~\eqref{eq:TC} is consistent if and only if 
\begin{equation*}
   C\Lambda C^+C = C\Lambda. 
    \label{eq:CLambda}
\end{equation*}
Thus, $C^+C=I$ is a sufficient criterion for Eq.~\eqref{eq:mat_appendix}. But $C^+C=I$ is satisfied if and only if the rank of $C$ is $\ell$. We therefore need to impose that $\ell = n$ and that $C$ is a non-singular matrix. 

All in all, we have proved that  if the  three conditions below are satisfied, then  Eq.~\eqref{eq:mat_appendix} has a solution:
\begin{enumerate}[label=(\arabic*)]
    \item  $M=CV$, where
    \item $C$ is a non-singular $n\times n$ real matrix and
    \item $V$ is a $n\times N$ real matrix composed of $n$ real\\ orthonormal row eigenvectors of $T$.
\end{enumerate}  
Moreover, the solution is unique and it is equal to
\begin{equation*}
   \mathcal{T} = C\Lambda C^{-1}. 
    \label{eq:solutionC_appendix}
\end{equation*}

\newpage

\onecolumngrid

\begin{table*}
\caption{\label{tab:table_reduced_dynamics}
The reduced dynamics for three phase models in terms of the synchronization observable $R_{\mu}$ and the phase observable $\Phi_{\mu}$. To simplify the equations, we have defined $\Phi_{xy, z} = \Phi_{x} + \Phi_{y} - \Phi_{z}$ where two indices separated by a comma indicate a difference between the corresponding phase observables and otherwise, the phase observables are summed.}
\begin{ruledtabular}
\begin{tabular}{l c c}
Model & Reduced phase dynamics \\
\colrule
Winfree & \parbox{14cm}{\begin{align}\dot{R}_{\mu} &= \sum_{\nu=1}^n \mathcal{W}_{\mu\nu}R_{\nu}\sin\Phi_{\mu,\nu} + \frac{\sigma\kappa_{\mu}}{2N}\cos{\Phi_{\mu}}-\frac{\sigma}{2N\kappa_{\mu}}\sum_{\xi,\tau = 1}^n \mathcal{K}_{\mu\xi}\mathcal{K}_{\mu\tau}R_{\xi}R_{\tau}\cos\Phi_{\xi\tau,\mu}\nonumber \\&\qquad+ \frac{\sigma}{2N}\sum_{\nu=1}^n\mathcal{A}_{\mu\nu}R_{\nu}\cos\Phi_{\mu}\cos\Phi_{\nu} - \frac{\sigma}{2N\kappa_{\mu}^2}\sum_{\nu, \xi,\tau = 1}^n \mathcal{A}_{\mu\nu}\mathcal{K}_{\mu\xi}\mathcal{K}_{\mu\tau}R_{\xi}R_{\tau}R_{\nu}\cos\Phi_{\nu}\cos\Phi_{\xi\tau,\mu} \label{reduced_winfree} \\\dot{\Phi}_{\mu} &= \sum_{\nu=1}^n \mathcal{W}_{\mu\nu}\frac{R_{\nu}}{R_{\mu}}\cos\Phi_{\nu,\mu} - \frac{\sigma\kappa_{\mu}}{2N}\sin{\Phi_{\mu}}-\frac{\sigma}{2N\kappa_{\mu}}\sum_{\xi,\tau = 1}^n \mathcal{K}_{\mu\xi}\mathcal{K}_{\mu\tau}\frac{R_{\xi}R_{\tau}}{R_{\mu}}\sin\Phi_{\xi\tau,\mu} \nonumber \\&\qquad- \frac{\sigma}{2N}\sum_{\nu=1}^n\mathcal{A}_{\mu\nu}\frac{R_{\nu}}{R_{\mu}}\sin\Phi_{\mu}\cos\Phi_{\nu} - \frac{\sigma}{2N\kappa_{\mu}^2}\sum_{\nu, \xi,\tau = 1}^n \mathcal{A}_{\mu\nu}\mathcal{K}_{\mu\xi}\mathcal{K}_{\mu\tau}\frac{R_{\xi}R_{\tau}R_{\nu}}{R_{\mu}}\cos\Phi_{\nu}\sin\Phi_{\xi\tau,\mu} \qquad\qquad\quad\nonumber\end{align}} \\
\colrule
Kuramoto & \parbox{14cm}{\begin{align}\dot{R}_{\mu} &= \sum_{\nu=1}^n \mathcal{W}_{\mu\nu}R_{\nu}\sin\Phi_{\mu,\nu} + \frac{\sigma}{2N}\sum_{\nu=1}^{n}\mathcal{A}_{\mu\nu}R_{\nu}\cos\Phi_{\nu,\mu} - \frac{\sigma}{2N\kappa_{\mu}^2}\sum_{\nu,\xi,\tau=1}^{n}\mathcal{A}_{\mu\nu}\mathcal{K}_{\mu\xi}\mathcal{K}_{\mu\tau}R_{\xi}R_{\tau}R_{\nu}\cos\Phi_{\mu\nu,\xi,\tau} \label{reduced_kuramoto} \\ \nonumber\dot{\Phi}_{\mu} &= \sum_{\nu=1}^n \mathcal{W}_{\mu\nu}\frac{R_{\nu}}{R_{\mu}}\cos\Phi_{\nu,\mu} - \frac{\sigma}{2N}\sum_{\nu=1}^{n}\mathcal{A}_{\mu\nu}\frac{R_{\nu}}{R_{\mu}}\sin\Phi_{\nu,\mu} - \frac{\sigma}{2N\kappa_{\mu}^2}\sum_{\nu,\xi,\tau=1}^{n}\mathcal{A}_{\mu\nu}\mathcal{K}_{\mu\xi}\mathcal{K}_{\mu\tau}\frac{R_{\xi}R_{\tau}R_{\nu}}{R_{\mu}}\sin\Phi_{\mu\nu,\xi,\tau} \quad\nonumber\end{align}} \\
\colrule
theta &  \parbox{14cm}{\begin{align}
\dot{R}_{\mu} &= \frac{\Omega_{\mu}}{2}\sin\Phi_{\mu}-\frac{\Omega_{\mu}^{-1}}{2}\sum_{\xi,\tau=1}^{n}\mathcal{W}_{\mu\xi}\mathcal{W}_{\mu\tau}R_{\xi}R_{\tau}\sin\Phi_{\xi\tau,\mu} - 2R_{\mu}^{-1}\sum_{\xi=1}^n\mathcal{W}_{\mu\xi}\sin\Phi_{\xi,\mu}\nonumber\\&\qquad -\left(\frac{1-R_{\mu}^2}{2}\right)\sin\Phi_{\mu}+\frac{\sigma \kappa_{\mu}}{2N}\sin\Phi_{\mu} - \frac{\sigma}{2N\kappa_{\mu}}\sum_{\xi,\tau=1}^{n}\mathcal{K}_{\mu\xi}\mathcal{K}_{\mu\tau}R_{\xi}R_{\tau}\sin\Phi_{\xi\tau,\mu}\nonumber\\&\qquad - \frac{\sigma}{N}\sum_{\xi=1}^{n}\mathcal{K}_{\mu\xi}R_{\xi}\sin\Phi_{\xi,\mu}+\frac{\sigma}{2N\kappa_{\mu}^2}\sum_{\nu,\xi,\tau=1}^{n}\mathcal{A}_{\mu\nu}\mathcal{K}_{\mu\xi}\mathcal{K}_{\mu\tau}R_{\xi}R_{\tau}R_{\nu}\cos\Phi_{\nu}\sin\Phi_{\xi\tau,\mu}\nonumber\\&\qquad+ \frac{\sigma}{N\kappa_{\mu}}\sum_{\nu,\xi=1}^{n}\mathcal{A}_{\mu\nu}\mathcal{K}_{\mu\xi}R_{\xi}R_{\nu}\cos\Phi_{\nu}\sin\Phi_{\xi,\mu} - \frac{\sigma}{2N}\sum_{\nu=1}^{n}\mathcal{A}_{\mu\nu}R_{\nu}\cos\Phi_{\nu}\sin\Phi_{\mu} \label{reduced_theta} \\
\dot{\Phi}_{\mu} &= 1- \left(\frac{1+R_{\mu}^2}{2R_{\mu}}\right)\cos\Phi_{\mu} - \frac{\sigma}{2N}\sum_{\nu=1}^{n}\mathcal{A}_{\mu\nu}\frac{R_{\nu}}{R_{\mu}}\cos\Phi_{\nu}\cos\Phi_{\mu} \nonumber\\&\qquad+ \frac{\Omega_{\mu}}{2R_{\mu}}\cos\Phi_{\mu} + \frac{\Omega_{\mu}^{-1}}{2}\sum_{\xi,\tau = 1}^n \mathcal{W}_{\mu\xi}\mathcal{W}_{\mu\tau}\frac{R_{\xi}R_{\tau}}{R_{\mu}}\cos\Phi_{\xi\tau,\mu} + \sum_{\xi=1}^n\mathcal{W}_{\mu\xi}\frac{R_{\xi}}{R_{\mu}}\cos\Phi_{\xi,\mu}\nonumber\\&\qquad + \frac{\sigma \kappa_{\mu}}{2NR_{\mu}}\cos\Phi_{\mu}  + \frac{\sigma}{2N\kappa_{\mu}}\sum_{\xi,\tau=1}^{n}\mathcal{K}_{\mu\xi}\mathcal{K}_{\mu\tau}\frac{R_{\xi}R_{\tau}}{R_{\mu}}\cos\Phi_{\xi\tau,\mu} + \frac{\sigma}{N}\sum_{\xi=1}^{n}\mathcal{K}_{\mu\xi}\frac{R_{\xi}}{R_{\mu}}\cos\Phi_{\xi,\mu}\nonumber\\&\qquad -\frac{\sigma}{2N\kappa_{\mu}^2}\sum_{\nu,\xi,\tau=1}^{n}\mathcal{A}_{\mu\nu}\mathcal{K}_{\mu\xi}\mathcal{K}_{\mu\tau}\frac{R_{\xi}R_{\tau}R_{\nu}}{R_{\mu}}\cos\Phi_{\nu}\cos\Phi_{\xi\tau,\mu} - \frac{\sigma}{N\kappa_{\mu}}\sum_{\nu,\xi=1}^{n}\mathcal{A}_{\mu\nu}\mathcal{K}_{\mu\xi}\frac{R_{\xi}R_{\nu}}{R_{\mu}}\cos\Phi_{\nu}\cos\Phi_{\xi,\mu} \nonumber
\end{align}}\\
\end{tabular}
\end{ruledtabular}
\end{table*}

\clearpage

\begin{table}[ht]
\caption{\label{tab:table4}
The reduced dynamics of the models introduced in Sec.~\ref{subsec:phase_dyn_synchro} in terms of the synchronization observable $R_{\mu}$ and the phase observable $\Phi_{\mu}$ when $\mathcal{K}_{\mu\nu} = \kappa_{\mu}\delta_{\mu\nu}$, $\mathcal{W}_{\mu\nu} = \Omega_{\mu}\delta_{\mu\nu}$, and $\kappa_{\mu} = \sum_{\nu = 1}^n\mathcal{A}_{\mu\nu}$.}
\begin{ruledtabular}
\begin{tabular}{l c c}
Model & Reduced phase dynamics \\
\colrule
Winfree & \parbox{14cm}{\begin{equation}\label{redmpwinfree_K_I}\begin{split}\dot{R}_{\mu} &= \sigma\left(\frac{1-R_{\mu}^2}{2N}\right)\cos\Phi_{\mu}\sum_{\nu=1}^{n}\mathcal{A}_{\mu\nu}\left(1+R_{\nu}\cos\Phi_\nu\right)\\\dot{\Phi}_{\mu} &= \Omega_{\mu} - \sigma\left(\frac{1+R_{\mu}^2}{2NR_{\mu}}\right)\sin\Phi_{\mu}\sum_{\nu=1}^{n}\mathcal{A}_{\mu\nu}\left(1+R_{\nu}\cos\Phi_\nu\right)\end{split}\end{equation}}\\
\colrule
Kuramoto  & \parbox{14cm}{\begin{equation}\label{redmpkuramoto_K_I}\begin{split}\dot{R}_{\mu} &= \sigma\left(\frac{1-R_{\mu}^2}{2N}\right)\sum_{\nu=1}^{n}\mathcal{A}_{\mu\nu}R_{\nu}\cos(\Phi_\nu - \Phi_{\mu}) \\\dot{\Phi}_{\mu} &= \Omega_{\mu} + \sigma\left(\frac{1+R_{\mu}^2}{2NR_{\mu}}\right)\sum_{\nu=1}^{n}\mathcal{A}_{\mu\nu}R_{\nu}\sin(\Phi_\nu - \Phi_{\mu})\qquad\,\, \end{split}\end{equation}} \\
\colrule
theta &  \parbox{14cm}{\begin{equation}\label{redmptheta_K_I}\begin{split}\dot{R}_{\mu} &= \left(\frac{1-R_{\mu}^2}{2}\right)\sin\Phi_{\mu}\left[\Omega_{\mu}-1+\frac{\sigma}{N}\sum_{\nu=1}^{n}\mathcal{A}_{\mu\nu}(1-R_{\nu}\cos\Phi_\nu)\right]\\
\dot{\Phi}_{\mu} &= 1-\left(\frac{1+R_{\mu}^2}{2R_{\mu}}\right)\cos\Phi_{\mu}+ \left[1+\left(\frac{1+R_{\mu}^2}{2R_{\mu}}\right)\cos\Phi_{\mu}\right]\left[\Omega_{\mu}+\frac{\sigma}{N} \sum_{\nu=1}^{n}\mathcal{A}_{\mu\nu}(1-R_{\nu}\cos\Phi_\nu)\right] \end{split}\end{equation}}\\
\end{tabular}
\end{ruledtabular}
\end{table}

\twocolumngrid

\section{\texorpdfstring{Calculation of $C_T$ and $V_T$\\\qquad in~Procedure~1}{Lg}}
\label{appendix:calculation}
We explain how to compute the matrices $C_T$ and $V_T$ of Procedure~\ref{proc:1}. The difficult part of the calculation is $C_T$, for which we propose two exact methods and an approximate method based on nonnegative matrix factorization, a special form of low-rank approximation of matrices.

 \vspace{-0.3cm}

\subsection{Eigenvector matrix \texorpdfstring{$V_T$}{Lg}}
Any row eigenvector of the target matrix $T$ can, in principle, be used to build the matrix $V_T$.  However, to get observables that describe the large-scale dynamics of the system, the eigenvectors of $V_T$ should have as many non-zero elements as possible. Moreover, to simplify the calculation of $C_T$, the matrix $V_T$ should already be close to a reduction matrix, which is required to satisfy Condition~\ref{itm:condB}. Hence, the sign of the eigenvectors should minimize the number of negative elements. This is easily done when the target matrix is either $W$ or $K$, since all eigenvectors can contain only nonnegative elements. When the target matrix is $A$, there is only one eigenvector with nonnegative elements, the Perron vector of $A$, associated with the largest eigenvalue of $A$. Thus, the Perron vector, once normalized, should always be included in $V_A$. 

 \vspace{-0.2cm}

\subsection{Exact methods to compute \texorpdfstring{$C_T$}{Lg} for a given \texorpdfstring{$V_T$}{Lg}}
There are two typical cases: (1) all elements of $V_T$ are nonnegative and (2) many elements of $V_T$ are negative, but one row of $V_T$ is nonnegative. 

If all the elements of $V_T$ are nonnegative, as for \mbox{$T=W$} or \mbox{$T=K$}, then the solution to Step~\ref{state:proc1_3} of Procedure \ref{proc:1} is 
\begin{equation}
C_T=\mathrm{diag}
\left( \frac{1}{\sum_{j}(V_T)_{1 j}}
,\ldots,\frac{1}{\sum_{j}(V_T)_{n j}}\right),
\label{eq:C_simple}
\end{equation}
which makes $M_T$ satisfy Condition~\ref{itm:condB}.

 If $V_T$ contains at least one row with nonnegative elements, as for the case $T=A$ when the Perron vector is included, then the solution to Step~\ref{state:proc1_3} is of the form
 \begin{equation}\label{eq:cde}
C_T= D E,
\end{equation}
with
\begin{equation}\label{eq:cde_d}
D=\mathrm{diag}\left(
\frac{1}{\sum_{j}(EV_T)_{1 j}},\ldots, \frac{1}{\sum_{j}(EV_T)_{n j}}\right),
\end{equation}
and 
\begin{equation}\label{eq:cde_e}
E = 
\begin{pmatrix}1&0&0&\cdots\\
\alpha_2&1&0&\cdots\\ 
\alpha_3&0&1&\ddots\\
\vdots & \vdots & \ddots &\ddots
\end{pmatrix},
\end{equation}
where we have placed the nonnegative eigenvector in the first row of $V_T$ without loss of generality.  The parameters $\alpha_\mu$ are chosen to ensure the non-negativity of all elements $(EV_T)_{\mu j}$.  A simple solution is
\begin{equation*}
\alpha_\mu = \max \left\{ \frac{(V_T)_{\mu 1}}{(V_T)_{11}},\ldots , \frac{(V_T)_{\mu N}}{(V_T)_{1N}}\right\}.
\end{equation*}

When $T = A$, the last method ensures the existence of at least one coefficient matrix $C_A$ which in turn implies that the reduction matrix $M_A$ is positive and solve exactly the compatibility equation $M_A A = \mathcal{A}M_A$. Thanks to Eqs.~(\ref{eq:cde}-\ref{eq:cde_e}), the method also provides an explicit formula for the calculation of $C_A$. However, the method does not confer any additional desirable property to $M_A$. For instance, $M_A$ could be far from being row-orthogonal [Condition~\ref{itm:condA'}] and could thus limit our capacity to interpret the linear observables $Z_1, \dots, Z_n$. As a consequence, we need to compute $C_A$ in another way. 

\vspace{-0.2cm}

\subsection{\texorpdfstring{Approximate method to compute\\ $C_T$ for a given $V_T$\qquad}{Lg}}\label{app:approximate}
In this paper, we rely on an approximate method that uses two matrix factorization algorithms.  
First, semi-nonnegative matrix factorization (SNMF) \cite{Ding2010, Aggarwal2013} decomposes $V_T$ as $X_1 X_2$, where $X_1$ is non-singular and $X_2$ is nonnegative [Condition~ \ref{itm:condB2}]. Actually, the algorithm minimizes the MSE between $V_T$ and $X_1 X_2$. We use the Python package \textit{PyMF} to perform SNMF. Second, we use orthogonal nonnegative matrix factorization (ONMF) \cite{Ding2006} to factorize the nonnegative matrix $X_2$ as $X_3 M_T$, where $X_3$ is non-singular and $M_T$ is as close as possible to a row-orthogonal nonnegative matrix, which is a desirable property for the reduction matrix [Condition~\ref{itm:condA'}]. To perform the calculations, we have modified and corrected the Python package \textit{iONMF}. Starting from eigenvector matrix $V_T$, we end up with a nonnegative matrix $M_T$, which is nonnegative and (or almost) row-orthogonal. The corresponding matrix $C_T$ is therefore $X_3^{-1}X_1^{-1}$. Elementary manipulations finally yield a row-normalized $M_T$ [Condition~\ref{itm:condB1}].

\section{Calculation of \texorpdfstring{$C_{T_{u}}$}{Lg} in Procedure 2}
\label{appendix:calculation_proc2}

Step 4 of Procedure~\ref{proc:2} does not specify how to find the coefficient matrix $C_{T_{u}}$.  The reason for this lack of specificity is that there is no unique method for the computation of $C_{T_{u}}$. We have chosen an approximate method, similar to the one of Appendix~\ref{appendix:calculation}, because it leads to reduction matrices $M_u$ that are almost row-orthogonal. The details are given below.  

Let us define 
\begin{equation*}
V_u=
\begin{cases}V_{T_1},& u=1,\\
     V_{T_2}V_{T_1}^+V_{T_1} , & u=2, \\ 
     V_{T_3}V_{T_1}^+V_{T_1}V_{T_2}^+V_{T_2}V_{T_1}^+V_{T_1}, & u=3. \\ 
\end{cases}
\end{equation*}
Therefore, the reduction matrix $M_u$ in Eq.~\eqref{eq:M_allcases} is
\begin{equation}
    M_u = C_{T_u} V_u, \quad u\in\{1,2,3\}.
    \label{eq:reduction_matrix_u}
\end{equation}
In general, it is not possible to find $C_{T_u}$ such that the reduction matrix $M_u$ in Eq.~\eqref{eq:reduction_matrix_u} is nonnegative, normalized, and as close as possible to a row-orthogonal matrix [Condition~\ref{itm:condB} and \ref{itm:condA'}]. 

However, we can consider $M_u$ as a factor of $V_u$ and find an approximate solution to Eq.~\eqref{eq:reduction_matrix_u}, i.e.,
\begin{equation}
    V_u \approx C_{T_u}^{-1} M_u,
    \label{eq:V_u_approx}
\end{equation}
which defines a matrix factorization problem where $C_{T_u}$ and $M_u$ are to be found.

First, to satisfy Condition~\ref{itm:condB2}, we factor out a nonnegative $n \times N$ matrix $X_1$ from $V_u$ using semi-nonnegative matrix factorization (SNMF) \cite{Ding2010, Aggarwal2013},
\begin{equation}
    V_u \approx C_1 X_1,
    \label{eq:V_u_snmf}
\end{equation}
where $C_1$ is a non-singular $n \times n$ matrix found with the SNMF algorithm.

Second, to satisfy Condition~\ref{itm:condA'}, we factor out a nonnegative and almost orthogonal $n \times N$ matrix $X_2$ from $X_1$ using orthogonal nonnegative matrix factorization (ONMF) \cite{Ding2006},
\begin{equation}
    X_1 \approx C_2 X_2,
    \label{eq:X1_onmf}
\end{equation}
where $C_2$ is a non-singular $n \times n$ matrix found with the ONMF algorithm. 

Third, to satisfy Condition~\ref{itm:condB1}, we factorize $X_2$ exactly to extract the reduction matrix $M_u$ (now nonnegative, almost orthogonal, and normalized)
\begin{equation}
    X_2 = C_3 M_u,
    \label{eq:X2_normalization}
\end{equation}
with a non-singular diagonal matrix $C_3$ where the $\mu$-th diagonal value is given by $\sum_{j=1}^N (X_2)_{\mu j}$.

Finally, substituting Eqs.~(\ref{eq:X1_onmf}-\ref{eq:X2_normalization}) into Eq.~\eqref{eq:V_u_snmf} and inverting the relation yields
\begin{equation*}
    M_u\approx C_3^{-1}C_2^{-1} C_1^{-1} V_u,
\end{equation*}
where we have used the non-singularity of $C_1$, $C_2$, and $C_3$. The coefficient matrix $C_{T_u}$ is therefore
\begin{equation*}
    C_{T_u} = C_3^{-1}C_2^{-1} C_1^{-1}.
\end{equation*}

Other NMF approaches could be useful for the computation of the coefficient matrix. For instance, convex nonnegative factorization (CNMF) seems particularly adapted to our problem. Indeed, as reported in \cite{Ding2010}, CNMF is able to approximate a matrix with both positive and negative elements as the product of two matrices: a first matrix that convexly combines the elements of the original matrix; a second nonnegative matrix whose rows are almost orthogonal. The impact of other NMF algorithms on the quality of DART will be investigated in another paper.

\section{\texorpdfstring{Equivalence between the complete and\\ \qquad\quad the reduced dynamics for $n = N$}{Lg}}
\label{appendix:n=N}

Let $n = N$. Then, $M = CV$ can be written as the identity matrix. Moreover, the reduced dynamics~\eqref{eq:general_reduced_equation} is equivalent to the complete dynamics~\eqref{eq:complex_complete_dyn} if and only if $H$ is affine with respect to its last two arguments, i.e., 
\begin{equation}\label{eq:affine_H}
    H(u, v, w, z) = E(u, v)w + Q(u, v) z + S(u,v),
\end{equation}
where $u,v,w,z \in \mathbb{C}$ and $E,Q,S$ are any complex-valued holomorphic functions. The rest of this appendix is used to prove the above statements.

First, since $W,K,A$ are real symmetric matrices, there exist an orthonormal eigenvector basis to construct the matrix $V$. For such basis, $V^\top$ is the inverse of $V$, so we can choose $C = V^\top$ to get $M = I$, the identity matrix. Note that the methods of Appendices~\ref{app:approximate}-\ref{appendix:calculation_proc2} don't necessarily return $M=I$, but rather $M=P$, where $P$ is a permutation binary matrix (with a single element 1 in each row and each column and 0 elsewhere). However, a simple relabelling of the variables, or the change $C\mapsto P^{-1}C$, yields the desired result $M=I$.

The equation $M=I$ readily implies that $Z_{\mu} = z_j$, $\Omega_{\mu} = \omega_j$, and $\kappa_{\mu} = k_j$ for all $j = \mu \in \{1,..., N\}$. Moreover, $\mathcal{D}_{\Omega}^{-1} = \mathcal{W} = W$, $\mathcal{D}_{\kappa}^{-1} = \mathcal{K} = K$, $\mathcal{A} = A$, and the compatibility equations are trivially satisfied. The reduced dynamics is therefore
\begin{equation*}
\dot{z}_j = F(z_j, \bar{z}_j) + \omega_j G(z_j, \bar{z}_j) + k_j H\left(z_j, \bar{z}_j, \epsilon_j, \bar{\epsilon}_j\right),
\end{equation*}
where $\epsilon_j = k_j^{-1}\sum_{\ell=1}^{N}A_{j\ell}z_{\ell}$.

Second, let us assume that Eq.~\eqref{eq:affine_H} is satisfied.  Then, according to the previous result, the reduced dynamics is 
\begin{multline*}
    \dot{z}_j = F(z_j, \bar{z}_j) + \omega_j G(z_j, \bar{z}_j)  \\+ k_j [E(z_j, \bar{z}_j) \epsilon_j+Q(z_j, \bar{z}_j)\bar{\epsilon}_j + S(z_j, \bar{z}_j)]
\end{multline*}
which is equivalent to the complete \mbox{dynamics~\eqref{eq:complex_complete_dyn}}.

Finally, let us assume that Eqs.~\eqref{eq:general_reduced_equation} and Eqs.~\eqref{eq:complex_complete_dyn} are equivalent.  Then, the equality
\begin{equation}\label{eq:comp_vs_red}
    k_j H\left(z_j, \bar{z}_j, \epsilon_j, \bar{\epsilon}_j\right) = \sum_{k=1}^N A_{jk} H(z_j, \bar{z}_j, z_k, \bar{z}_k)
\end{equation}
must be true for all $j \in \{1,...,N\}$ and $(z_1,...,z_N) \in \mathbb{C}^N$. Since $H$ is a holomorphic function, we can develop the right-hand side function $H$ in Taylor series around $\epsilon_j$ and $\bar{\epsilon}_j$ in the third and fourth arguments respectively:
\begin{align*}
    H(z_j, \bar{z}_j, z_k, \bar{z}_k) = H\left(z_j, \bar{z}_j, \epsilon_j, \bar{\epsilon}_j\right)&+ (z_k - \epsilon_j)H_3 \\ &+ (\bar{z}_k - \bar{\epsilon}_j)H_4+ s_{jk}^H.
\end{align*}
where $s_{jk}^H$ contains the higher-order terms, i.e.,
\begin{equation*}
    s_{jk}^H = \sum_{\ell_1, \ell_2= 2}^\infty \frac{(z_k - \epsilon_j)^{\ell_1}(\bar{z}_k - \bar{\epsilon}_j)^{\ell_2}}{\ell_1 !\,\, \ell_2!}\frac{\partial^{\ell_1 + \ell_2} H}{\partial z_k^{\ell_1} \partial \bar{z}_k^{\ell_2}}\Bigg|_{(z_k, \bar{z}_k) = (\epsilon_j, \bar{\epsilon}_j)}.
\end{equation*}
Substituting the expansion into Eq.~\eqref{eq:comp_vs_red}, the zeroth order term cancels with the left-hand side and the sum over $k$ of $A_{jk}$ times the first order terms gives zero. Hence, 
\begin{align}\label{eq:higher_order_terms}
    \sum_{k=1}^NA_{jk}s_{jk}^H = 0,
\end{align}
which tells us that all the higher-order ($\geq 2$) terms of the Taylor series of $H$ must be zero. This is true if and only if $H$ is affine with respect to its third and fourth argument as in Eq.~\eqref{eq:affine_H}. This concludes the proof.

It is straightforward to show that the coupling function $H$ of the Winfree, Kuramoto, and theta models has the form of Eq.~\eqref{eq:affine_H}. Indeed, for the Winfree model,
\begin{equation*}
E(z_j, \bar{z}_j) = Q(z_j, \bar{z}_j) =  \frac{1}{2}S(z_j, \bar{z}_j)  = -\frac{\sigma}{4N}(z_j^2-1),
\end{equation*}
for the Kuramoto model,
\begin{equation*}
    E(z_j, \bar{z}_j)=\frac{\sigma}{2N}, \,\,Q(z_j, \bar{z}_j) =  -\frac{\sigma z_j^2}{2N}, \,\,\,\,S(z_j, \bar{z}_j) = 0,
\end{equation*}
and for the theta model,
\begin{equation*} 
    E(z_j, \bar{z}_j) = Q(z_j, \bar{z}_j) =  -\frac{1}{2}S(z_j, \bar{z}_j)  = \frac{\sigma}{4N}(z_j+1)^2.
\end{equation*}

\section{Initial conditions to get chimeras}
\label{appendix:initial_conditions_chimeras}

One important fact about the simulations of the Kuramoto-Sakaguchi dynamics is that the initial conditions to get a chimera state are not known \textit{a priori}. Apart from Ref.~\cite{martens2016}, there is not much information in the literature about the initial conditions that lead to chimera states.
In this paper, we drew multiple initial conditions from a uniform distribution for each point in the density space of the mean SBM. After some numerical exploration, we found that the number of observed chimeras per initial condition is not equal everywhere in the density space as reported in \cite{martens2016}. These experiments raised the question: What are the initial conditions that allow the emergence of chimeras ?

\begin{figure}[b] 
 \centering
   \includegraphics[width=0.95\linewidth]{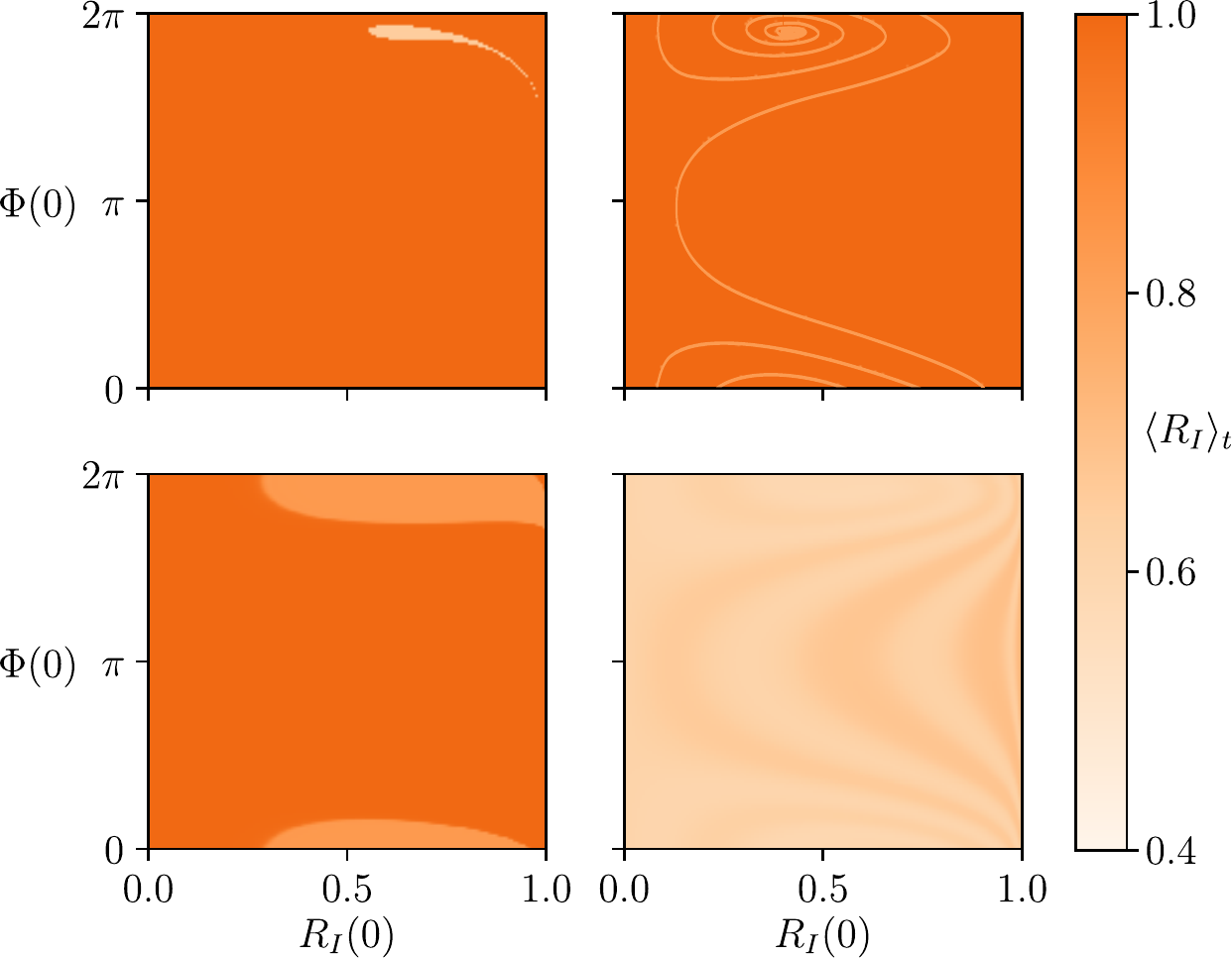}
   
   \vspace{-0.4cm}
   
   \caption{(Color online) Basin of attraction of the chimera states in the Kuramoto-Sakaguchi model (\ref{rired}-\ref{phired}) on the mean SBM for different $f$, $\alpha$, $\rho$, $\Delta$. (Top left) Stable chimera region: $f = 1.0$, $\alpha = 1.4$, $\rho = 0.57$, $\Delta = 0.33$. (Top right) Breathing chimera region: $f = 1.1$, $\alpha =1.4$, $\rho = 0.5$, $\Delta = 0.4$. The line of the spiral pattern has been thickened for visualization purposes. (Bottom left) Stable-chimera region: $f = 1.3$, $\alpha = 1.45$, $\rho = 0.8$, $\Delta = 0.13$ (Bottom right) Breathing chimera region: $f = 1.5$, $\alpha = 1.45$, $\rho = 0.6$, $\Delta = 0.2$.
   }
   \label{fig:inichimap}
 \end{figure}

For given $\rho, \Delta, f$, and $\alpha$, the initial conditions that lead to chimera states vary considerably. This is shown in Fig.~\ref{fig:inichimap}. The heatmaps of initial conditions show diverse non-trivial patterns (for instance, spiral patterns) and the size of the regions of initial conditions giving chimera can be big as the complete map or only a very small part of it. It is therefore a hard problem to make an appropriate choice of initial conditions to find chimeras.

\clearpage


\begin{thebibliography}{123}%
\makeatletter
\providecommand \@ifxundefined [1]{%
 \@ifx{#1\undefined}
}%
\providecommand \@ifnum [1]{%
 \ifnum #1\expandafter \@firstoftwo
 \else \expandafter \@secondoftwo
 \fi
}%
\providecommand \@ifx [1]{%
 \ifx #1\expandafter \@firstoftwo
 \else \expandafter \@secondoftwo
 \fi
}%
\providecommand \natexlab [1]{#1}%
\providecommand \enquote  [1]{``#1''}%
\providecommand \bibnamefont  [1]{#1}%
\providecommand \bibfnamefont [1]{#1}%
\providecommand \citenamefont [1]{#1}%
\providecommand \href@noop [0]{\@secondoftwo}%
\providecommand \href [0]{\begingroup \@sanitize@url \@href}%
\providecommand \@href[1]{\@@startlink{#1}\@@href}%
\providecommand \@@href[1]{\endgroup#1\@@endlink}%
\providecommand \@sanitize@url [0]{\catcode `\\12\catcode `\$12\catcode
  `\&12\catcode `\#12\catcode `\^12\catcode `\_12\catcode `\%12\relax}%
\providecommand \@@startlink[1]{}%
\providecommand \@@endlink[0]{}%
\providecommand \url  [0]{\begingroup\@sanitize@url \@url }%
\providecommand \@url [1]{\endgroup\@href {#1}{\urlprefix }}%
\providecommand \urlprefix  [0]{URL }%
\providecommand \Eprint [0]{\href }%
\providecommand \doibase [0]{http://dx.doi.org/}%
\providecommand \selectlanguage [0]{\@gobble}%
\providecommand \bibinfo  [0]{\@secondoftwo}%
\providecommand \bibfield  [0]{\@secondoftwo}%
\providecommand \translation [1]{[#1]}%
\providecommand \BibitemOpen [0]{}%
\providecommand \bibitemStop [0]{}%
\providecommand \bibitemNoStop [0]{.\EOS\space}%
\providecommand \EOS [0]{\spacefactor3000\relax}%
\providecommand \BibitemShut  [1]{\csname bibitem#1\endcsname}%
\let\auto@bib@innerbib\@empty
\bibitem [{\citenamefont {Mitchell}(2009)}]{Mitchell2009}%
  \BibitemOpen
  \bibfield  {author} {\bibinfo {author} {\bibfnamefont {M.}~\bibnamefont
  {Mitchell}},\ }\href@noop {} {\emph {\bibinfo {title} {Complexity: A Guided
  Tour}}}\ (\bibinfo  {publisher} {Oxford University Press},\ \bibinfo {year}
  {2009})\BibitemShut {NoStop}%
\bibitem [{\citenamefont {Charbonneau}(2017)}]{Charbonneau2017}%
  \BibitemOpen
  \bibfield  {author} {\bibinfo {author} {\bibfnamefont {P.}~\bibnamefont
  {Charbonneau}},\ }\href@noop {} {\emph {\bibinfo {title} {Natural Complexity:
  A Modeling Handbook}}}\ (\bibinfo  {publisher} {Princeton University Press},\
  \bibinfo {year} {2017})\BibitemShut {NoStop}%
\bibitem [{\citenamefont {Boccaletti}\ \emph {et~al.}(2018)\citenamefont
  {Boccaletti}, \citenamefont {Pisarchik}, \citenamefont {Del~Genio},\ and\
  \citenamefont {Amann}}]{Boccaletti2018}%
  \BibitemOpen
  \bibfield  {author} {\bibinfo {author} {\bibfnamefont {S.}~\bibnamefont
  {Boccaletti}}, \bibinfo {author} {\bibfnamefont {A.~N.}\ \bibnamefont
  {Pisarchik}}, \bibinfo {author} {\bibfnamefont {C.~I.}\ \bibnamefont
  {Del~Genio}}, \ and\ \bibinfo {author} {\bibfnamefont {A.}~\bibnamefont
  {Amann}},\ }\href@noop {} {\emph {\bibinfo {title} {Synchronization: From
  Coupled Systems to Complex Networks}}}\ (\bibinfo  {publisher} {Cambridge
  University Press},\ \bibinfo {year} {2018})\BibitemShut {NoStop}%
\bibitem [{\citenamefont {Pikovsky}\ \emph {et~al.}(2003)\citenamefont
  {Pikovsky}, \citenamefont {Rosenblum},\ and\ \citenamefont
  {Kurths}}]{Pikovsky2003}%
  \BibitemOpen
  \bibfield  {author} {\bibinfo {author} {\bibfnamefont {A.}~\bibnamefont
  {Pikovsky}}, \bibinfo {author} {\bibfnamefont {M.}~\bibnamefont {Rosenblum}},
  \ and\ \bibinfo {author} {\bibfnamefont {J.}~\bibnamefont {Kurths}},\
  }\href@noop {} {\emph {\bibinfo {title} {Synchronization: a Universal Concept
  in Nonlinear Sciences}}}\ (\bibinfo  {publisher} {Cambridge University
  press},\ \bibinfo {year} {2003})\BibitemShut {NoStop}%
\bibitem [{\citenamefont {Strogatz}(2003)}]{strogatz2003}%
  \BibitemOpen
  \bibfield  {author} {\bibinfo {author} {\bibfnamefont {S.~H.}\ \bibnamefont
  {Strogatz}},\ }\href@noop {} {\emph {\bibinfo {title} {Sync: How Order
  Emerges from Chaos in the Universe, Nature, and Daily Life}}}\ (\bibinfo
  {publisher} {Penguin UK},\ \bibinfo {year} {2003})\BibitemShut {NoStop}%
\bibitem [{\citenamefont {Turtle}\ \emph {et~al.}(2017)\citenamefont {Turtle},
  \citenamefont {Buono}, \citenamefont {Palacios}, \citenamefont {Dabrowski},
  \citenamefont {In},\ and\ \citenamefont {Longhini}}]{Turtle2017}%
  \BibitemOpen
  \bibfield  {author} {\bibinfo {author} {\bibfnamefont {J.}~\bibnamefont
  {Turtle}}, \bibinfo {author} {\bibfnamefont {P.-L.}\ \bibnamefont {Buono}},
  \bibinfo {author} {\bibfnamefont {A.}~\bibnamefont {Palacios}}, \bibinfo
  {author} {\bibfnamefont {C.}~\bibnamefont {Dabrowski}}, \bibinfo {author}
  {\bibfnamefont {V.}~\bibnamefont {In}}, \ and\ \bibinfo {author}
  {\bibfnamefont {P.}~\bibnamefont {Longhini}},\ }\href
  {http://dx.doi.org/10.1103/PhysRevB.95.144412} {\bibfield  {journal}
  {\bibinfo  {journal} {Phys. Rev. B}\ }\textbf {\bibinfo {volume} {95}},\
  \bibinfo {pages} {144412} (\bibinfo {year} {2017})}\BibitemShut {NoStop}%
\bibitem [{\citenamefont {Izhikevich}(2007)}]{Izhikevich2007}%
  \BibitemOpen
  \bibfield  {author} {\bibinfo {author} {\bibfnamefont {E.~M.}\ \bibnamefont
  {Izhikevich}},\ }\href@noop {} {\emph {\bibinfo {title} {Dynamical Systems in
  Neuroscience}}}\ (\bibinfo  {publisher} {MIT Press},\ \bibinfo {year}
  {2007})\BibitemShut {NoStop}%
\bibitem [{\citenamefont {Laurent}(2002)}]{laurent2002}%
  \BibitemOpen
  \bibfield  {author} {\bibinfo {author} {\bibfnamefont {G.}~\bibnamefont
  {Laurent}},\ }\href {https://www.nature.com/articles/nrn964} {\bibfield
  {journal} {\bibinfo  {journal} {Nat. Rev. Neurosci.}\ }\textbf {\bibinfo
  {volume} {3}},\ \bibinfo {pages} {884} (\bibinfo {year} {2002})}\BibitemShut
  {NoStop}%
\bibitem [{\citenamefont {Fell}\ and\ \citenamefont
  {Axmacher}(2011)}]{Fell2011}%
  \BibitemOpen
  \bibfield  {author} {\bibinfo {author} {\bibfnamefont {J.}~\bibnamefont
  {Fell}}\ and\ \bibinfo {author} {\bibfnamefont {N.}~\bibnamefont
  {Axmacher}},\ }\href {https://www.nature.com/articles/nrn2979} {\bibfield
  {journal} {\bibinfo  {journal} {Nat. Rev. Neurosci.}\ }\textbf {\bibinfo
  {volume} {12}},\ \bibinfo {pages} {105} (\bibinfo {year} {2011})}\BibitemShut
  {NoStop}%
\bibitem [{\citenamefont {di~Santo}\ \emph {et~al.}(2018)\citenamefont
  {di~Santo}, \citenamefont {Villegas}, \citenamefont {Burioni},\ and\
  \citenamefont {Mu{\~{n}}oz}}]{DiSanto2018}%
  \BibitemOpen
  \bibfield  {author} {\bibinfo {author} {\bibfnamefont {S.}~\bibnamefont
  {di~Santo}}, \bibinfo {author} {\bibfnamefont {P.}~\bibnamefont {Villegas}},
  \bibinfo {author} {\bibfnamefont {R.}~\bibnamefont {Burioni}}, \ and\
  \bibinfo {author} {\bibfnamefont {M.~A.}\ \bibnamefont {Mu{\~{n}}oz}},\
  }\href {https://doi.org/10.1073/pnas.1712989115} {\bibfield  {journal}
  {\bibinfo  {journal} {Proc. Natl. Acad. Sci. U.S.A.}\ }\textbf {\bibinfo
  {volume} {115}},\ \bibinfo {pages} {1356} (\bibinfo {year}
  {2018})}\BibitemShut {NoStop}%
\bibitem [{\citenamefont {Lynn}\ and\ \citenamefont
  {Bassett}(2019)}]{Lynn2019}%
  \BibitemOpen
  \bibfield  {author} {\bibinfo {author} {\bibfnamefont {C.~W.}\ \bibnamefont
  {Lynn}}\ and\ \bibinfo {author} {\bibfnamefont {D.~S.}\ \bibnamefont
  {Bassett}},\ }\href {\doibase 10.1038/s42254-019-0040-8} {\bibfield
  {journal} {\bibinfo  {journal} {Nat. Rev. Phys.}\ }\textbf {\bibinfo {volume}
  {1}},\ \bibinfo {pages} {318} (\bibinfo {year} {2019})}\BibitemShut {NoStop}%
\bibitem [{\citenamefont {Potts}(1984)}]{potts1984}%
  \BibitemOpen
  \bibfield  {author} {\bibinfo {author} {\bibfnamefont {W.~K.}\ \bibnamefont
  {Potts}},\ }\href {https://www.nature.com/articles/309344a0} {\bibfield
  {journal} {\bibinfo  {journal} {Nature}\ }\textbf {\bibinfo {volume} {309}},\
  \bibinfo {pages} {344} (\bibinfo {year} {1984})}\BibitemShut {NoStop}%
\bibitem [{\citenamefont {Vasseur}\ and\ \citenamefont
  {Fox}(2009)}]{Vasseur2009}%
  \BibitemOpen
  \bibfield  {author} {\bibinfo {author} {\bibfnamefont {D.~A.}\ \bibnamefont
  {Vasseur}}\ and\ \bibinfo {author} {\bibfnamefont {J.~W.}\ \bibnamefont
  {Fox}},\ }\href {http://dx.doi.org/10.1038/nature08208} {\bibfield  {journal}
  {\bibinfo  {journal} {Nature}\ }\textbf {\bibinfo {volume} {460}},\ \bibinfo
  {pages} {1007} (\bibinfo {year} {2009})}\BibitemShut {NoStop}%
\bibitem [{\citenamefont {Vicsek}\ and\ \citenamefont
  {Zafeiris}(2012)}]{Vicsek2012}%
  \BibitemOpen
  \bibfield  {author} {\bibinfo {author} {\bibfnamefont {T.}~\bibnamefont
  {Vicsek}}\ and\ \bibinfo {author} {\bibfnamefont {A.}~\bibnamefont
  {Zafeiris}},\ }\href {\doibase https://doi.org/10.1016/j.physrep.2012.03.004}
  {\bibfield  {journal} {\bibinfo  {journal} {Phys. Rep.}\ }\textbf {\bibinfo
  {volume} {517}},\ \bibinfo {pages} {71} (\bibinfo {year} {2012})}\BibitemShut
  {NoStop}%
\bibitem [{\citenamefont {Deacy}\ \emph {et~al.}(2017)\citenamefont {Deacy},
  \citenamefont {Armstrong}, \citenamefont {Leacock}, \citenamefont {Robbins},\
  and\ \citenamefont {Gustine}}]{Deacy2017}%
  \BibitemOpen
  \bibfield  {author} {\bibinfo {author} {\bibfnamefont {W.~W.}\ \bibnamefont
  {Deacy}}, \bibinfo {author} {\bibfnamefont {J.~B.}\ \bibnamefont
  {Armstrong}}, \bibinfo {author} {\bibfnamefont {W.~B.}\ \bibnamefont
  {Leacock}}, \bibinfo {author} {\bibfnamefont {C.~T.}\ \bibnamefont
  {Robbins}}, \ and\ \bibinfo {author} {\bibfnamefont {D.~D.}\ \bibnamefont
  {Gustine}},\ }\href {https://doi.org/10.1073/pnas.1705248114} {\bibfield
  {journal} {\bibinfo  {journal} {Proc. Natl. Acad. Sci. U.S.A.}\ }\textbf
  {\bibinfo {volume} {114}},\ \bibinfo {pages} {10432} (\bibinfo {year}
  {2017})}\BibitemShut {NoStop}%
\bibitem [{\citenamefont {Couzin}(2018)}]{Couzin2018}%
  \BibitemOpen
  \bibfield  {author} {\bibinfo {author} {\bibfnamefont {I.~D.}\ \bibnamefont
  {Couzin}},\ }\href {https://doi.org/10.1016/j.tics.2018.08.001} {\bibfield
  {journal} {\bibinfo  {journal} {Trends Cogn. Sci.}\ }\textbf {\bibinfo
  {volume} {22}},\ \bibinfo {pages} {844} (\bibinfo {year} {2018})}\BibitemShut
  {NoStop}%
\bibitem [{\citenamefont {Newman}(2018)}]{Newman2018}%
  \BibitemOpen
  \bibfield  {author} {\bibinfo {author} {\bibfnamefont {M.}~\bibnamefont
  {Newman}},\ }\href@noop {} {\emph {\bibinfo {title} {Networks}}},\ \bibinfo
  {edition} {2nd}\ ed.\ (\bibinfo  {publisher} {Oxford University Press},\
  \bibinfo {year} {2018})\BibitemShut {NoStop}%
\bibitem [{\citenamefont {Watanabe}\ and\ \citenamefont
  {Strogatz}(1994)}]{Watanabe1994}%
  \BibitemOpen
  \bibfield  {author} {\bibinfo {author} {\bibfnamefont {S.}~\bibnamefont
  {Watanabe}}\ and\ \bibinfo {author} {\bibfnamefont {S.~H.}\ \bibnamefont
  {Strogatz}},\ }\href {https://doi.org/10.1016/0167-2789(94)90196-1}
  {\bibfield  {journal} {\bibinfo  {journal} {Physica D}\ }\textbf {\bibinfo
  {volume} {74}},\ \bibinfo {pages} {197} (\bibinfo {year} {1994})}\BibitemShut
  {NoStop}%
\bibitem [{\citenamefont {Ott}\ and\ \citenamefont {Antonsen}(2008)}]{Ott2008}%
  \BibitemOpen
  \bibfield  {author} {\bibinfo {author} {\bibfnamefont {E.}~\bibnamefont
  {Ott}}\ and\ \bibinfo {author} {\bibfnamefont {T.~M.}\ \bibnamefont
  {Antonsen}},\ }\href {\doibase 10.1063/1.2930766} {\bibfield  {journal}
  {\bibinfo  {journal} {Chaos}\ }\textbf {\bibinfo {volume} {18}},\ \bibinfo
  {pages} {037113} (\bibinfo {year} {2008})}\BibitemShut {NoStop}%
\bibitem [{\citenamefont {Luke}\ \emph {et~al.}(2013)\citenamefont {Luke},
  \citenamefont {Barreto},\ and\ \citenamefont {So}}]{Luke2013}%
  \BibitemOpen
  \bibfield  {author} {\bibinfo {author} {\bibfnamefont {T.~B.}\ \bibnamefont
  {Luke}}, \bibinfo {author} {\bibfnamefont {E.}~\bibnamefont {Barreto}}, \
  and\ \bibinfo {author} {\bibfnamefont {P.}~\bibnamefont {So}},\ }\href
  {\doibase 10.1162/NECO\_a\_00525} {\bibfield  {journal} {\bibinfo  {journal}
  {Neural Comput.}\ }\textbf {\bibinfo {volume} {25}},\ \bibinfo {pages} {3207}
  (\bibinfo {year} {2013})}\BibitemShut {NoStop}%
\bibitem [{\citenamefont {Bick}\ \emph {et~al.}(2020)\citenamefont {Bick},
  \citenamefont {Goodfellow}, \citenamefont {Laing},\ and\ \citenamefont
  {Martens}}]{Bick2020}%
  \BibitemOpen
  \bibfield  {author} {\bibinfo {author} {\bibfnamefont {C.}~\bibnamefont
  {Bick}}, \bibinfo {author} {\bibfnamefont {M.}~\bibnamefont {Goodfellow}},
  \bibinfo {author} {\bibfnamefont {C.~R.}\ \bibnamefont {Laing}}, \ and\
  \bibinfo {author} {\bibfnamefont {E.~A.}\ \bibnamefont {Martens}},\ }\href
  {\doibase 10.1186/s13408-020-00086-9} {\bibfield  {journal} {\bibinfo
  {journal} {J. Math. Neurosc.}\ }\textbf {\bibinfo {volume} {10}},\ \bibinfo
  {pages} {1} (\bibinfo {year} {2020})}\BibitemShut {NoStop}%
\bibitem [{\citenamefont {Tyulkina}\ \emph {et~al.}(2018)\citenamefont
  {Tyulkina}, \citenamefont {Goldobin}, \citenamefont {Klimenko},\ and\
  \citenamefont {Pikovsky}}]{Tyulkina2018}%
  \BibitemOpen
  \bibfield  {author} {\bibinfo {author} {\bibfnamefont {I.~V.}\ \bibnamefont
  {Tyulkina}}, \bibinfo {author} {\bibfnamefont {D.~S.}\ \bibnamefont
  {Goldobin}}, \bibinfo {author} {\bibfnamefont {L.~S.}\ \bibnamefont
  {Klimenko}}, \ and\ \bibinfo {author} {\bibfnamefont {A.}~\bibnamefont
  {Pikovsky}},\ }\href {\doibase 10.1103/PhysRevLett.120.264101} {\bibfield
  {journal} {\bibinfo  {journal} {Phys. Rev. Lett.}\ }\textbf {\bibinfo
  {volume} {120}},\ \bibinfo {pages} {264101} (\bibinfo {year}
  {2018})}\BibitemShut {NoStop}%
\bibitem [{\citenamefont {Rodrigues}\ \emph {et~al.}(2016)\citenamefont
  {Rodrigues}, \citenamefont {Peron}, \citenamefont {Ji},\ and\ \citenamefont
  {Kurths}}]{Rodrigues2016}%
  \BibitemOpen
  \bibfield  {author} {\bibinfo {author} {\bibfnamefont {F.~A.}\ \bibnamefont
  {Rodrigues}}, \bibinfo {author} {\bibfnamefont {T.}~\bibnamefont {Peron}},
  \bibinfo {author} {\bibfnamefont {P.}~\bibnamefont {Ji}}, \ and\ \bibinfo
  {author} {\bibfnamefont {J.}~\bibnamefont {Kurths}},\ }\href
  {http://dx.doi.org/10.1016/j.physrep.2015.10.008} {\bibfield  {journal}
  {\bibinfo  {journal} {Phys. Rep.}\ }\textbf {\bibinfo {volume} {610}},\
  \bibinfo {pages} {1} (\bibinfo {year} {2016})}\BibitemShut {NoStop}%
\bibitem [{\citenamefont {Gao}\ \emph {et~al.}(2016{\natexlab{a}})\citenamefont
  {Gao}, \citenamefont {Barzel},\ and\ \citenamefont {Barab{\'a}si}}]{Gao2016}%
  \BibitemOpen
  \bibfield  {author} {\bibinfo {author} {\bibfnamefont {J.}~\bibnamefont
  {Gao}}, \bibinfo {author} {\bibfnamefont {B.}~\bibnamefont {Barzel}}, \ and\
  \bibinfo {author} {\bibfnamefont {A.-L.}\ \bibnamefont {Barab{\'a}si}},\
  }\href {https://www.nature.com/articles/nature16948} {\bibfield  {journal}
  {\bibinfo  {journal} {Nature}\ }\textbf {\bibinfo {volume} {530}},\ \bibinfo
  {pages} {307} (\bibinfo {year} {2016}{\natexlab{a}})}\BibitemShut {NoStop}%
\bibitem [{\citenamefont {Jiang}\ \emph {et~al.}(2018)\citenamefont {Jiang},
  \citenamefont {Huang}, \citenamefont {Seager}, \citenamefont {Lin},
  \citenamefont {Grebogi}, \citenamefont {Hastings},\ and\ \citenamefont
  {Lai}}]{Jiang2018}%
  \BibitemOpen
  \bibfield  {author} {\bibinfo {author} {\bibfnamefont {J.}~\bibnamefont
  {Jiang}}, \bibinfo {author} {\bibfnamefont {Z.-G.}\ \bibnamefont {Huang}},
  \bibinfo {author} {\bibfnamefont {T.~P.}\ \bibnamefont {Seager}}, \bibinfo
  {author} {\bibfnamefont {W.}~\bibnamefont {Lin}}, \bibinfo {author}
  {\bibfnamefont {C.}~\bibnamefont {Grebogi}}, \bibinfo {author} {\bibfnamefont
  {A.}~\bibnamefont {Hastings}}, \ and\ \bibinfo {author} {\bibfnamefont
  {Y.-C.}\ \bibnamefont {Lai}},\ }\href {\doibase 10.1073/pnas.1714958115}
  {\bibfield  {journal} {\bibinfo  {journal} {Proc. Natl. Acad. Sci. U.S.A.}\
  }\textbf {\bibinfo {volume} {115}},\ \bibinfo {pages} {E639} (\bibinfo {year}
  {2018})}\BibitemShut {NoStop}%
\bibitem [{\citenamefont {Cho}\ \emph {et~al.}(2017)\citenamefont {Cho},
  \citenamefont {Nishikawa},\ and\ \citenamefont {Motter}}]{Cho2017}%
  \BibitemOpen
  \bibfield  {author} {\bibinfo {author} {\bibfnamefont {Y.~S.}\ \bibnamefont
  {Cho}}, \bibinfo {author} {\bibfnamefont {T.}~\bibnamefont {Nishikawa}}, \
  and\ \bibinfo {author} {\bibfnamefont {A.~E.}\ \bibnamefont {Motter}},\
  }\href {\doibase 10.1103/PhysRevLett.119.084101} {\bibfield  {journal}
  {\bibinfo  {journal} {Phys. Rev. Lett.}\ }\textbf {\bibinfo {volume} {119}},\
  \bibinfo {pages} {084101} (\bibinfo {year} {2017})}\BibitemShut {NoStop}%
\bibitem [{\citenamefont {Laurence}\ \emph {et~al.}(2019)\citenamefont
  {Laurence}, \citenamefont {Doyon}, \citenamefont {Dub{\'{e}}},\ and\
  \citenamefont {Desrosiers}}]{Laurence2019}%
  \BibitemOpen
  \bibfield  {author} {\bibinfo {author} {\bibfnamefont {E.}~\bibnamefont
  {Laurence}}, \bibinfo {author} {\bibfnamefont {N.}~\bibnamefont {Doyon}},
  \bibinfo {author} {\bibfnamefont {L.~J.}\ \bibnamefont {Dub{\'{e}}}}, \ and\
  \bibinfo {author} {\bibfnamefont {P.}~\bibnamefont {Desrosiers}},\ }\href
  {https://doi.org/10.1103/PhysRevX.9.011042} {\bibfield  {journal} {\bibinfo
  {journal} {Phys. Rev. X}\ }\textbf {\bibinfo {volume} {9}},\ \bibinfo {pages}
  {011042} (\bibinfo {year} {2019})}\BibitemShut {NoStop}%
\bibitem [{\citenamefont {Moon}\ \emph {et~al.}(2006)\citenamefont {Moon},
  \citenamefont {Ghanem},\ and\ \citenamefont {Kevrekidis}}]{Moon2006}%
  \BibitemOpen
  \bibfield  {author} {\bibinfo {author} {\bibfnamefont {S.~J.}\ \bibnamefont
  {Moon}}, \bibinfo {author} {\bibfnamefont {R.}~\bibnamefont {Ghanem}}, \ and\
  \bibinfo {author} {\bibfnamefont {I.~G.}\ \bibnamefont {Kevrekidis}},\ }\href
  {https://doi.org/10.1103/PhysRevLett.96.144101} {\bibfield  {journal}
  {\bibinfo  {journal} {Phys. Rev. Lett.}\ }\textbf {\bibinfo {volume} {96}},\
  \bibinfo {pages} {144101} (\bibinfo {year} {2006})}\BibitemShut {NoStop}%
\bibitem [{\citenamefont {Rajendran}\ and\ \citenamefont
  {Kevrekidis}(2011)}]{Rajendran2011}%
  \BibitemOpen
  \bibfield  {author} {\bibinfo {author} {\bibfnamefont {K.}~\bibnamefont
  {Rajendran}}\ and\ \bibinfo {author} {\bibfnamefont {I.~G.}\ \bibnamefont
  {Kevrekidis}},\ }\href {https://doi.org/10.1103/PhysRevE.84.036708}
  {\bibfield  {journal} {\bibinfo  {journal} {Phys. Rev. E}\ }\textbf {\bibinfo
  {volume} {84}},\ \bibinfo {pages} {036708} (\bibinfo {year}
  {2011})}\BibitemShut {NoStop}%
\bibitem [{\citenamefont {Gfeller}\ and\ \citenamefont
  {De~Los~Rios}(2007)}]{Gfeller2007}%
  \BibitemOpen
  \bibfield  {author} {\bibinfo {author} {\bibfnamefont {D.}~\bibnamefont
  {Gfeller}}\ and\ \bibinfo {author} {\bibfnamefont {P.}~\bibnamefont
  {De~Los~Rios}},\ }\href {\doibase 10.1103/PhysRevLett.99.038701} {\bibfield
  {journal} {\bibinfo  {journal} {Phys. Rev. Lett.}\ }\textbf {\bibinfo
  {volume} {99}},\ \bibinfo {pages} {038701} (\bibinfo {year}
  {2007})}\BibitemShut {NoStop}%
\bibitem [{\citenamefont {Gfeller}\ and\ \citenamefont
  {De~Los~Rios}(2008)}]{Gfeller2008}%
  \BibitemOpen
  \bibfield  {author} {\bibinfo {author} {\bibfnamefont {D.}~\bibnamefont
  {Gfeller}}\ and\ \bibinfo {author} {\bibfnamefont {P.}~\bibnamefont
  {De~Los~Rios}},\ }\href {\doibase 10.1103/PhysRevLett.100.174104} {\bibfield
  {journal} {\bibinfo  {journal} {Phys. Rev. Lett.}\ }\textbf {\bibinfo
  {volume} {100}},\ \bibinfo {pages} {174104} (\bibinfo {year}
  {2008})}\BibitemShut {NoStop}%
\bibitem [{\citenamefont {Antoulas}(2005)}]{Antoulas2005}%
  \BibitemOpen
  \bibfield  {author} {\bibinfo {author} {\bibfnamefont {A.~C.}\ \bibnamefont
  {Antoulas}},\ }\href@noop {} {\emph {\bibinfo {title} {{Approximation of
  Large-Scale Dynamical System}}}}\ (\bibinfo  {publisher} {SIAM},\ \bibinfo
  {year} {2005})\BibitemShut {NoStop}%
\bibitem [{\citenamefont {Kramer}\ and\ \citenamefont
  {Willcox}(2019)}]{Kramer2019}%
  \BibitemOpen
  \bibfield  {author} {\bibinfo {author} {\bibfnamefont {B.}~\bibnamefont
  {Kramer}}\ and\ \bibinfo {author} {\bibfnamefont {K.~E.}\ \bibnamefont
  {Willcox}},\ }\href {\doibase 10.2514/1.J057791} {\bibfield  {journal}
  {\bibinfo  {journal} {AIAA J.}\ }\textbf {\bibinfo {volume} {57}},\ \bibinfo
  {pages} {2297} (\bibinfo {year} {2019})}\BibitemShut {NoStop}%
\bibitem [{\citenamefont {Condon}\ and\ \citenamefont
  {Karp}(2001)}]{Condon2001}%
  \BibitemOpen
  \bibfield  {author} {\bibinfo {author} {\bibfnamefont {A.}~\bibnamefont
  {Condon}}\ and\ \bibinfo {author} {\bibfnamefont {R.~M.}\ \bibnamefont
  {Karp}},\ }\href
  {https://doi.org/10.1002/1098-2418(200103)18:2<116::AID-RSA1001>3.0.CO;2-2}
  {\bibfield  {journal} {\bibinfo  {journal} {Random Struct. Alg.,}\ }\textbf
  {\bibinfo {volume} {18}},\ \bibinfo {pages} {116} (\bibinfo {year}
  {2001})}\BibitemShut {NoStop}%
\bibitem [{\citenamefont {Young}\ \emph {et~al.}(2017)\citenamefont {Young},
  \citenamefont {Desrosiers}, \citenamefont {H\'ebert-Dufresne}, \citenamefont
  {Laurence},\ and\ \citenamefont {Dub\'e}}]{Young2017}%
  \BibitemOpen
  \bibfield  {author} {\bibinfo {author} {\bibfnamefont {J.-G.}\ \bibnamefont
  {Young}}, \bibinfo {author} {\bibfnamefont {P.}~\bibnamefont {Desrosiers}},
  \bibinfo {author} {\bibfnamefont {L.}~\bibnamefont {H\'ebert-Dufresne}},
  \bibinfo {author} {\bibfnamefont {E.}~\bibnamefont {Laurence}}, \ and\
  \bibinfo {author} {\bibfnamefont {L.~J.}\ \bibnamefont {Dub\'e}},\ }\href
  {\doibase 10.1103/PhysRevE.95.062304} {\bibfield  {journal} {\bibinfo
  {journal} {Phys. Rev. E}\ }\textbf {\bibinfo {volume} {95}},\ \bibinfo
  {pages} {062304} (\bibinfo {year} {2017})}\BibitemShut {NoStop}%
\bibitem [{\citenamefont {Pietras}\ and\ \citenamefont
  {Daffertshofer}(2019)}]{Pietras2019}%
  \BibitemOpen
  \bibfield  {author} {\bibinfo {author} {\bibfnamefont {B.}~\bibnamefont
  {Pietras}}\ and\ \bibinfo {author} {\bibfnamefont {A.}~\bibnamefont
  {Daffertshofer}},\ }\href {https://doi.org/10.1016/j.physrep.2019.06.001}
  {\bibfield  {journal} {\bibinfo  {journal} {Phys. Rep.}\ }\textbf {\bibinfo
  {volume} {819}},\ \bibinfo {pages} {1} (\bibinfo {year} {2019})}\BibitemShut
  {NoStop}%
\bibitem [{\citenamefont {Winfree}(1967)}]{Winfree1967}%
  \BibitemOpen
  \bibfield  {author} {\bibinfo {author} {\bibfnamefont {A.~T.}\ \bibnamefont
  {Winfree}},\ }\href {https://doi.org/10.1016/0022-5193(67)90051-3} {\bibfield
   {journal} {\bibinfo  {journal} {J. Theoret. Biol.}\ }\textbf {\bibinfo
  {volume} {16}},\ \bibinfo {pages} {15} (\bibinfo {year} {1967})}\BibitemShut
  {NoStop}%
\bibitem [{\citenamefont {Gallego}\ \emph {et~al.}(2017)\citenamefont
  {Gallego}, \citenamefont {Montbri{\'{o}}},\ and\ \citenamefont
  {Paz{\'{o}}}}]{Gallego2017}%
  \BibitemOpen
  \bibfield  {author} {\bibinfo {author} {\bibfnamefont {R.}~\bibnamefont
  {Gallego}}, \bibinfo {author} {\bibfnamefont {E.}~\bibnamefont
  {Montbri{\'{o}}}}, \ and\ \bibinfo {author} {\bibfnamefont {D.}~\bibnamefont
  {Paz{\'{o}}}},\ }\href {https://doi.org/10.1103/PhysRevE.96.042208}
  {\bibfield  {journal} {\bibinfo  {journal} {Phys. Rev. E}\ }\textbf {\bibinfo
  {volume} {96}},\ \bibinfo {pages} {042208} (\bibinfo {year}
  {2017})}\BibitemShut {NoStop}%
\bibitem [{\citenamefont {Ariaratnam}\ and\ \citenamefont
  {Strogatz}(2001)}]{Ariaratnam2001}%
  \BibitemOpen
  \bibfield  {author} {\bibinfo {author} {\bibfnamefont {J.~T.}\ \bibnamefont
  {Ariaratnam}}\ and\ \bibinfo {author} {\bibfnamefont {S.~H.}\ \bibnamefont
  {Strogatz}},\ }\href {https://doi.org/10.1103/PhysRevLett.86.4278} {\bibfield
   {journal} {\bibinfo  {journal} {Phys. Rev. Lett.}\ }\textbf {\bibinfo
  {volume} {86}},\ \bibinfo {pages} {4278} (\bibinfo {year}
  {2001})}\BibitemShut {NoStop}%
\bibitem [{\citenamefont {Kuramoto}(1975)}]{Kuramoto1975}%
  \BibitemOpen
  \bibfield  {author} {\bibinfo {author} {\bibfnamefont {Y.}~\bibnamefont
  {Kuramoto}},\ }in\ \href@noop {} {\emph {\bibinfo {booktitle} {Int. Symp.
  Math. Probl. Theor. Phys.}}}\ (\bibinfo {year} {1975})\ p.\ \bibinfo {pages}
  {420}\BibitemShut {NoStop}%
\bibitem [{\citenamefont {Wiesenfeld}\ \emph {et~al.}(1996)\citenamefont
  {Wiesenfeld}, \citenamefont {Colet},\ and\ \citenamefont
  {Strogatz}}]{Wiesenfeld1996}%
  \BibitemOpen
  \bibfield  {author} {\bibinfo {author} {\bibfnamefont {K.}~\bibnamefont
  {Wiesenfeld}}, \bibinfo {author} {\bibfnamefont {P.}~\bibnamefont {Colet}}, \
  and\ \bibinfo {author} {\bibfnamefont {S.~H.}\ \bibnamefont {Strogatz}},\
  }\href {https://journals.aps.org/prl/abstract/10.1103/PhysRevLett.76.404}
  {\bibfield  {journal} {\bibinfo  {journal} {Phys. Rev. Lett.}\ }\textbf
  {\bibinfo {volume} {76}},\ \bibinfo {pages} {404} (\bibinfo {year}
  {1996})}\BibitemShut {NoStop}%
\bibitem [{\citenamefont {Matheny}\ \emph {et~al.}(2019)\citenamefont
  {Matheny}, \citenamefont {Emenheiser}, \citenamefont {Fon}, \citenamefont
  {Chapman}, \citenamefont {Salova}, \citenamefont {Rohden}, \citenamefont
  {Li}, \citenamefont {Hudoba De~Badyn}, \citenamefont {P{\'{o}}sfai},
  \citenamefont {Duenas-Osorio}, \citenamefont {Mesbahi}, \citenamefont
  {Crutchfield}, \citenamefont {Cross}, \citenamefont {D'Souza},\ and\
  \citenamefont {Roukes}}]{Matheny2019}%
  \BibitemOpen
  \bibfield  {author} {\bibinfo {author} {\bibfnamefont {M.~H.}\ \bibnamefont
  {Matheny}}, \bibinfo {author} {\bibfnamefont {J.}~\bibnamefont {Emenheiser}},
  \bibinfo {author} {\bibfnamefont {W.}~\bibnamefont {Fon}}, \bibinfo {author}
  {\bibfnamefont {A.}~\bibnamefont {Chapman}}, \bibinfo {author} {\bibfnamefont
  {A.}~\bibnamefont {Salova}}, \bibinfo {author} {\bibfnamefont
  {M.}~\bibnamefont {Rohden}}, \bibinfo {author} {\bibfnamefont
  {J.}~\bibnamefont {Li}}, \bibinfo {author} {\bibfnamefont {M.}~\bibnamefont
  {Hudoba De~Badyn}}, \bibinfo {author} {\bibfnamefont {M.}~\bibnamefont
  {P{\'{o}}sfai}}, \bibinfo {author} {\bibfnamefont {L.}~\bibnamefont
  {Duenas-Osorio}}, \bibinfo {author} {\bibfnamefont {M.}~\bibnamefont
  {Mesbahi}}, \bibinfo {author} {\bibfnamefont {J.~P.}\ \bibnamefont
  {Crutchfield}}, \bibinfo {author} {\bibfnamefont {M.~C.}\ \bibnamefont
  {Cross}}, \bibinfo {author} {\bibfnamefont {R.~M.}\ \bibnamefont {D'Souza}},
  \ and\ \bibinfo {author} {\bibfnamefont {M.~L.}\ \bibnamefont {Roukes}},\
  }\href {\doibase 10.1126/science.aav7932} {\bibfield  {journal} {\bibinfo
  {journal} {Science}\ }\textbf {\bibinfo {volume} {363}},\ \bibinfo {pages}
  {1057} (\bibinfo {year} {2019})}\BibitemShut {NoStop}%
\bibitem [{\citenamefont {Sakaguchi}\ and\ \citenamefont
  {Kuramoto}(1986)}]{Sakaguchi1986}%
  \BibitemOpen
  \bibfield  {author} {\bibinfo {author} {\bibfnamefont {H.}~\bibnamefont
  {Sakaguchi}}\ and\ \bibinfo {author} {\bibfnamefont {Y.}~\bibnamefont
  {Kuramoto}},\ }\href {https://doi.org/10.1143/PTP.76.576} {\bibfield
  {journal} {\bibinfo  {journal} {Prog. Theor. Phys.}\ }\textbf {\bibinfo
  {volume} {76}},\ \bibinfo {pages} {576} (\bibinfo {year} {1986})}\BibitemShut
  {NoStop}%
\bibitem [{\citenamefont {Abrams}\ \emph {et~al.}(2008)\citenamefont {Abrams},
  \citenamefont {Mirollo}, \citenamefont {Strogatz},\ and\ \citenamefont
  {Wiley}}]{Abrams2008}%
  \BibitemOpen
  \bibfield  {author} {\bibinfo {author} {\bibfnamefont {D.~M.}\ \bibnamefont
  {Abrams}}, \bibinfo {author} {\bibfnamefont {R.}~\bibnamefont {Mirollo}},
  \bibinfo {author} {\bibfnamefont {S.~H.}\ \bibnamefont {Strogatz}}, \ and\
  \bibinfo {author} {\bibfnamefont {D.~A.}\ \bibnamefont {Wiley}},\ }\href
  {\doibase 10.1103/PhysRevLett.101.084103} {\bibfield  {journal} {\bibinfo
  {journal} {Phys. Rev. Lett.}\ }\textbf {\bibinfo {volume} {101}},\ \bibinfo
  {pages} {084103} (\bibinfo {year} {2008})}\BibitemShut {NoStop}%
\bibitem [{\citenamefont {Izhikevich}(2004)}]{Izhikevich2004}%
  \BibitemOpen
  \bibfield  {author} {\bibinfo {author} {\bibfnamefont {E.~M.}\ \bibnamefont
  {Izhikevich}},\ }\href {\doibase 10.1109/TNN.2004.832719} {\bibfield
  {journal} {\bibinfo  {journal} {IEEE Trans. Neural Netw.}\ }\textbf {\bibinfo
  {volume} {15}},\ \bibinfo {pages} {1063} (\bibinfo {year}
  {2004})}\BibitemShut {NoStop}%
\bibitem [{\citenamefont {Ermentrout}\ and\ \citenamefont
  {Terman}(2010)}]{Ermentrout2010}%
  \BibitemOpen
  \bibfield  {author} {\bibinfo {author} {\bibfnamefont {G.~B.}\ \bibnamefont
  {Ermentrout}}\ and\ \bibinfo {author} {\bibfnamefont {D.~H.}\ \bibnamefont
  {Terman}},\ }\href@noop {} {\emph {\bibinfo {title} {Mathematical Foundations
  of Neuroscience}}}\ (\bibinfo  {publisher} {Springer},\ \bibinfo {year}
  {2010})\BibitemShut {NoStop}%
\bibitem [{\citenamefont {Ermentrout}\ and\ \citenamefont
  {Kopell}(1986)}]{Ermentrout1986}%
  \BibitemOpen
  \bibfield  {author} {\bibinfo {author} {\bibfnamefont {G.~B.}\ \bibnamefont
  {Ermentrout}}\ and\ \bibinfo {author} {\bibfnamefont {N.}~\bibnamefont
  {Kopell}},\ }\href {https://doi.org/10.1137/0146017} {\bibfield  {journal}
  {\bibinfo  {journal} {SIAM J. Appl. Math.}\ }\textbf {\bibinfo {volume}
  {46}},\ \bibinfo {pages} {233} (\bibinfo {year} {1986})}\BibitemShut
  {NoStop}%
\bibitem [{\citenamefont {Montbri{\'{o}}}\ \emph {et~al.}(2015)\citenamefont
  {Montbri{\'{o}}}, \citenamefont {Paz{\'{o}}},\ and\ \citenamefont
  {Roxin}}]{Montbrio2015}%
  \BibitemOpen
  \bibfield  {author} {\bibinfo {author} {\bibfnamefont {E.}~\bibnamefont
  {Montbri{\'{o}}}}, \bibinfo {author} {\bibfnamefont {D.}~\bibnamefont
  {Paz{\'{o}}}}, \ and\ \bibinfo {author} {\bibfnamefont {A.}~\bibnamefont
  {Roxin}},\ }\href {\doibase 10.1103/PhysRevX.5.021028} {\bibfield  {journal}
  {\bibinfo  {journal} {Phys. Rev. X}\ }\textbf {\bibinfo {volume} {5}},\
  \bibinfo {pages} {021028} (\bibinfo {year} {2015})}\BibitemShut {NoStop}%
\bibitem [{\citenamefont {Budi{\v{s}}i{\'{c}}}\ \emph
  {et~al.}(2012)\citenamefont {Budi{\v{s}}i{\'{c}}}, \citenamefont {Mohr},\
  and\ \citenamefont {Mezi{\'{c}}}}]{Budisic2012}%
  \BibitemOpen
  \bibfield  {author} {\bibinfo {author} {\bibfnamefont {M.}~\bibnamefont
  {Budi{\v{s}}i{\'{c}}}}, \bibinfo {author} {\bibfnamefont {R.}~\bibnamefont
  {Mohr}}, \ and\ \bibinfo {author} {\bibfnamefont {I.}~\bibnamefont
  {Mezi{\'{c}}}},\ }\href {\doibase 10.1063/1.4772195} {\bibfield  {journal}
  {\bibinfo  {journal} {Chaos}\ }\textbf {\bibinfo {volume} {22}},\ \bibinfo
  {pages} {047510} (\bibinfo {year} {2012})}\BibitemShut {NoStop}%
\bibitem [{\citenamefont {Schr{\"{o}}der}\ \emph {et~al.}(2017)\citenamefont
  {Schr{\"{o}}der}, \citenamefont {Timme},\ and\ \citenamefont
  {Witthaut}}]{Schroder2017}%
  \BibitemOpen
  \bibfield  {author} {\bibinfo {author} {\bibfnamefont {M.}~\bibnamefont
  {Schr{\"{o}}der}}, \bibinfo {author} {\bibfnamefont {M.}~\bibnamefont
  {Timme}}, \ and\ \bibinfo {author} {\bibfnamefont {D.}~\bibnamefont
  {Witthaut}},\ }\href {https://doi.org/10.1063/1.4995963} {\bibfield
  {journal} {\bibinfo  {journal} {Chaos}\ }\textbf {\bibinfo {volume} {27}},\
  \bibinfo {pages} {073119} (\bibinfo {year} {2017})}\BibitemShut {NoStop}%
\bibitem [{Note1()}]{Note1}%
  \BibitemOpen
  \bibinfo {note} {For $u = 3$, one can also set \unhbox \voidb@x \hbox {$M_3 =
  C_{T_3}V_{T_3}V_{T_2}^+V_{T_2}V_{T_1}^+V_{T_1}$}, but Eq.~\protect \textup
  {\hbox {\mathsurround \z@ \protect \normalfont (\ignorespaces \ref
  {eq:M_allcases}\unskip \@@italiccorr )}} has been favored based on its better
  performance in numerical experiments.}\BibitemShut {Stop}%
\bibitem [{Note2()}]{Note2}%
  \BibitemOpen
  \bibinfo {note} {One may wonder if solving the compact compatibility equation
  $\protect \mathcal {Q}M = MQ$, where $Q = aW + bK + cA$, helps to satisfy the
  compatibility equations. If the compatibility equations are satisfied, then
  $\protect \mathcal {Q} = a\protect \mathcal {W} + b \protect \mathcal {K} + c
  \protect \mathcal {A}$. Otherwise, we find $|| MQ - \protect \mathcal {Q} M
  || \leq |a| \protect \tmspace +\thinmuskip {.1667em}|| MW - \protect \mathcal
  {W} M || + |b| \protect \tmspace +\thinmuskip {.1667em}|| MK - \protect
  \mathcal {K} M || + |c| \protect \tmspace +\thinmuskip {.1667em}|| MA -
  \protect \mathcal {A} M ||$ using the triangle inequality. However, recall
  that to minimize the first-order errors in DART, we must minimize $|| MW -
  \protect \mathcal {W} M ||$, $|| MK - \protect \mathcal {K} M ||$, and $|| MA
  - \protect \mathcal {A} M ||$. Since $|| MQ - \protect \mathcal {Q} M ||$ is
  a lower bound on these errors, minimizing the error between $MQ$ and
  $\protect \mathcal {Q} M$ is not helpful.}\BibitemShut {Stop}%
\bibitem [{\citenamefont {Van~Mieghem}(2011)}]{VanMieghem2011}%
  \BibitemOpen
  \bibfield  {author} {\bibinfo {author} {\bibfnamefont {P.}~\bibnamefont
  {Van~Mieghem}},\ }\href@noop {} {\emph {\bibinfo {title} {Graph Spectra for
  Complex Networks}}}\ (\bibinfo  {publisher} {Cambridge University Press},\
  \bibinfo {year} {2011})\BibitemShut {NoStop}%
\bibitem [{\citenamefont {May}(2001)}]{May2001}%
  \BibitemOpen
  \bibfield  {author} {\bibinfo {author} {\bibfnamefont {R.~M.}\ \bibnamefont
  {May}},\ }\href@noop {} {\emph {\bibinfo {title} {Stability and Complexity in
  Model Ecosystems}}}\ (\bibinfo  {publisher} {Princeton University Press},\
  \bibinfo {year} {2001})\BibitemShut {NoStop}%
\bibitem [{\citenamefont {Larremore}\ \emph {et~al.}(2011)\citenamefont
  {Larremore}, \citenamefont {Shew},\ and\ \citenamefont
  {Restrepo}}]{Larremore2011}%
  \BibitemOpen
  \bibfield  {author} {\bibinfo {author} {\bibfnamefont {D.~B.}\ \bibnamefont
  {Larremore}}, \bibinfo {author} {\bibfnamefont {W.~L.}\ \bibnamefont {Shew}},
  \ and\ \bibinfo {author} {\bibfnamefont {J.~G.}\ \bibnamefont {Restrepo}},\
  }\href {\doibase 10.1103/PhysRevLett.106.058101} {\bibfield  {journal}
  {\bibinfo  {journal} {Phys. Rev. Lett.}\ }\textbf {\bibinfo {volume} {106}},\
  \bibinfo {pages} {058101} (\bibinfo {year} {2011})}\BibitemShut {NoStop}%
\bibitem [{\citenamefont {Restrepo}\ \emph {et~al.}(2005)\citenamefont
  {Restrepo}, \citenamefont {Ott},\ and\ \citenamefont {Hunt}}]{Restrepo2005}%
  \BibitemOpen
  \bibfield  {author} {\bibinfo {author} {\bibfnamefont {J.~G.}\ \bibnamefont
  {Restrepo}}, \bibinfo {author} {\bibfnamefont {E.}~\bibnamefont {Ott}}, \
  and\ \bibinfo {author} {\bibfnamefont {B.~R.}\ \bibnamefont {Hunt}},\ }\href
  {\doibase 10.1103/PhysRevE.71.036151} {\bibfield  {journal} {\bibinfo
  {journal} {Phys. Rev. E}\ }\textbf {\bibinfo {volume} {71}},\ \bibinfo
  {pages} {036151} (\bibinfo {year} {2005})}\BibitemShut {NoStop}%
\bibitem [{\citenamefont {Van~Mieghem}(2012)}]{VanMieghem2012}%
  \BibitemOpen
  \bibfield  {author} {\bibinfo {author} {\bibfnamefont {P.}~\bibnamefont
  {Van~Mieghem}},\ }\href {\doibase 10.1209/0295-5075/97/48004} {\bibfield
  {journal} {\bibinfo  {journal} {EPL}\ }\textbf {\bibinfo {volume} {97}},\
  \bibinfo {pages} {48004} (\bibinfo {year} {2012})}\BibitemShut {NoStop}%
\bibitem [{\citenamefont {Castellano}\ and\ \citenamefont
  {Pastor-Satorras}(2017)}]{Castellano2017a}%
  \BibitemOpen
  \bibfield  {author} {\bibinfo {author} {\bibfnamefont {C.}~\bibnamefont
  {Castellano}}\ and\ \bibinfo {author} {\bibfnamefont {R.}~\bibnamefont
  {Pastor-Satorras}},\ }\href {\doibase 10.1103/PhysRevX.7.041024} {\bibfield
  {journal} {\bibinfo  {journal} {Phys. Rev. X}\ }\textbf {\bibinfo {volume}
  {7}},\ \bibinfo {pages} {041024} (\bibinfo {year} {2017})}\BibitemShut
  {NoStop}%
\bibitem [{\citenamefont {Lei}\ and\ \citenamefont {Rinaldo}(2015)}]{Lei2015}%
  \BibitemOpen
  \bibfield  {author} {\bibinfo {author} {\bibfnamefont {J.}~\bibnamefont
  {Lei}}\ and\ \bibinfo {author} {\bibfnamefont {A.}~\bibnamefont {Rinaldo}},\
  }\href {\doibase 10.1214/14-AOS1274} {\bibfield  {journal} {\bibinfo
  {journal} {Ann. Statist.}\ }\textbf {\bibinfo {volume} {43}},\ \bibinfo
  {pages} {215} (\bibinfo {year} {2015})}\BibitemShut {NoStop}%
\bibitem [{\citenamefont {Schaeffer}(2007)}]{Schaeffer2007}%
  \BibitemOpen
  \bibfield  {author} {\bibinfo {author} {\bibfnamefont {S.~E.}\ \bibnamefont
  {Schaeffer}},\ }\href {\doibase 10.1016/j.cosrev.2007.05.001} {\bibfield
  {journal} {\bibinfo  {journal} {Comput. Sci. Rev.}\ }\textbf {\bibinfo
  {volume} {1}},\ \bibinfo {pages} {27} (\bibinfo {year} {2007})}\BibitemShut
  {NoStop}%
\bibitem [{\citenamefont {Fiedler}(1975)}]{Fiedler1975}%
  \BibitemOpen
  \bibfield  {author} {\bibinfo {author} {\bibfnamefont {M.}~\bibnamefont
  {Fiedler}},\ }\href {https://dml.cz/handle/10338.dmlcz/101357} {\bibfield
  {journal} {\bibinfo  {journal} {Czechoslov. Math. J.}\ }\textbf {\bibinfo
  {volume} {25}},\ \bibinfo {pages} {619} (\bibinfo {year} {1975})}\BibitemShut
  {NoStop}%
\bibitem [{\citenamefont {Barnes}(1982)}]{Barnes1982}%
  \BibitemOpen
  \bibfield  {author} {\bibinfo {author} {\bibfnamefont {E.~R.}\ \bibnamefont
  {Barnes}},\ }\href {\doibase https://doi.org/10.1137/0603056} {\bibfield
  {journal} {\bibinfo  {journal} {SIAM J. Alg. Disc. Meth.}\ }\textbf {\bibinfo
  {volume} {3}},\ \bibinfo {pages} {541} (\bibinfo {year} {1982})}\BibitemShut
  {NoStop}%
\bibitem [{\citenamefont {Barnes}\ and\ \citenamefont
  {Hoffman}(1984)}]{Barnes1984}%
  \BibitemOpen
  \bibfield  {author} {\bibinfo {author} {\bibfnamefont {E.~R.}\ \bibnamefont
  {Barnes}}\ and\ \bibinfo {author} {\bibfnamefont {A.~J.}\ \bibnamefont
  {Hoffman}},\ }in\ \href {\doibase 10.2307/2008156} {\emph {\bibinfo
  {booktitle} {Prog. Comb. Optim.}}},\ Vol.~\bibinfo {volume} {45},\ \bibinfo
  {editor} {edited by\ \bibinfo {editor} {\bibfnamefont {W.~R.}\ \bibnamefont
  {Pulleyblank}}}\ (\bibinfo  {publisher} {Academic Press},\ \bibinfo {year}
  {1984})\ p.~\bibinfo {pages} {13}\BibitemShut {NoStop}%
\bibitem [{\citenamefont {Powers}(1988)}]{Powers1988}%
  \BibitemOpen
  \bibfield  {author} {\bibinfo {author} {\bibfnamefont {D.~L.}\ \bibnamefont
  {Powers}},\ }\href {\doibase 10.1016/0024-3795(88)90147-4} {\bibfield
  {journal} {\bibinfo  {journal} {Linear Algebr. Appl.}\ }\textbf {\bibinfo
  {volume} {101}},\ \bibinfo {pages} {121} (\bibinfo {year}
  {1988})}\BibitemShut {NoStop}%
\bibitem [{\citenamefont {Barlev}\ \emph {et~al.}(2011)\citenamefont {Barlev},
  \citenamefont {Antonsen},\ and\ \citenamefont {Ott}}]{Barlev2011}%
  \BibitemOpen
  \bibfield  {author} {\bibinfo {author} {\bibfnamefont {G.}~\bibnamefont
  {Barlev}}, \bibinfo {author} {\bibfnamefont {T.~M.}\ \bibnamefont
  {Antonsen}}, \ and\ \bibinfo {author} {\bibfnamefont {E.}~\bibnamefont
  {Ott}},\ }\href {\doibase 10.1063/1.3596711} {\bibfield  {journal} {\bibinfo
  {journal} {Chaos}\ }\textbf {\bibinfo {volume} {21}},\ \bibinfo {pages}
  {025103} (\bibinfo {year} {2011})}\BibitemShut {NoStop}%
\bibitem [{Note3()}]{Note3}%
  \BibitemOpen
  \bibinfo {note} {For instance, choosing $\protect \bm {v}_1 = (1/2, 1/2, 0,
  0, 1/2, 1/2)$ as the first row of $V_K$ would create singular matrices (e.g.,
  $V_KV_A^+$ would be singular). Note that if $V_K$ includes $\protect \bm
  {v}_1$, the predictions in Fig.~\ref {fig:small_graphs_reduction} when
  targeting $K$ are not improved and therefore, the conclusion drawn in this
  section would still be valid.}\BibitemShut {Stop}%
\bibitem [{Note4()}]{Note4}%
  \BibitemOpen
  \bibinfo {note} {Comparable predictions can also be achieved with the targets
  $L \to W$, where $L$ is the Laplacian matrix, and the compatibility equations
  $\protect \mathcal {W}M = MW$, $\protect \mathcal {L}M = ML$. For $n=2$, we
  can build the Laplacian eigenvector matrix $V_L$ with the eigenvectors
  $\protect \bm {v}_1$ and $\protect \bm {v}_2$ corresponding to the two lowest
  eigenvalues: $\lambda _1 = 0$ and the Fiedler eigenvalue $\lambda _2$ \cite
  {Fiedler1973}. The uniformity of $\protect \bm {v}_1$ helps to get a positive
  reduction matrix, while $\protect \bm {v}_2$ is useful for graph partitioning
  \cite {Hall1970} and community detection \cite {Newman2006}.}\BibitemShut
  {Stop}%
\bibitem [{\citenamefont {Fortunato}\ and\ \citenamefont
  {Hric}(2016)}]{Fortunato2016}%
  \BibitemOpen
  \bibfield  {author} {\bibinfo {author} {\bibfnamefont {S.}~\bibnamefont
  {Fortunato}}\ and\ \bibinfo {author} {\bibfnamefont {D.}~\bibnamefont
  {Hric}},\ }\href {\doibase 10.1016/j.physrep.2016.09.002} {\bibfield
  {journal} {\bibinfo  {journal} {Phys. Rep.}\ }\textbf {\bibinfo {volume}
  {659}},\ \bibinfo {pages} {1} (\bibinfo {year} {2016})}\BibitemShut {NoStop}%
\bibitem [{\citenamefont {Alpert}\ and\ \citenamefont
  {Yao}(1995)}]{Alpert1995}%
  \BibitemOpen
  \bibfield  {author} {\bibinfo {author} {\bibfnamefont {C.~J.}\ \bibnamefont
  {Alpert}}\ and\ \bibinfo {author} {\bibfnamefont {S.-Z.}\ \bibnamefont
  {Yao}},\ }in\ \href@noop {} {\emph {\bibinfo {booktitle} {Proc. 32nd Annu.
  ACM/IEEE Des. Autom. Conf.}}}\ (\bibinfo {year} {1995})\ p.\ \bibinfo {pages}
  {195}\BibitemShut {NoStop}%
\bibitem [{\citenamefont {Acebr{\'{o}}n}\ \emph {et~al.}(2005)\citenamefont
  {Acebr{\'{o}}n}, \citenamefont {Bonilla}, \citenamefont
  {P{\'{e}}rez-Vicente}, \citenamefont {Ritort},\ and\ \citenamefont
  {Spigler}}]{Acebron2005}%
  \BibitemOpen
  \bibfield  {author} {\bibinfo {author} {\bibfnamefont {J.~A.}\ \bibnamefont
  {Acebr{\'{o}}n}}, \bibinfo {author} {\bibfnamefont {L.~L.}\ \bibnamefont
  {Bonilla}}, \bibinfo {author} {\bibfnamefont {C.~J.}\ \bibnamefont
  {P{\'{e}}rez-Vicente}}, \bibinfo {author} {\bibfnamefont {F.}~\bibnamefont
  {Ritort}}, \ and\ \bibinfo {author} {\bibfnamefont {R.}~\bibnamefont
  {Spigler}},\ }\href {https://doi.org/10.1103/RevModPhys.77.137} {\bibfield
  {journal} {\bibinfo  {journal} {Rev. Mod. Phys.}\ }\textbf {\bibinfo {volume}
  {77}},\ \bibinfo {pages} {137} (\bibinfo {year} {2005})}\BibitemShut
  {NoStop}%
\bibitem [{Note5()}]{Note5}%
  \BibitemOpen
  \bibinfo {note} {When the oscillators in the two layers of the bipartite
  graph have opposite natural frequencies [which is almost the case in
  Fig.~\ref {fig:all_trans} (b)], they are called Janus oscillators \cite
  {Nicolaou2019a, Peron2020}. These oscillators exhibit an impressive diversity
  of oscillatory phenomena and DART could be useful to get further analytical
  insights.}\BibitemShut {Stop}%
\bibitem [{Note6()}]{Note6}%
  \BibitemOpen
  \bibinfo {note} {Note that $P(d)$ can be computed numerically or analytically
  from the probability distribution $P(A)$ of adjacency matrices given in
  Eq.~\protect \textup {\hbox {\mathsurround \z@ \protect \normalfont
  (\ignorespaces \ref {eq:probA}\unskip \@@italiccorr )}}.}\BibitemShut {Stop}%
\bibitem [{\citenamefont {Kotwal}\ \emph {et~al.}(2017)\citenamefont {Kotwal},
  \citenamefont {Jiang},\ and\ \citenamefont {Abrams}}]{Kotwal2017}%
  \BibitemOpen
  \bibfield  {author} {\bibinfo {author} {\bibfnamefont {T.}~\bibnamefont
  {Kotwal}}, \bibinfo {author} {\bibfnamefont {X.}~\bibnamefont {Jiang}}, \
  and\ \bibinfo {author} {\bibfnamefont {D.~M.}\ \bibnamefont {Abrams}},\
  }\href {\doibase 10.1103/PhysRevLett.119.264101} {\bibfield  {journal}
  {\bibinfo  {journal} {Phys. Rev. Lett.}\ }\textbf {\bibinfo {volume} {119}},\
  \bibinfo {pages} {264101} (\bibinfo {year} {2017})}\BibitemShut {NoStop}%
\bibitem [{\citenamefont {Chen}\ \emph {et~al.}(2017)\citenamefont {Chen},
  \citenamefont {Sun}, \citenamefont {Gao}, \citenamefont {Xu},\ and\
  \citenamefont {Zheng}}]{Chen2017}%
  \BibitemOpen
  \bibfield  {author} {\bibinfo {author} {\bibfnamefont {H.}~\bibnamefont
  {Chen}}, \bibinfo {author} {\bibfnamefont {Y.}~\bibnamefont {Sun}}, \bibinfo
  {author} {\bibfnamefont {J.}~\bibnamefont {Gao}}, \bibinfo {author}
  {\bibfnamefont {C.}~\bibnamefont {Xu}}, \ and\ \bibinfo {author}
  {\bibfnamefont {Z.}~\bibnamefont {Zheng}},\ }\href {\doibase
  10.1007/s11467-017-0651-4} {\bibfield  {journal} {\bibinfo  {journal} {Front.
  Phys.}\ }\textbf {\bibinfo {volume} {12}},\ \bibinfo {pages} {120504}
  (\bibinfo {year} {2017})}\BibitemShut {NoStop}%
\bibitem [{\citenamefont {Xu}\ \emph {et~al.}(2018)\citenamefont {Xu},
  \citenamefont {Boccaletti}, \citenamefont {Guan},\ and\ \citenamefont
  {Zheng}}]{Xu2018b}%
  \BibitemOpen
  \bibfield  {author} {\bibinfo {author} {\bibfnamefont {C.}~\bibnamefont
  {Xu}}, \bibinfo {author} {\bibfnamefont {S.}~\bibnamefont {Boccaletti}},
  \bibinfo {author} {\bibfnamefont {S.}~\bibnamefont {Guan}}, \ and\ \bibinfo
  {author} {\bibfnamefont {Z.}~\bibnamefont {Zheng}},\ }\href {\doibase
  10.1103/PhysRevE.98.050202} {\bibfield  {journal} {\bibinfo  {journal} {Phys.
  Rev. E}\ }\textbf {\bibinfo {volume} {98}},\ \bibinfo {pages} {050202(R)}
  (\bibinfo {year} {2018})}\BibitemShut {NoStop}%
\bibitem [{\citenamefont {Kuramoto}\ and\ \citenamefont
  {Battogtokh}(2002)}]{Kuramoto2002}%
  \BibitemOpen
  \bibfield  {author} {\bibinfo {author} {\bibfnamefont {Y.}~\bibnamefont
  {Kuramoto}}\ and\ \bibinfo {author} {\bibfnamefont {D.}~\bibnamefont
  {Battogtokh}},\ }\href {https://arxiv.org/abs/cond-mat/0210694} {\bibfield
  {journal} {\bibinfo  {journal} {Nonlinear Phenom. Complex Syst.}\ }\textbf
  {\bibinfo {volume} {5}},\ \bibinfo {pages} {380} (\bibinfo {year}
  {2002})}\BibitemShut {NoStop}%
\bibitem [{\citenamefont {Abrams}\ and\ \citenamefont
  {Strogatz}(2004)}]{Abrams2004}%
  \BibitemOpen
  \bibfield  {author} {\bibinfo {author} {\bibfnamefont {D.~M.}\ \bibnamefont
  {Abrams}}\ and\ \bibinfo {author} {\bibfnamefont {S.~H.}\ \bibnamefont
  {Strogatz}},\ }\href {\doibase 10.1103/PhysRevLett.93.174102} {\bibfield
  {journal} {\bibinfo  {journal} {Phys. Rev. Lett.}\ }\textbf {\bibinfo
  {volume} {93}},\ \bibinfo {pages} {174102} (\bibinfo {year}
  {2004})}\BibitemShut {NoStop}%
\bibitem [{\citenamefont {Totz}\ \emph {et~al.}(2018)\citenamefont {Totz},
  \citenamefont {Rode}, \citenamefont {Tinsley}, \citenamefont {Showalter},\
  and\ \citenamefont {Engel}}]{Totz2018}%
  \BibitemOpen
  \bibfield  {author} {\bibinfo {author} {\bibfnamefont {J.~F.}\ \bibnamefont
  {Totz}}, \bibinfo {author} {\bibfnamefont {J.}~\bibnamefont {Rode}}, \bibinfo
  {author} {\bibfnamefont {M.~R.}\ \bibnamefont {Tinsley}}, \bibinfo {author}
  {\bibfnamefont {K.}~\bibnamefont {Showalter}}, \ and\ \bibinfo {author}
  {\bibfnamefont {H.}~\bibnamefont {Engel}},\ }\href
  {https://www.nature.com/articles/s41567-017-0005-8} {\bibfield  {journal}
  {\bibinfo  {journal} {Nat. Phys.}\ }\textbf {\bibinfo {volume} {14}},\
  \bibinfo {pages} {282} (\bibinfo {year} {2018})}\BibitemShut {NoStop}%
\bibitem [{\citenamefont {Tinsley}\ \emph {et~al.}(2012)\citenamefont
  {Tinsley}, \citenamefont {Nkomo},\ and\ \citenamefont
  {Showalter}}]{Tinsley2012}%
  \BibitemOpen
  \bibfield  {author} {\bibinfo {author} {\bibfnamefont {M.~R.}\ \bibnamefont
  {Tinsley}}, \bibinfo {author} {\bibfnamefont {S.}~\bibnamefont {Nkomo}}, \
  and\ \bibinfo {author} {\bibfnamefont {K.}~\bibnamefont {Showalter}},\ }\href
  {https://www.nature.com/articles/nphys2371} {\bibfield  {journal} {\bibinfo
  {journal} {Nat. Phys.}\ }\textbf {\bibinfo {volume} {8}},\ \bibinfo {pages}
  {662} (\bibinfo {year} {2012})}\BibitemShut {NoStop}%
\bibitem [{\citenamefont {Wojewoda}\ \emph {et~al.}(2016)\citenamefont
  {Wojewoda}, \citenamefont {Czolczynski}, \citenamefont {Maistrenko},\ and\
  \citenamefont {Kapitaniak}}]{Wojewoda2016}%
  \BibitemOpen
  \bibfield  {author} {\bibinfo {author} {\bibfnamefont {J.}~\bibnamefont
  {Wojewoda}}, \bibinfo {author} {\bibfnamefont {K.}~\bibnamefont
  {Czolczynski}}, \bibinfo {author} {\bibfnamefont {Y.}~\bibnamefont
  {Maistrenko}}, \ and\ \bibinfo {author} {\bibfnamefont {T.}~\bibnamefont
  {Kapitaniak}},\ }\href {https://www.nature.com/articles/srep34329} {\bibfield
   {journal} {\bibinfo  {journal} {Sci. Rep.}\ }\textbf {\bibinfo {volume}
  {6}},\ \bibinfo {pages} {34329} (\bibinfo {year} {2016})}\BibitemShut
  {NoStop}%
\bibitem [{\citenamefont {Kapitaniak}\ \emph {et~al.}(2014)\citenamefont
  {Kapitaniak}, \citenamefont {Kuzma}, \citenamefont {Wojewoda}, \citenamefont
  {Czolczynski},\ and\ \citenamefont {Maistrenko}}]{Kapitaniak2014}%
  \BibitemOpen
  \bibfield  {author} {\bibinfo {author} {\bibfnamefont {T.}~\bibnamefont
  {Kapitaniak}}, \bibinfo {author} {\bibfnamefont {P.}~\bibnamefont {Kuzma}},
  \bibinfo {author} {\bibfnamefont {J.}~\bibnamefont {Wojewoda}}, \bibinfo
  {author} {\bibfnamefont {K.}~\bibnamefont {Czolczynski}}, \ and\ \bibinfo
  {author} {\bibfnamefont {Y.}~\bibnamefont {Maistrenko}},\ }\href
  {https://www.nature.com/articles/srep06379} {\bibfield  {journal} {\bibinfo
  {journal} {Sci. Rep.}\ }\textbf {\bibinfo {volume} {4}},\ \bibinfo {pages}
  {6379} (\bibinfo {year} {2014})}\BibitemShut {NoStop}%
\bibitem [{\citenamefont {Martens}\ \emph {et~al.}(2013)\citenamefont
  {Martens}, \citenamefont {Thutupalli}, \citenamefont {Fourri{\`e}re},\ and\
  \citenamefont {Hallatschek}}]{Martens2013}%
  \BibitemOpen
  \bibfield  {author} {\bibinfo {author} {\bibfnamefont {E.~A.}\ \bibnamefont
  {Martens}}, \bibinfo {author} {\bibfnamefont {S.}~\bibnamefont {Thutupalli}},
  \bibinfo {author} {\bibfnamefont {A.}~\bibnamefont {Fourri{\`e}re}}, \ and\
  \bibinfo {author} {\bibfnamefont {O.}~\bibnamefont {Hallatschek}},\ }\href
  {http://www.pnas.org/content/110/26/10563.short} {\bibfield  {journal}
  {\bibinfo  {journal} {Proc. Natl. Acad. Sci. U.S.A.}\ }\textbf {\bibinfo
  {volume} {110}},\ \bibinfo {pages} {10563} (\bibinfo {year}
  {2013})}\BibitemShut {NoStop}%
\bibitem [{\citenamefont {Hagerstrom}\ \emph {et~al.}(2012)\citenamefont
  {Hagerstrom}, \citenamefont {Murphy}, \citenamefont {Roy}, \citenamefont
  {H{\"o}vel}, \citenamefont {Omelchenko},\ and\ \citenamefont
  {Sch{\"o}ll}}]{Hagerstrom2012}%
  \BibitemOpen
  \bibfield  {author} {\bibinfo {author} {\bibfnamefont {A.~M.}\ \bibnamefont
  {Hagerstrom}}, \bibinfo {author} {\bibfnamefont {T.~E.}\ \bibnamefont
  {Murphy}}, \bibinfo {author} {\bibfnamefont {R.}~\bibnamefont {Roy}},
  \bibinfo {author} {\bibfnamefont {P.}~\bibnamefont {H{\"o}vel}}, \bibinfo
  {author} {\bibfnamefont {I.}~\bibnamefont {Omelchenko}}, \ and\ \bibinfo
  {author} {\bibfnamefont {E.}~\bibnamefont {Sch{\"o}ll}},\ }\href
  {https://www.nature.com/articles/nphys2372} {\bibfield  {journal} {\bibinfo
  {journal} {Nat. Phys.}\ }\textbf {\bibinfo {volume} {8}},\ \bibinfo {pages}
  {658} (\bibinfo {year} {2012})}\BibitemShut {NoStop}%
\bibitem [{\citenamefont {Bansal}\ \emph {et~al.}(2019)\citenamefont {Bansal},
  \citenamefont {Garcia}, \citenamefont {Tompson}, \citenamefont {Verstynen},
  \citenamefont {Vettel},\ and\ \citenamefont {Muldoon}}]{Bansal2019}%
  \BibitemOpen
  \bibfield  {author} {\bibinfo {author} {\bibfnamefont {K.}~\bibnamefont
  {Bansal}}, \bibinfo {author} {\bibfnamefont {J.~O.}\ \bibnamefont {Garcia}},
  \bibinfo {author} {\bibfnamefont {S.~H.}\ \bibnamefont {Tompson}}, \bibinfo
  {author} {\bibfnamefont {T.}~\bibnamefont {Verstynen}}, \bibinfo {author}
  {\bibfnamefont {J.~M.}\ \bibnamefont {Vettel}}, \ and\ \bibinfo {author}
  {\bibfnamefont {S.~F.}\ \bibnamefont {Muldoon}},\ }\href
  {https://advances.sciencemag.org/content/5/4/eaau8535?intcmp=trendmd-adv}
  {\bibfield  {journal} {\bibinfo  {journal} {Sci. Adv.}\ }\textbf {\bibinfo
  {volume} {5}},\ \bibinfo {pages} {8535} (\bibinfo {year} {2019})}\BibitemShut
  {NoStop}%
\bibitem [{\citenamefont {Calim}\ \emph {et~al.}(2018)\citenamefont {Calim},
  \citenamefont {H{\"{o}}vel}, \citenamefont {Ozer},\ and\ \citenamefont
  {Uzuntarla}}]{Calim2018}%
  \BibitemOpen
  \bibfield  {author} {\bibinfo {author} {\bibfnamefont {A.}~\bibnamefont
  {Calim}}, \bibinfo {author} {\bibfnamefont {P.}~\bibnamefont {H{\"{o}}vel}},
  \bibinfo {author} {\bibfnamefont {M.}~\bibnamefont {Ozer}}, \ and\ \bibinfo
  {author} {\bibfnamefont {M.}~\bibnamefont {Uzuntarla}},\ }\href
  {http://dx.doi.org/10.1103/PhysRevE.98.062217} {\bibfield  {journal}
  {\bibinfo  {journal} {Phys. Rev. E}\ }\textbf {\bibinfo {volume} {98}},\
  \bibinfo {pages} {062217} (\bibinfo {year} {2018})}\BibitemShut {NoStop}%
\bibitem [{\citenamefont {Andrzejak}\ \emph {et~al.}(2016)\citenamefont
  {Andrzejak}, \citenamefont {Rummel}, \citenamefont {Mormann},\ and\
  \citenamefont {Schindler}}]{Andrzejak2016}%
  \BibitemOpen
  \bibfield  {author} {\bibinfo {author} {\bibfnamefont {R.~G.}\ \bibnamefont
  {Andrzejak}}, \bibinfo {author} {\bibfnamefont {C.}~\bibnamefont {Rummel}},
  \bibinfo {author} {\bibfnamefont {F.}~\bibnamefont {Mormann}}, \ and\
  \bibinfo {author} {\bibfnamefont {K.}~\bibnamefont {Schindler}},\ }\href
  {https://www.nature.com/articles/srep23000} {\bibfield  {journal} {\bibinfo
  {journal} {Sci. Rep.}\ }\textbf {\bibinfo {volume} {6}},\ \bibinfo {pages}
  {23000} (\bibinfo {year} {2016})}\BibitemShut {NoStop}%
\bibitem [{\citenamefont {Hizanidis}\ \emph {et~al.}(2016)\citenamefont
  {Hizanidis}, \citenamefont {Kouvaris}, \citenamefont {Zamora-L{\'o}pez},
  \citenamefont {D{\'\i}az-Guilera},\ and\ \citenamefont
  {Antonopoulos}}]{Hizanidis2016}%
  \BibitemOpen
  \bibfield  {author} {\bibinfo {author} {\bibfnamefont {J.}~\bibnamefont
  {Hizanidis}}, \bibinfo {author} {\bibfnamefont {N.~E.}\ \bibnamefont
  {Kouvaris}}, \bibinfo {author} {\bibfnamefont {G.}~\bibnamefont
  {Zamora-L{\'o}pez}}, \bibinfo {author} {\bibfnamefont {A.}~\bibnamefont
  {D{\'\i}az-Guilera}}, \ and\ \bibinfo {author} {\bibfnamefont {C.~G.}\
  \bibnamefont {Antonopoulos}},\ }\href
  {https://www.nature.com/articles/srep19845} {\bibfield  {journal} {\bibinfo
  {journal} {Sci. Rep.}\ }\textbf {\bibinfo {volume} {6}},\ \bibinfo {pages}
  {19845} (\bibinfo {year} {2016})}\BibitemShut {NoStop}%
\bibitem [{\citenamefont {Saha}\ \emph {et~al.}(2019)\citenamefont {Saha},
  \citenamefont {Bairagi},\ and\ \citenamefont {Dana}}]{Saha2019}%
  \BibitemOpen
  \bibfield  {author} {\bibinfo {author} {\bibfnamefont {S.}~\bibnamefont
  {Saha}}, \bibinfo {author} {\bibfnamefont {N.}~\bibnamefont {Bairagi}}, \
  and\ \bibinfo {author} {\bibfnamefont {S.~K.}\ \bibnamefont {Dana}},\ }\href
  {https://doi.org/10.3389/fams.2019.00015} {\bibfield  {journal} {\bibinfo
  {journal} {Front. Appl. Math. Stat.}\ }\textbf {\bibinfo {volume} {5}},\
  \bibinfo {pages} {15} (\bibinfo {year} {2019})}\BibitemShut {NoStop}%
\bibitem [{\citenamefont {Kemeth}\ \emph {et~al.}(2016)\citenamefont {Kemeth},
  \citenamefont {Haugland}, \citenamefont {Schmidt}, \citenamefont
  {Kevrekidis},\ and\ \citenamefont {Krischer}}]{Kemeth2016}%
  \BibitemOpen
  \bibfield  {author} {\bibinfo {author} {\bibfnamefont {F.~P.}\ \bibnamefont
  {Kemeth}}, \bibinfo {author} {\bibfnamefont {S.~W.}\ \bibnamefont
  {Haugland}}, \bibinfo {author} {\bibfnamefont {L.}~\bibnamefont {Schmidt}},
  \bibinfo {author} {\bibfnamefont {I.~G.}\ \bibnamefont {Kevrekidis}}, \ and\
  \bibinfo {author} {\bibfnamefont {K.}~\bibnamefont {Krischer}},\ }\href
  {https://aip.scitation.org/doi/abs/10.1063/1.4959804} {\bibfield  {journal}
  {\bibinfo  {journal} {Chaos}\ }\textbf {\bibinfo {volume} {26}},\ \bibinfo
  {pages} {094815} (\bibinfo {year} {2016})}\BibitemShut {NoStop}%
\bibitem [{\citenamefont {Maistrenko}\ \emph {et~al.}(2017)\citenamefont
  {Maistrenko}, \citenamefont {Brezetsky}, \citenamefont {Jaros}, \citenamefont
  {Levchenko},\ and\ \citenamefont {Kapitaniak}}]{Maistrenko2017}%
  \BibitemOpen
  \bibfield  {author} {\bibinfo {author} {\bibfnamefont {Y.}~\bibnamefont
  {Maistrenko}}, \bibinfo {author} {\bibfnamefont {S.}~\bibnamefont
  {Brezetsky}}, \bibinfo {author} {\bibfnamefont {P.}~\bibnamefont {Jaros}},
  \bibinfo {author} {\bibfnamefont {R.}~\bibnamefont {Levchenko}}, \ and\
  \bibinfo {author} {\bibfnamefont {T.}~\bibnamefont {Kapitaniak}},\ }\href
  {https://journals.aps.org/pre/abstract/10.1103/PhysRevE.95.010203} {\bibfield
   {journal} {\bibinfo  {journal} {Phys. Rev. E}\ }\textbf {\bibinfo {volume}
  {95}},\ \bibinfo {pages} {010203(R)} (\bibinfo {year} {2017})}\BibitemShut
  {NoStop}%
\bibitem [{\citenamefont {Bera}\ \emph {et~al.}(2017)\citenamefont {Bera},
  \citenamefont {Majhi}, \citenamefont {Ghosh},\ and\ \citenamefont
  {Perc}}]{bera2017}%
  \BibitemOpen
  \bibfield  {author} {\bibinfo {author} {\bibfnamefont {B.~K.}\ \bibnamefont
  {Bera}}, \bibinfo {author} {\bibfnamefont {S.}~\bibnamefont {Majhi}},
  \bibinfo {author} {\bibfnamefont {D.}~\bibnamefont {Ghosh}}, \ and\ \bibinfo
  {author} {\bibfnamefont {M.}~\bibnamefont {Perc}},\ }\href
  {http://iopscience.iop.org/article/10.1209/0295-5075/118/10001/meta}
  {\bibfield  {journal} {\bibinfo  {journal} {EPL}\ }\textbf {\bibinfo {volume}
  {118}},\ \bibinfo {pages} {10001} (\bibinfo {year} {2017})}\BibitemShut
  {NoStop}%
\bibitem [{\citenamefont {Sethia}\ and\ \citenamefont
  {Sen}(2014)}]{Sethia2014}%
  \BibitemOpen
  \bibfield  {author} {\bibinfo {author} {\bibfnamefont {G.~C.}\ \bibnamefont
  {Sethia}}\ and\ \bibinfo {author} {\bibfnamefont {A.}~\bibnamefont {Sen}},\
  }\href {\doibase 10.1103/PhysRevLett.112.144101} {\bibfield  {journal}
  {\bibinfo  {journal} {Phys. Rev. Lett.}\ }\textbf {\bibinfo {volume} {112}},\
  \bibinfo {pages} {144101} (\bibinfo {year} {2014})}\BibitemShut {NoStop}%
\bibitem [{\citenamefont {Laing}(2015)}]{Laing2015}%
  \BibitemOpen
  \bibfield  {author} {\bibinfo {author} {\bibfnamefont {C.~R.}\ \bibnamefont
  {Laing}},\ }\href {\doibase 10.1103/PhysRevE.92.050904} {\bibfield  {journal}
  {\bibinfo  {journal} {Phys. Rev. E}\ }\textbf {\bibinfo {volume} {92}},\
  \bibinfo {pages} {050904(R)} (\bibinfo {year} {2015})}\BibitemShut {NoStop}%
\bibitem [{\citenamefont {Yeldesbay}\ \emph {et~al.}(2014)\citenamefont
  {Yeldesbay}, \citenamefont {Pikovsky},\ and\ \citenamefont
  {Rosenblum}}]{Yeldesbay2014}%
  \BibitemOpen
  \bibfield  {author} {\bibinfo {author} {\bibfnamefont {A.}~\bibnamefont
  {Yeldesbay}}, \bibinfo {author} {\bibfnamefont {A.}~\bibnamefont {Pikovsky}},
  \ and\ \bibinfo {author} {\bibfnamefont {M.}~\bibnamefont {Rosenblum}},\
  }\href {\doibase 10.1103/PhysRevLett.112.144103} {\bibfield  {journal}
  {\bibinfo  {journal} {Phys. Rev. Lett.}\ }\textbf {\bibinfo {volume} {112}},\
  \bibinfo {pages} {144103} (\bibinfo {year} {2014})}\BibitemShut {NoStop}%
\bibitem [{\citenamefont {Ashwin}\ and\ \citenamefont
  {Burylko}(2015)}]{Ashwin2015}%
  \BibitemOpen
  \bibfield  {author} {\bibinfo {author} {\bibfnamefont {P.}~\bibnamefont
  {Ashwin}}\ and\ \bibinfo {author} {\bibfnamefont {O.}~\bibnamefont
  {Burylko}},\ }\href {\doibase 10.1063/1.4905197} {\bibfield  {journal}
  {\bibinfo  {journal} {Chaos}\ }\textbf {\bibinfo {volume} {25}},\ \bibinfo
  {pages} {013106} (\bibinfo {year} {2015})}\BibitemShut {NoStop}%
\bibitem [{Note7()}]{Note7}%
  \BibitemOpen
  \bibinfo {note} {Note that {$\delimiter "426830A A \delimiter "526930B
  _{\protect \text {SBM}}$} is the mean adjacency matrix of a SBM for which
  self-loops are allowed. Indeed, a node has a probability $p_{\protect \text
  {in}}$ to have a link with itself.}\BibitemShut {Stop}%
\bibitem [{Note8()}]{Note8}%
  \BibitemOpen
  \bibinfo {note} {Indeed, this choice gives the same differential equations as
  if we had made the substitution $\theta _j = \phi _j + \omega t$, where $\phi
  _j$ is the phase variable in the center of mass referential.}\BibitemShut
  {Stop}%
\bibitem [{Note9()}]{Note9}%
  \BibitemOpen
  \bibinfo {note} {Our analysis is different from previous studies where the
  bifurcation diagram for the parameters $\Delta $ and $\alpha $ is
  investigated by imposing that the sum of the in and out coupling values is
  equal one \cite {Abrams2008, Kotwal2017, Martens2010}. In our approach, this
  choice is equivalent to fixing the density at $\rho = 0.5$.}\BibitemShut
  {Stop}%
\bibitem [{\citenamefont {D'Souza}\ \emph {et~al.}(2019)\citenamefont
  {D'Souza}, \citenamefont {G{\'{o}}mez-Garde{\~{n}}es}, \citenamefont
  {Nagler},\ and\ \citenamefont {Arenas}}]{DSouza2019}%
  \BibitemOpen
  \bibfield  {author} {\bibinfo {author} {\bibfnamefont {R.~M.}\ \bibnamefont
  {D'Souza}}, \bibinfo {author} {\bibfnamefont {J.}~\bibnamefont
  {G{\'{o}}mez-Garde{\~{n}}es}}, \bibinfo {author} {\bibfnamefont
  {J.}~\bibnamefont {Nagler}}, \ and\ \bibinfo {author} {\bibfnamefont
  {A.}~\bibnamefont {Arenas}},\ }\href
  {https://www.tandfonline.com/doi/abs/10.1080/00018732.2019.1650450}
  {\bibfield  {journal} {\bibinfo  {journal} {Adv. Phys.}\ }\textbf {\bibinfo
  {volume} {68}} (\bibinfo {year} {2019})}\BibitemShut {NoStop}%
\bibitem [{\citenamefont {Boccaletti}\ \emph {et~al.}(2016)\citenamefont
  {Boccaletti}, \citenamefont {Almendral}, \citenamefont {Guan}, \citenamefont
  {Leyva}, \citenamefont {Liu}, \citenamefont {Sendi{\~{n}}a-Nadal},
  \citenamefont {Wang},\ and\ \citenamefont {Zou}}]{Boccaletti2016}%
  \BibitemOpen
  \bibfield  {author} {\bibinfo {author} {\bibfnamefont {S.}~\bibnamefont
  {Boccaletti}}, \bibinfo {author} {\bibfnamefont {J.~A.}\ \bibnamefont
  {Almendral}}, \bibinfo {author} {\bibfnamefont {S.}~\bibnamefont {Guan}},
  \bibinfo {author} {\bibfnamefont {I.}~\bibnamefont {Leyva}}, \bibinfo
  {author} {\bibfnamefont {Z.}~\bibnamefont {Liu}}, \bibinfo {author}
  {\bibfnamefont {I.}~\bibnamefont {Sendi{\~{n}}a-Nadal}}, \bibinfo {author}
  {\bibfnamefont {Z.}~\bibnamefont {Wang}}, \ and\ \bibinfo {author}
  {\bibfnamefont {Y.}~\bibnamefont {Zou}},\ }\href {\doibase
  10.1016/j.physrep.2016.10.004} {\bibfield  {journal} {\bibinfo  {journal}
  {Phys. Rep.}\ }\textbf {\bibinfo {volume} {660}},\ \bibinfo {pages} {1}
  (\bibinfo {year} {2016})}\BibitemShut {NoStop}%
\bibitem [{\citenamefont {Vlasov}\ \emph {et~al.}(2015)\citenamefont {Vlasov},
  \citenamefont {Zou},\ and\ \citenamefont {Pereira}}]{Vlasov2015}%
  \BibitemOpen
  \bibfield  {author} {\bibinfo {author} {\bibfnamefont {V.}~\bibnamefont
  {Vlasov}}, \bibinfo {author} {\bibfnamefont {Y.}~\bibnamefont {Zou}}, \ and\
  \bibinfo {author} {\bibfnamefont {T.}~\bibnamefont {Pereira}},\ }\href
  {\doibase 10.1103/PhysRevE.92.012904} {\bibfield  {journal} {\bibinfo
  {journal} {Phys. Rev. E}\ }\textbf {\bibinfo {volume} {92}},\ \bibinfo
  {pages} {012904} (\bibinfo {year} {2015})}\BibitemShut {NoStop}%
\bibitem [{\citenamefont {G{\'{o}}mez-Garde{\~{n}}es}\ \emph
  {et~al.}(2011)\citenamefont {G{\'{o}}mez-Garde{\~{n}}es}, \citenamefont
  {Gomez}, \citenamefont {Arenas},\ and\ \citenamefont
  {Moreno}}]{Gomez-Gardenes2011}%
  \BibitemOpen
  \bibfield  {author} {\bibinfo {author} {\bibfnamefont {J.}~\bibnamefont
  {G{\'{o}}mez-Garde{\~{n}}es}}, \bibinfo {author} {\bibfnamefont
  {S.}~\bibnamefont {Gomez}}, \bibinfo {author} {\bibfnamefont
  {A.}~\bibnamefont {Arenas}}, \ and\ \bibinfo {author} {\bibfnamefont
  {Y.}~\bibnamefont {Moreno}},\ }\href {\doibase
  10.1103/PhysRevLett.106.128701} {\bibfield  {journal} {\bibinfo  {journal}
  {Phys. Rev. Lett.}\ }\textbf {\bibinfo {volume} {106}},\ \bibinfo {pages}
  {128701} (\bibinfo {year} {2011})}\BibitemShut {NoStop}%
\bibitem [{\citenamefont {Zou}\ \emph {et~al.}(2014)\citenamefont {Zou},
  \citenamefont {Pereira}, \citenamefont {Small}, \citenamefont {Liu},\ and\
  \citenamefont {Kurths}}]{Zou2014}%
  \BibitemOpen
  \bibfield  {author} {\bibinfo {author} {\bibfnamefont {Y.}~\bibnamefont
  {Zou}}, \bibinfo {author} {\bibfnamefont {T.}~\bibnamefont {Pereira}},
  \bibinfo {author} {\bibfnamefont {M.}~\bibnamefont {Small}}, \bibinfo
  {author} {\bibfnamefont {Z.}~\bibnamefont {Liu}}, \ and\ \bibinfo {author}
  {\bibfnamefont {J.}~\bibnamefont {Kurths}},\ }\href {\doibase
  10.1103/PhysRevLett.112.114102} {\bibfield  {journal} {\bibinfo  {journal}
  {Phys. Rev. Lett.}\ }\textbf {\bibinfo {volume} {112}},\ \bibinfo {pages}
  {114102} (\bibinfo {year} {2014})}\BibitemShut {NoStop}%
\bibitem [{\citenamefont {Gao}\ \emph {et~al.}(2016{\natexlab{b}})\citenamefont
  {Gao}, \citenamefont {Xu}, \citenamefont {Sun},\ and\ \citenamefont
  {Zheng}}]{Gao2016synchro}%
  \BibitemOpen
  \bibfield  {author} {\bibinfo {author} {\bibfnamefont {J.}~\bibnamefont
  {Gao}}, \bibinfo {author} {\bibfnamefont {C.}~\bibnamefont {Xu}}, \bibinfo
  {author} {\bibfnamefont {Y.}~\bibnamefont {Sun}}, \ and\ \bibinfo {author}
  {\bibfnamefont {Z.}~\bibnamefont {Zheng}},\ }\href {\doibase
  10.1038/srep30184} {\bibfield  {journal} {\bibinfo  {journal} {Sci. Rep.}\
  }\textbf {\bibinfo {volume} {6}},\ \bibinfo {pages} {30184} (\bibinfo {year}
  {2016}{\natexlab{b}})}\BibitemShut {NoStop}%
\bibitem [{Note10()}]{Note10}%
  \BibitemOpen
  \bibinfo {note} {The equation of Ref.~\cite {Chen2017} that corresponds to
  our Eq.~\protect \textup {\hbox {\mathsurround \z@ \protect \normalfont
  (\ignorespaces \ref {kuramoto_star_R}\unskip \@@italiccorr )}} contains the
  factor $\protect \qopname \relax o{cos}(\Phi +\alpha )$, instead of $\protect
  \qopname \relax o{cos}(\Phi - \alpha )$, which is the correct
  factor.}\BibitemShut {Stop}%
\bibitem [{Note11()}]{Note11}%
  \BibitemOpen
  \bibinfo {note} {There is a non-trivial part in the procedure where we must
  find $\Phi $ such that $A\protect \qopname \relax o{sin}\Phi + B\protect
  \qopname \relax o{cos}\Phi = C$. To solve the equation, we divide both sides
  by $D = \protect \sqrt {A^2 + B^2}$ and define an angle $\Theta = \protect
  \qopname \relax o{arccos}(A/D)$. Hence, $\protect \qopname \relax
  o{cos}^2\Theta + \protect \qopname \relax o{sin}^2\Theta = 1$. Moreover,
  $\protect \qopname \relax o{cos}\Theta \protect \qopname \relax o{sin}\Phi +
  \protect \qopname \relax o{sin}\Theta \protect \qopname \relax o{cos}\Phi =
  \protect \qopname \relax o{sin}(\Phi + \Theta ) = C/D$. The last equality
  then allows to express $\Phi $ in terms of $A$, $B$, and $C$.}\BibitemShut
  {Stop}%
\bibitem [{\citenamefont {Huang}\ \emph {et~al.}(2016)\citenamefont {Huang},
  \citenamefont {Gao}, \citenamefont {Sun}, \citenamefont {Zheng},\ and\
  \citenamefont {Xu}}]{Huang2016}%
  \BibitemOpen
  \bibfield  {author} {\bibinfo {author} {\bibfnamefont {X.}~\bibnamefont
  {Huang}}, \bibinfo {author} {\bibfnamefont {J.}~\bibnamefont {Gao}}, \bibinfo
  {author} {\bibfnamefont {Y.}~\bibnamefont {Sun}}, \bibinfo {author}
  {\bibfnamefont {Z.}~\bibnamefont {Zheng}}, \ and\ \bibinfo {author}
  {\bibfnamefont {C.}~\bibnamefont {Xu}},\ }\href {\doibase
  10.1007/s11467-016-0597-y} {\bibfield  {journal} {\bibinfo  {journal} {Front.
  Phys.}\ }\textbf {\bibinfo {volume} {11}},\ \bibinfo {pages} {110504}
  (\bibinfo {year} {2016})}\BibitemShut {NoStop}%
\bibitem [{\citenamefont {Pan}\ \emph {et~al.}(2020)\citenamefont {Pan},
  \citenamefont {Yang}, \citenamefont {Wang}, \citenamefont {Cai},
  \citenamefont {Zhou},\ and\ \citenamefont {Lai}}]{Pan2020}%
  \BibitemOpen
  \bibfield  {author} {\bibinfo {author} {\bibfnamefont {L.}~\bibnamefont
  {Pan}}, \bibinfo {author} {\bibfnamefont {D.}~\bibnamefont {Yang}}, \bibinfo
  {author} {\bibfnamefont {W.}~\bibnamefont {Wang}}, \bibinfo {author}
  {\bibfnamefont {S.}~\bibnamefont {Cai}}, \bibinfo {author} {\bibfnamefont
  {T.}~\bibnamefont {Zhou}}, \ and\ \bibinfo {author} {\bibfnamefont {Y.-C.}\
  \bibnamefont {Lai}},\ }\href {\doibase 10.1103/physrevresearch.2.023233}
  {\bibfield  {journal} {\bibinfo  {journal} {Phys. Rev. Res.}\ }\textbf
  {\bibinfo {volume} {2}},\ \bibinfo {pages} {023233} (\bibinfo {year}
  {2020})}\BibitemShut {NoStop}%
\bibitem [{\citenamefont {Strogatz}(2018)}]{Strogatz2018a}%
  \BibitemOpen
  \bibfield  {author} {\bibinfo {author} {\bibfnamefont {S.~H.}\ \bibnamefont
  {Strogatz}},\ }\href {\doibase 10.1007/s11356-018-1771-2} {\emph {\bibinfo
  {title} {{Nonlinear Dynamics and Chaos}}}}\ (\bibinfo  {publisher} {CRC
  Press},\ \bibinfo {year} {2018})\BibitemShut {NoStop}%
\bibitem [{\citenamefont {Osipov}\ \emph {et~al.}(2007)\citenamefont {Osipov},
  \citenamefont {Kurths},\ and\ \citenamefont {Zhou}}]{Osipov2007}%
  \BibitemOpen
  \bibfield  {author} {\bibinfo {author} {\bibfnamefont {G.~V.}\ \bibnamefont
  {Osipov}}, \bibinfo {author} {\bibfnamefont {J.}~\bibnamefont {Kurths}}, \
  and\ \bibinfo {author} {\bibfnamefont {C.}~\bibnamefont {Zhou}},\ }\href@noop
  {} {\emph {\bibinfo {title} {{Synchronization in Oscillatory Networks}}}}\
  (\bibinfo  {publisher} {Springer},\ \bibinfo {year} {2007})\BibitemShut
  {NoStop}%
\bibitem [{Note12()}]{Note12}%
  \BibitemOpen
  \bibinfo {note} {One could still be interested in the reduced dynamics of
  linear observables and get good results in some cases.}\BibitemShut {Stop}%
\bibitem [{\citenamefont {Thibeault}(2020)}]{Thibeault2020}%
  \BibitemOpen
  \bibfield  {author} {\bibinfo {author} {\bibfnamefont {V.}~\bibnamefont
  {Thibeault}},\ }\emph {\bibinfo {title} {{R{\'{e}}duire la dimension des
  syst{\`{e}}mes complexes : un regard sur l'{\'{e}}mergence de la
  synchronisation}}},\ \href@noop {} {Master's thesis},\ \bibinfo  {school}
  {Universit{\'{e}} Laval} (\bibinfo {year} {2020})\BibitemShut {NoStop}%
\bibitem [{\citenamefont {Penrose}(1955)}]{Penrose1955}%
  \BibitemOpen
  \bibfield  {author} {\bibinfo {author} {\bibfnamefont {R.}~\bibnamefont
  {Penrose}},\ }\href {https://doi.org/10.1017/S0305004100030401} {\bibfield
  {journal} {\bibinfo  {journal} {Math. Proc. Camb. Philos. Soc.}\ }\textbf
  {\bibinfo {volume} {51}},\ \bibinfo {pages} {406} (\bibinfo {year}
  {1955})}\BibitemShut {NoStop}%
\bibitem [{\citenamefont {Ding}\ \emph {et~al.}(2010)\citenamefont {Ding},
  \citenamefont {Li},\ and\ \citenamefont {Jordan}}]{Ding2010}%
  \BibitemOpen
  \bibfield  {author} {\bibinfo {author} {\bibfnamefont {C.}~\bibnamefont
  {Ding}}, \bibinfo {author} {\bibfnamefont {T.}~\bibnamefont {Li}}, \ and\
  \bibinfo {author} {\bibfnamefont {M.~I.}\ \bibnamefont {Jordan}},\ }\href
  {\doibase 10.1109/TPAMI.2008.277} {\bibfield  {journal} {\bibinfo  {journal}
  {IEEE Trans. Pattern Anal. Mach. Intell.}\ }\textbf {\bibinfo {volume}
  {32}},\ \bibinfo {pages} {45} (\bibinfo {year} {2010})}\BibitemShut {NoStop}%
\bibitem [{\citenamefont {Aggarwal}\ and\ \citenamefont
  {Reddy}(2014)}]{Aggarwal2013}%
  \BibitemOpen
  \bibfield  {author} {\bibinfo {author} {\bibfnamefont {C.~C.}\ \bibnamefont
  {Aggarwal}}\ and\ \bibinfo {author} {\bibfnamefont {C.~K.}\ \bibnamefont
  {Reddy}},\ }\href@noop {} {\emph {\bibinfo {title} {Data Clustering:
  Algorithms and Applications}}}\ (\bibinfo  {publisher} {Chapman {\&}
  Hall/CRC},\ \bibinfo {year} {2014})\ p.\ \bibinfo {pages} {652}\BibitemShut
  {NoStop}%
\bibitem [{\citenamefont {Ding}\ \emph {et~al.}(2006)\citenamefont {Ding},
  \citenamefont {Li}, \citenamefont {Peng},\ and\ \citenamefont
  {Park}}]{Ding2006}%
  \BibitemOpen
  \bibfield  {author} {\bibinfo {author} {\bibfnamefont {C.}~\bibnamefont
  {Ding}}, \bibinfo {author} {\bibfnamefont {T.}~\bibnamefont {Li}}, \bibinfo
  {author} {\bibfnamefont {W.}~\bibnamefont {Peng}}, \ and\ \bibinfo {author}
  {\bibfnamefont {H.}~\bibnamefont {Park}},\ }\href {\doibase
  10.1145/1150402.1150420} {\bibfield  {journal} {\bibinfo  {journal} {Proc.
  12th ACM SIGKDD Int. Conf. Knowl. Discov. data Min.~}\ ,\ \bibinfo {pages}
  {126}} (\bibinfo {year} {2006})}\BibitemShut {NoStop}%
\bibitem [{\citenamefont {Martens}\ \emph {et~al.}(2016)\citenamefont
  {Martens}, \citenamefont {Panaggio},\ and\ \citenamefont
  {Abrams}}]{martens2016}%
  \BibitemOpen
  \bibfield  {author} {\bibinfo {author} {\bibfnamefont {E.~A.}\ \bibnamefont
  {Martens}}, \bibinfo {author} {\bibfnamefont {M.~J.}\ \bibnamefont
  {Panaggio}}, \ and\ \bibinfo {author} {\bibfnamefont {D.~M.}\ \bibnamefont
  {Abrams}},\ }\href
  {http://iopscience.iop.org/article/10.1088/1367-2630/18/2/022002/meta}
  {\bibfield  {journal} {\bibinfo  {journal} {New. J. Phys.}\ }\textbf
  {\bibinfo {volume} {18}},\ \bibinfo {pages} {022002} (\bibinfo {year}
  {2016})}\BibitemShut {NoStop}%
\bibitem [{\citenamefont {Fiedler}(1973)}]{Fiedler1973}%
  \BibitemOpen
  \bibfield  {author} {\bibinfo {author} {\bibfnamefont {M.}~\bibnamefont
  {Fiedler}},\ }\href@noop {} {\bibfield  {journal} {\bibinfo  {journal}
  {Czechoslov. Math. J.}\ }\textbf {\bibinfo {volume} {23}},\ \bibinfo {pages}
  {298} (\bibinfo {year} {1973})}\BibitemShut {NoStop}%
\bibitem [{\citenamefont {Hall}(1970)}]{Hall1970}%
  \BibitemOpen
  \bibfield  {author} {\bibinfo {author} {\bibfnamefont {K.~M.}\ \bibnamefont
  {Hall}},\ }\href {\doibase 10.1287/mnsc.17.3.219} {\bibfield  {journal}
  {\bibinfo  {journal} {Manag. Sci.}\ }\textbf {\bibinfo {volume} {17}},\
  \bibinfo {pages} {219} (\bibinfo {year} {1970})}\BibitemShut {NoStop}%
\bibitem [{\citenamefont {Newman}(2006)}]{Newman2006}%
  \BibitemOpen
  \bibfield  {author} {\bibinfo {author} {\bibfnamefont {M.~E.~J.}\
  \bibnamefont {Newman}},\ }\href {\doibase 10.1103/PhysRevE.74.036104}
  {\bibfield  {journal} {\bibinfo  {journal} {Phys. Rev. E}\ }\textbf {\bibinfo
  {volume} {74}},\ \bibinfo {pages} {036104} (\bibinfo {year}
  {2006})}\BibitemShut {NoStop}%
\bibitem [{\citenamefont {Nicolaou}\ \emph {et~al.}(2019)\citenamefont
  {Nicolaou}, \citenamefont {Eroglu},\ and\ \citenamefont
  {Motter}}]{Nicolaou2019a}%
  \BibitemOpen
  \bibfield  {author} {\bibinfo {author} {\bibfnamefont {Z.~G.}\ \bibnamefont
  {Nicolaou}}, \bibinfo {author} {\bibfnamefont {D.}~\bibnamefont {Eroglu}}, \
  and\ \bibinfo {author} {\bibfnamefont {A.~E.}\ \bibnamefont {Motter}},\
  }\href {\doibase 10.1103/PhysRevX.9.011017} {\bibfield  {journal} {\bibinfo
  {journal} {Phys. Rev. X}\ }\textbf {\bibinfo {volume} {9}},\ \bibinfo {pages}
  {11017} (\bibinfo {year} {2019})}\BibitemShut {NoStop}%
\bibitem [{\citenamefont {Peron}\ \emph {et~al.}(2020)\citenamefont {Peron},
  \citenamefont {Eroglu}, \citenamefont {Rodrigues},\ and\ \citenamefont
  {Moreno}}]{Peron2020}%
  \BibitemOpen
  \bibfield  {author} {\bibinfo {author} {\bibfnamefont {T.}~\bibnamefont
  {Peron}}, \bibinfo {author} {\bibfnamefont {D.}~\bibnamefont {Eroglu}},
  \bibinfo {author} {\bibfnamefont {F.~A.}\ \bibnamefont {Rodrigues}}, \ and\
  \bibinfo {author} {\bibfnamefont {Y.}~\bibnamefont {Moreno}},\ }\href
  {\doibase 10.1103/physrevresearch.2.013255} {\bibfield  {journal} {\bibinfo
  {journal} {Phys. Rev. Res.}\ }\textbf {\bibinfo {volume} {2}},\ \bibinfo
  {pages} {013255} (\bibinfo {year} {2020})}\BibitemShut {NoStop}%
\bibitem [{\citenamefont {Martens}(2010)}]{Martens2010}%
  \BibitemOpen
  \bibfield  {author} {\bibinfo {author} {\bibfnamefont {E.~A.}\ \bibnamefont
  {Martens}},\ }\href {\doibase 10.1063/1.3499502} {\bibfield  {journal}
  {\bibinfo  {journal} {Chaos}\ }\textbf {\bibinfo {volume} {20}},\ \bibinfo
  {pages} {043122} (\bibinfo {year} {2010})}\BibitemShut {NoStop}%
\end{thebibliography}

%

\end{document}